\documentclass[11pt]{article}
\pdfoutput=1
\usepackage{jheppub}
\makeatletter
\def\@fpheader{\relax}
\makeatother
\usepackage{amssymb,amsfonts,amsmath,amsthm,lineno}
\usepackage{enumerate}
\usepackage{mathrsfs}
\usepackage{slashed}
\usepackage{pgfplots}
\pgfplotsset{compat=1.15}
\usepgfplotslibrary{fillbetween}
\usepackage{bbold}
\usepackage{verbatim}
\usepackage{xargs}
\usepackage{hyperref}
\usepackage{tikz, adjustbox}
\usepackage{microtype}

\makeatletter
\newcommand{\Vast}{\bBigg@{4.75}}
\makeatother

\newcommand{\be}{\begin{equation}}
\newcommand{\ee}{\end{equation}}
\newcommand{\bea}{\begin{eqnarray}}
\newcommand{\eea}{\end{eqnarray}}

\newcommand{\CA}{\mathcal{A}}

\newcommand{\CE}{\mathcal{E}}
\newcommand{\CF}{\mathcal{F}}

\newcommand{\CL}{\mathcal{L}}
\newcommand{\CN}{\mathcal{N}}

\newcommand{\CO}{\mathcal{O}}
\newcommand{\CP}{\mathcal{P}}

\newcommand{\CR}{\mathcal{R}}
\newcommand{\CT}{\mathcal{T}}

\newcommand{\lr}{\left (}
\newcommand{\rr}{\right )}
\newcommand{\ls}{\left [}
\newcommand{\rs}{\right ]}

\newcommand\qt\tau

\newcommand{\p}{\partial}

\renewcommand{\tilde}[1]{\widetilde{#1}}
\newcommand{\tr}{\text{tr}}
\renewcommand{\emph}[1]{\textit{#1}}

\newcommand{\wa}{\alpha}
\newcommand{\wb}{\beta} 
\newcommand{\wc}{\gamma}
\newcommand{\wvd}{\delta} 

\makeatletter
\renewcommand{\@seccntformat}[1]{\csname the#1\endcsname.\,\,}
\makeatother
\allowdisplaybreaks

\let \savenumberline \numberline
\def \numberline#1{\savenumberline{#1.}}

\usetikzlibrary{shapes,arrows,chains}
\usetikzlibrary{decorations.markings}
\usetikzlibrary{decorations.pathmorphing}
\usetikzlibrary{shapes.multipart}
\usetikzlibrary{positioning}
\tikzset{snake it/.style={decorate, decoration=snake}}

\newcommand{\dd}{\mathrm{d}}

\definecolor{vub}{RGB}{0,52,154}
\definecolor{vubo}{RGB}{255,102,0}
\definecolor{redd}{RGB}{255,40,40}
\definecolor{r}{RGB}{228,32,20}
\definecolor{o}{RGB}{238,69,4}
\definecolor{y}{RGB}{253,228,1}
\definecolor{g}{RGB}{108,160,0}
\definecolor{b}{RGB}{0,162,203}
\definecolor{i}{RGB}{120,42,117}
\definecolor{vred}{rgb}{0.78, 0.03, 0.08}
\definecolor{DarkRed}{rgb}{0.75, 0.2, 0}
\definecolor{DarkBlue}{rgb}{0, 0.2, 0.75}
\definecolor{LightGray}{gray}{0.8}
\definecolor{olivegreen}{rgb}{0.2, 0.6, 0.0}

\newcommand{\lambdap}{\boldsymbol{t}}
\newcommand{\lbb}{\tilde{\ell}}

\title{Matrix Theory Reloaded: A BPS Road to Holography}
\author[a]{Chris D. A. Blair,}
\author[b]{Johannes Lahnsteiner,}
\author[b,c]{Niels A. Obers,}
\author[b]{and Ziqi Yan\medskip}
\emailAdd{c.blair@csic.es}
\emailAdd{j.m.lahnsteiner@outlook.com}
\emailAdd{obers@nbi.ku.dk}
\emailAdd{ziqi.yan@su.se}
\affiliation[a]{Instituto de F\'{i}sica Te\'{o}rica UAM/CSIC, Universidad Aut\'{o}noma de Madrid, Cantoblanco, Madrid 28049, Spain \smallskip}
\affiliation[b]{Nordita, KTH Royal Institute of Technology and Stockholm University,
Hannes Alfv\'{e}ns v\"{a}g 12, SE-106 91 Stockholm, Sweden \smallskip}
\affiliation[c]{Niels Bohr International Academy, The Niels Bohr Institute, Copenhagen University, Blegdamsvej 17,
DK-2100 Copenhagen \O, Denmark}

\preprint{IFT-UAM/CSIC-24-139, NORDITA 2024-033}

\abstract{
We revisit the decoupling limits that lead to matrix theories on D-branes.
We highlight the BPS nature of these limits, in which the target space geometry becomes non-Lorentzian and wrapped D-branes experience instantaneous gravitational forces.
Applied to curved D-brane geometries, we show that a single BPS decoupling limit
induces the bulk near-horizon limit leading to AdS/CFT. 
By consecutively applying two such limits, we systematically generate further examples of holography, including novel versions with non-Lorentzian bulk geometry.
Uplifted to M-theory, we are led to a unified framework where each BPS decoupling limit corresponds to a Discrete Light Cone Quantisation (DLCQ).
We conjecture that a DLCQ${}^n$/DLCQ${}^{m}$ correspondence, with $m>n$\,, captures the notion of holography in string theory. 
In particular, AdS${}_5$/CFT${}_4$ can be viewed as an example of DLCQ${}^0$/DLCQ${}^{1}$\,, with the extra DLCQ on the field theory side corresponding to the near-horizon limit in the bulk geometry. 
We further show that undoing these BPS decoupling limits can be viewed as deformations of matrix theories. We explain how these deformations are related to the $T \bar T$ deformation in two dimensions. 
In the context of holography, this allows us to view the ten-dimensional near-horizon brane geometry as an intrinsic deformation of the flat non-Lorentzian geometry that arises asymptotically.
In field theoretic terms, these generalisations lead to $T\bar{T}$-like flow equations for the D$p$-brane DBI action.

}

\begin{document}%%%%%%%%%%%%%%%%%%%%%%%%%%%%%%%%%%%%%%%%%%%%%%%%%%%%%%%%%%%%%%%%%%%%%%%%%%%%%%%
%%%%%%%%%%%%%%%%%%%%%%%%%%%%%%%%%%%%%%%%%%%%%%%%%%%%%%%%%%%%%%%%%%%%%%%%%%%%%%%

\maketitle
\vfill\eject

\section{Introduction}

A key breakthrough of the second superstring revolution was understanding the importance of D-branes \cite{Polchinski:1995mt}. As BPS states, D-branes are crucial for studying strong-coupling behaviour and dualities. Dual descriptions of string theory in the presence of D-branes underlie the original formulation of the AdS/CFT correspondence \cite{Maldacena:1997re}, while through the matrix theory proposal, a large $N$ limit of the D-brane worldvolume theory is conjectured to be sufficient to describe M-theory in flat spacetime~\cite{Banks:1996vh}.
Both the AdS/CFT and matrix theory conjectures make use of decoupling limits, zooming in on particular states and regimes of string and M-theory.
These examples show that decoupling limits allow us to probe surprisingly rich physics, uncovering non-perturbative aspects of string theory and therefore offering hints towards reverse engineering M-theory itself from its limits.

The AdS/CFT correspondence and matrix theory are two complementary and connected approaches to quantum gravity. 
Through the past decades, we have witnessed remarkable progress in accumulating an overwhelming amount of evidence for the AdS/CFT conjecture, and understanding its wide applications.  
    In contrast, it is fair to say that, after the initial flurry of excitement, progress on the matrix theory perspective of M-theory as originally proposed has been less far-reaching. However, in recent years, new techniques developed in the bootstrap program~\cite{Han:2020bkb, Lin:2023owt, Komatsu:2024vnb, Cho:2024kxn, Lin:2024vvg} and in the study of amplitudes~\cite{Miller:2022fvc, Tropper:2023fjr,Herderschee:2023pza,Herderschee:2023bnc} have been applied to understand the dynamics of matrix theory, while also intriguing relations to non-Lorentzian corners of string theory~\cite{Andringa:2012uz, Harmark:2017rpg, Bergshoeff:2018yvt} have been uncovered (see~\cite{Oling:2022fft, Demulder:2023bux} for reviews).
It is timely and beneficial to revisit the original decoupling limits used to propose these conjectures, as well as some of the foundational questions, from a modern perspective. 

One important question is how to generalise these decoupling limits to derive other holographic duals embedded in string theory, especially in the light of the active programs of constructing flat space and non-Lorentzian holography, which are typically approached from a bottom-up way. It is therefore tempting to ask,
\begin{itemize}

\item

\emph{\textbf{Question 1:} What are the guiding principles for mapping out self-consistent decoupling limits in string theory?}

\end{itemize}
Useful insights for this question come from how matrix theory arises from a limit of the D-brane worldvolume theory: this is a BPS decoupling limit, relying on the fact that the brane tension equals its charge, which involves fine-tuning the Ramond-Ramond (RR) gauge field coupled to the brane. Logically, such a BPS decoupling limit should be applied not only to the D-brane but also to the full string theory. 
Furthermore, we can combine several such limits, and generate new ones using dualities, to discover a series of new corners of string theory.
This logic was set out as a program to unify, classify and apply the BPS decoupling limits of string and M-theory in~\cite{Blair:2023noj, Gomis:2023eav}. A complementary perspective using the BPS mass spectrum to classify the U-dual orbits of various decoupling limits was explored in~\cite{bpslimits}.

In this paper, we flesh out the decoupling limits that are of most direct relevance to matrix theory, using the framework studied in~\cite{Blair:2023noj} and then reinforced from the string worldsheet perspective in~\cite{Gomis:2023eav}. One probably counter-intuitive finding is that the ten-dimensional target space geometry in such decoupling limits is generically \emph{non-Lorentzian}, as it develops a foliation structure adapted to the choice of D-brane that is singled out by the BPS decoupling limit. This observation might appear to be surprising at first sight, especially in the context of the AdS/CFT correspondence. In particular, as is well known, matrix quantum mechanics on a three-torus is T-dual to $\CN = 4$ super Yang-Mills (SYM) theory, which is further dual to the bulk AdS${}^{}_5 \times S^5$ geometry. This bulk geometry is of course perfectly Lorentzian. We are then led to a second question that is of imminent importance:
\begin{itemize}

\item

\emph{\textbf{Question 2:} In the context of holography, what is the role of the ten-dimensional non-Lorentzian geometry coupled to matrix theory on the D-branes?}

\end{itemize}
The answer to this question begs for a sharp understanding of the precise relation between matrix theory and the AdS/CFT correspondence. The BPS decoupling limit related to matrix theory is applied to the asymptotic infinity of the bulk D-brane geometry. We will show that, while this BPS decoupling limit leads to a ten-dimensional flat non-Lorentzian geometry at the asymptotic infinity, it is mapped to the near-horizon limit in the bulk, where the latter leads to the desired bulk AdS geometry.\,\footnote{See also related discussions in the (dual) context of non-relativistic string theory in~\cite{Danielsson:2000mu,Avila:2023aey, Guijosa:2023qym, soliton}.} We will also see that this perspective is more than just a refinement of the original Maldacena decoupling limit~\cite{Maldacena:1997re}: this careful examination of the BPS decoupling limit leads to a systematic mechanism for generating possible holographic duals that generalise the original AdS/CFT correspondence. 

Finally, it is tempting to ask whether there is any `intrinsic' way (that is independent of the BPS decoupling limits) to relate the non-Lorentzian spacetime at the asymptotic infinity to the Lorentzian bulk geometry. 
We thus pose the following further question:
\begin{itemize}

\item

\emph{\textbf{Question 3:} How is the ten-dimensional bulk geometry generated intrinsically?} 
    
\end{itemize}
We will find that the bulk geometry is generated via a D-brane generalisation of the $T\bar{T}$ deformation: namely, the ten-dimensional curved bulk geometry is generated by deforming the field theory dual in a way that is U-dual to the well-known $T\bar{T}$ deformation \cite{Zamolodchikov:2004ce,Smirnov:2016lqw} in two dimensions. 
This approach would allow us to build the bulk geometry without resorting to any limiting procedure.
Our answer to this question will also pave the way towards new generalisations of the $T\bar{T}$ deformation in other dimensions. 

In this current paper, we provide precise answers to the above foundational questions. 
In the remainder of this Introduction, we provide a summary of our approach and findings. This may be viewed as a short survey of the rather extensive contents contained in this paper, structured according to the three questions that we posed above. 
One may also wonder about how the perspective to be developed in the current paper could help us improve our understanding of the conjectured correspondence between matrix theory and M-theory or eleven-dimensional supergravity. We will discuss this in the Outlook in Section~\ref{sec:ol}, together with some further insights into holography and the $T\bar{T}$ deformation that arise from our explorations.

\subsection{Guided by a BPS Road...}

In connection to \emph{\textbf{Question 1}} regarding the guiding principle for classifying the decoupling limits in string theory, we first revisit the BPS decoupling limit of a stack of D0-branes that leads to the 
Banks-Fischler-Shenker-Susskind (BFSS) matrix quantum mechanics~\cite{deWit:1988wri, Banks:1996vh,Susskind:1997cw}.
This BPS decoupling limit should be viewed not merely as a limit of the D0-branes themselves but of the full type IIA superstring theory.
This leads to a corner of the IIA theory that we refer to as \emph{matrix 0-brane theory} (M0T). Using T-duality, we map the defining prescription for M0T to define general \emph{matrix $p$-brane theory} (M$p$T)~\cite{Blair:2023noj, Gomis:2023eav}.  

\vspace{3mm}

\noindent $\bullet$~\emph{Matrix theory in matrix $p$-brane theory.}
Historically, matrix theory was initially studied in the context of the quantisation of the eleven-dimensional supermembrane~\cite{deWit:1988wri}. It was later shown that matrix theory arises from M-theory in the infinite momentum frame \cite{Banks:1996vh} or in the Discrete Light Cone Quantisation (DLCQ)~\cite{Susskind:1997cw,Seiberg:1997ad}, and it in fact describes a second quantisation of the membranes. 
A formulation of the associated decoupling limit more directly in terms of D0-branes was provided by Sen~\cite{Sen:1997we}. A little later the decoupling limit reappears implicitly and explicitly in various `open brane' decoupling limits \emph{e.g.}~\cite{Gopakumar:2000ep, Harmark:2000ff, Hyun:2000cw} and in the `closed string/brane'  decoupling limits of~\cite{Klebanov:2000pp, Gomis:2000bd, Danielsson:2000gi, Danielsson:2000mu}.
The latter limits produce a target space which is non-Lorentzian. 
Much more recently, the nature of these string and brane non-Lorentzian geometries has been the focus of renewed interest from a variety of angles, and some of them are reviewed in~\cite{Oling:2022fft, Demulder:2023bux}.

The fundamental degrees of freedom in M$p$T are captured by D$p$-branes instead of the fundamental string, and their dynamics is described by matrix (gauge) theories~\cite{Obers:1998fb}, generalising the matrix quantum mechanics of M0T to higher dimensions. These D$p$-branes on which ($p+1$)-dimensional matrix theory lives are wrapped on a $p$-dimensional toroidal compactification. Shrinking the $p$-torus decompactifies the cycles in the T-dual frame, mapping matrix theory to BFSS matrix quantum mechanics. In this sense, the wrapped D$p$-branes behave like non-relativistic particles as in matrix quantum mechanics, experiencing instantaneous Newton-like gravitational forces. 

It might not be surprising that matrix theory, viewed as SYM 
on a toroidal compactification, exhibits non-relativistic behaviour; however, as we will emphasise, upon the decompactification of the $p$-torus in M$p$T, the ten-dimensional target space still develops a non-Lorentzian geometric structure, equipped with a codimension-$(p+1)$ foliation structure~\cite{Blair:2023noj}. We will take a target space perspective here, which is complementary to the worldsheet derivation in~\cite{Gomis:2023eav}. This target space approach has the advantage that the RR potentials, which play an essential role in the BPS decoupling limits, are much easier to access.

Note that a subtlety arises when $p > 3$\,, where the matrix theories are not perturbatively defined and one would have to resort to a (possibly S-dual) strong coupling description \cite{Rozali:1997cb, Berkooz:1997cq, Seiberg:1997ad,Obers:1997kk}. Implicitly, our discussion throughout this paper is mostly restricted to $p \leq 3$\,. Nevertheless, the discussions on the geometric aspects are independent of this field-theoretical subtlety, and they are applicable to all $p$\,. 

\vspace{3mm}
\noindent $\bullet$~\emph{Aspects of non-Lorentzian geometry.}
It is worthwhile to illustrate at this point some basic features of the non-Lorentzian geometries that appear.
Let us show how to obtain the simplest flat non-Lorentzian $p$-brane geometry appearing in M$p$T.
Start with the following flat relativistic (string frame) metric, 
\be \label{intro_limit}
    \dd s^2 = \omega \, \dd x^A \, \dd x^B \, \eta^{}_{AB} + \omega^{-1} \, \dd x^{A'} \, \dd x^{A'}\,,
\ee
where $A=0\,, \, \dots, \, p$ are longitudinal directions and $A' = p+1\,, \, \dots, \, 9$ are transverse.
We have introduced (here by rescaling the Minkowski coordinates) a dimensionless parameter $\omega$ and we take the limit $\omega \rightarrow \infty$\,. 
This breaks the ten-dimensional Lorentz symmetry.
After taking the limit, the coordinates $(x^A, \,x^{A'})$ parametrise a non-Lorentzian flat spacetime with a codimension-($p$+1) foliation structure.
In addition to local $\mathrm{SO}(1\,,\,p)$ and $\mathrm{SO}(9-p)$ symmetries acting on $x^A$ and $x^{A'}$, respectively, this spacetime admits a \emph{p-brane boost symmetry} such that
\be \label{intro_boost}
    \delta^{}_\text{G} x^A = 0 \,,
        \qquad% 
    \delta^{}_\text{G} x^{A'} = \Lambda^{A'}{}_{\!A} \, x^A \,,
\ee
which for $p=0$ is the standard Galilean boost transformation.
We then generalise this limit to curved backgrounds by upgrading the coordinate one-forms to `vielbeins', thus $\dd x^A \rightarrow \tau^{}_\mu{}^A \, \dd x^\mu$ and $\dd x^{A'} \rightarrow E^{}_\mu{}^{A'} \, \dd x^\mu$\,.
These vielbeins of the non-Lorentzian geometry replace the usual metric structure on a Lorentzian manifold. 
This vielbein structure leads to a generalised version of the Newton-Cartan geometry, where the latter refers to the geometric covariantisation of Newtonian gravity (see~\cite{Hartong:2022lsy} for a modern review).\footnote{Versions of such $p$-brane generalisations of Newton-Cartan geometries have appeared in~\cite{Kluson:2017abm,Pereniguez:2019eoq, Blair:2021waq, Ebert:2021mfu, Novosad:2021tlq, Bergshoeff:2023rkk, Ebert:2023hba,Bergshoeff:2024ipq}. In the current paper we will sometimes use the phrase `M$p$T geometry' to refer to the target space geometry in M$p$T, which is implicitly a generalisation of the $p$-brane Newton-Cartan geometry for type II superstring theory.} The appearance of such a structure is necessary as the usual Lorentzian metric description is invalidated in the $\omega \rightarrow \infty$ limit. 
In addition, even in a flat background, we have to specify certain $\omega$-dependent redefinitions of the other massless fields in order to realise a finite limit, in particular involving the dilaton field and the $(p+1)$-form RR potential which is tuned to a `critical' value~\cite{Gopakumar:2000ep, Gomis:2000bd}. The latter is chosen such that the limit of the D$p$-brane worldvolume theory is manifestly finite, and reduces to (in a flat background) SYM. This notion of BPS decoupling limits underlies the unified framework of decoupling limits in string theory pursued in~\cite{Blair:2023noj, Gomis:2023eav} -- note that \cite{Gomis:2023eav} contains a more intrinsic worldsheet derivation of the M$p$T theory, without resorting to the limiting procedure. 

\vspace{3mm}

We will develop the details of what has been outlined here in Section~\ref{sec:MxThBPS}. There, we discuss the origin of matrix theory as a BPS decoupling limit, motivating it as an infinite speed of light limit of a charged relativistic particle in Section~\ref{sec:nonrelpp}, emphasising the (curved) non-Lorentzian geometry that appears in Section~\ref{sec:mzbt}, as well as obtaining the underlying superalgebra in Section \ref{sec:alg}.
We then discuss how to T-dualise between different $p$-brane decoupling limits in the presence of isometries, in Section \ref{sec:stdmpt}.
Finally, in Section \ref{sec:dmgt} we briefly survey the fate of D$q$-branes in M$p$T, and in particular show how matrix theory arises on D$p$-branes in the limit. 

\subsection{To Holography...} 

We are now faced with \emph{\textbf{Question 2}}, which asks what role the non-Lorentzian geometry coupled to matrix theory plays in holography.\,\footnote{Along other lines but still in the context of holography, the first instance of non-Lorentzian geometry on the boundary was the discovery that a torsional generalisation of Newton-Cartan geometry appears as the boundary geometry for non-AdS (Lifshitz) bulk spacetimes \cite{Christensen:2013lma,Christensen:2013rfa}. 
There are also examples of  holographic setups with non-relativistic gravity theories in the bulk and non-relativistic field theories on the boundary \cite{Kachru:2008yh,Griffin:2012qx,Hofman:2014loa,Hartong:2016yrf}. Finally, Ref.~\cite{Hartong:2017bwq} exhibits a case with a non-Lorentzian bulk Chern-Simons theory and a relativistic CFT on the boundary.} 
The story starts with the well-known fact that BFSS matrix theory on a three-torus is dual to $\CN = 4$ SYM on a three-torus~\cite{Taylor:2001vb}. In the decompactification limit, we are led to $\CN = 4$ SYM residing within the longitudinal sector of M3T. However, the complete ten-dimensional target space geometry in M3T is non-Lorentzian, as the longitudinal and transverse sectors are related to each other via a 3-brane Galilean boost \eqref{intro_boost}. This M3T geometry 
does \emph{not} admit any ten-dimensional metric description. Instead, as indicated above, the appropriate description of the target space geometry in M3T, to which $\CN = 4$ SYM couples, requires the vielbein fields of the generalised Newton-Cartan formalism.

In the AdS/CFT correspondence, four-dimensional strongly coupled $\CN = 4$ SYM on a stack of D3-branes in the open string sector is dual to AdS${}_5 \times S^5$ geometry, which is Lorentzian. 
The bulk AdS geometry is obtained by taking the Maldacena decoupling limit of the D3-brane solution in IIB supergravity~\cite{Maldacena:1997re}. 
However, $\CN = 4$ SYM also arises from the M3T decoupling limit, whereupon it appears coupled to non-Lorentzian M3T geometry.
This splits \emph{\textbf{Question 2}} into two more precise subquestions.
Firstly, how are these two decoupling limits related?
Secondly, how are these two geometries, namely, the Lorentzian AdS geometry in the bulk and the non-Lorentzian M3T geometry, to which $\mathcal{N} = 4$ SYM couples at the asymptotic infinity, mapped to each other? We will find precise answers to these questions.

\vspace{3mm}

\noindent $\bullet$~\emph{Near-horizon limit revisited.} In Section~\ref{sec:hnhbpsl},
we will argue that
the relation between the two geometries can be understood by examining the mapping of the decoupling limits on both sides of the AdS/CFT correspondence. The bulk AdS${}_5$ geometry arises from a near-horizon limit of the D3-brane solution in IIB supergravity, while the M3T geometry at the asymptotic infinity arises from a BPS decoupling limit that zooms in on a background D3-brane in type IIB superstring theory. The mapping between these two limits underlies the original proposal of the AdS/CFT correspondence. 
The crucial point turns out to be that the near-horizon limit in the bulk is generated from the \emph{same} asymptotic BPS decoupling limit, now applied to the brane geometry.
This will in fact apply to general D$p$-brane solutions, whose analogous decoupling limits dual to appropriate regimes of $(p+1)$-dimensional SYM follow from the analysis of Itzhaki-Maldacena-Sonnenschein-Yankielowicz (IMSY)~\cite{Itzhaki:1998dd}.
Additionally, the near-horizon geometries obtained can be seen to themselves \emph{asymptotically approach an M$p$T limit}, yielding back the non-Lorentzian geometry seen by the SYM at asymptotic infinity.

These observations clarify in a very elegant -- and geometric -- manner the relationship between the matrix theory and AdS/CFT decoupling limits.
In part this is to be expected: for instance, see the detailed discussion in \cite{Polchinski:1999br} of the $p=0$ case emphasising the role of the back-reacted D0-brane geometry \cite{Hyun:1997zt,Balasubramanian:1997kd,Hyun:1998bi,Hyun:1998iq} as well as the D0-brane treatment in IMSY \cite{Itzhaki:1998dd}. For a variety of other complementary contemporaneous discussions, see~\cite{Boonstra:1997dy,
Hyun:1998bi,Hyun:1998iq,Li:1998jy,deAlwis:1998ki,Martinec:1998ja,Lifschytz:1998pk,Townsend:1998qp,Yoneya:1999bb, Chepelev:1999pm} (for some more recent discussion see~\cite{Biggs:2023sqw,Maldacena:2023acv}).
Our treatment provides an improved understanding of how to define the self-same (asymptotic) matrix theory or near-horizon decoupling limits intrinsically for general $p$, emphasising the geometric effects in different regions of spacetime.

\vspace{3mm}

\noindent $\bullet$~\emph{Landscape of holographic duals.}
Having established the link between asymptotic M$p$T limits and near-horizon limits of D$p$-brane geometries, we then explore how to use this to generate new examples of holographic bulk geometries.
In full, we show how to apply these limits in two cases:

\begin{enumerate}[(1)]

\item

\emph{First}, in the simplest setup, the standard AdS${}_{5}$/CFT${}_{4}$ 
correspondence is generated from a single asymptotic BPS decoupling limit, or, more generally, D$p$-brane near-horizon limits are obtained from the M$p$T limit whose longitudinal sector is aligned with the D$p$-brane sourcing the bulk brane geometry. 
This we show explicitly in Section \ref{sec:badsmt}.

\item

\emph{Second}, interesting non-Lorentzian geometries arise when an asymptotic M$p$T limit is applied to a bulk D$q$-brane geometry, with $p \neq q$\,. In this latter case, the asymptotic M$p$T limit is mapped to a bulk M$p$T limit instead of a near-horizon limit. These off-aligned brane configurations give rise to a non-Lorentzian brane geometry, for which a further near-horizon limit can be applied. At the asymptotic infinity, the bulk near-horizon limit is mapped to a second BPS decoupling limit of the M$q$T type. This double BPS decoupling limit of type II superstring theory at the asymptotic infinity essentially zooms in on two intersecting background branes. 
This is the focus of Sections \ref{sec:nlbdal} and \ref{sec:morebranegeos}.

\end{enumerate}

\noindent As we already explained, the first of these cases reproduces the original AdS/CFT decoupling limit~\cite{Maldacena:1997re} for $p=3$, or more generally the near-horizon decoupling limit that leads to geometries that are dual (in the appropriate regime) to the SYM (or matrix theory) living on the D$p$-brane worldvolume in the limit~\cite{Itzhaki:1998dd}.
From the second, we find examples of non-Lorentzian holography,\footnote{In this paper we use the terminology \emph{non-Lorentzian holography} when the ten-dimensional bulk geometry is non-Lorentzian.} relating non-Lorentzian geometries in M$p$T to non-relativistic limits of field theories.
These examples include those proposed recently in \cite{Lambert:2024yjk}.\footnote{\emph{Note added:} Following the appearance of this paper, these examples and others were further elaborated on in Ref.~\cite{Lambert:2024ncn}.} See~\cite{Lambert:2024uue, Fontanella:2024rvn, Fontanella:2024kyl}
for similar examples using membrane and string, rather than D-brane decoupling limits, and \cite{Gomis:2005pg} for an earlier non-relativistic string limit of AdS${}_5 \times S^5$. In general, one expects such generalised holographic dualities beyond the original AdS/CFT correspondence~\cite{Maldacena:1997re} to be along the lines of open/closed string duality (see \emph{e.g.}~\cite{Niarchos:2015moa} and references therein), which goes beyond the correspondence between geometry and field theory.

\vspace{3mm}

\noindent $\bullet$~\emph{Holography as a DLCQ${}^{\,n}$/DLCQ${}^{\,m}$ correspondence.}
In Section~\ref{sec:chdlcq}, we will realise a unified picture using M-theory. We show that the proposed holographic duals in Section~\ref{sec:hnhbpsl} are related to the original AdS/CFT correspondence via a classification into (U-)duality orbits of M-theory compactified consecutively over one or (after further dualisation) a series of lightlike circles, in a manner we will make precise in Section \ref{sec:chdlcq}. We refer to each of the lightlike compactifications as a DLCQ following the terminology of \emph{e.g.}~\cite{Susskind:1997cw}. 
We will formalise this DLCQ perspective by being systematic about the classification of BPS decoupling limits following \cite{Blair:2023noj, Gomis:2023eav}. More precisely, we can classify all BPS decoupling limits into duality orbits linked to one or multiple DLCQs~\cite{bpslimits}.

This perspective can then be used to reveal a wide range of possible holographic duals with intriguing bulk geometries, including not only AdS and but also various classes of non-Lorentzian ones. 
The non-Lorentzian examples that we will encounter in this paper are all Galilei-like geometries, but more generally Carrollian counterparts can also be generated \cite{Blair:2023noj, Gomis:2023eav}.
We will argue that all the holographic dualities discussed here are unified as a \emph{DLCQ${}^{\,n}$/DLCQ${}^{\,m}$ correspondence}, with $m > n$\,. 

We now give a prelude for how this conjecture arises in the simplest case for the \emph{DLCQ${}^{\,n}$/DLCQ${}^{\,n+1}$ correspondence}. 
One essential upshot of Section~\ref{sec:hnhbpsl} is that we extrapolate AdS/CFT to correspondences between: (1) a bulk geometry 
arising from $n$ BPS decoupling limits plus an extra near-horizon limit, and (2) a field theory from applying $n+1$ BPS decoupling limits. 
The AdS${}_5$/CFT${}_4$ correspondence and its IMSY generalisations in~\cite{Itzhaki:1998dd} are associated with $n=0$\,, while $n=1$ corresponds to the non-Lorentzian versions of Section~\ref{sec:hnhbpsl} and~\cite{Lambert:2024yjk, Lambert:2024uue, Fontanella:2024rvn, Fontanella:2024kyl}. In Section~\ref{sec:chdlcq}, we will see that each BPS decoupling limit is associated with performing a DLCQ in M-theory. This observation then leads to the conjecture that, for a nonnegative integer $n$\,, the bulk geometry arises from the duality orbit after performing $n$ DLCQs, while the dual field theory lives in the DLCQ${}^{n+1}$ orbit. 
The discrepancy in $n$ is accounted for by the fact that the near-horizon decoupling limit does \emph{not} alter the nature of the geometry in the bulk.

The occurrence of the DLCQ here is directly linked to how it appears in matrix theory, and can be traced back (by duality) to the D0-brane near-horizon geometry.
For this case, the uplifted M-theory pp-wave geometry only gives asymptotically a null compactification \cite{Hyun:1997zt, Balasubramanian:1997kd, Polchinski:1999br}, connecting to the DLCQ which leads to matrix theory.
Back in ten dimensions, and for more general D$p$-brane near-horizon geometries, this statement translates quite precisely to the appearance of asymptotic non-Lorentzian geometry.

We will first develop this conjecture of the DLCQ${}^{n}$/DLCQ${}^{n+1}$ correspondence in Section~\ref{sec:chdlcq}. In Section~\ref{sec:mptmtd}, we review the relationship between matrix theory as M0T and the DLCQ of M-theory. 
Then in Sections~\ref{sec:adsdlcq} and \ref{sec:furtherdlcqs} we discuss the first and second layers of the duality web of DLCQs and the induced holographic correspondences. In Section~\ref{sec:dlcqmn}, we discuss the generalisation to the DLCQ${}^{n}$/DLCQ${}^{m}$ correspondence with $m > n+1$\,, when the bulk geometry is associated with intersecting branes. 

\subsection{And Back Via \texorpdfstring{$T\bar{T}$}{TTbar}}

After analysing the BPS decoupling limits that lead to matrix theory and holographic duals, we will be in a position to address \emph{\textbf{Question 3}} and provide an intrinsic way to generate the ten-dimensional bulk geometry from the asymptotic flat non-Lorentzian spacetime. 

\vspace{3mm}

\noindent $\bullet$~\emph{Generating near-horizon bulk geometries: an intrinsic perspective.} The expressions for the near-horizon geometries obtained in Section \ref{sec:hnhbpsl} reveal an intriguing intrinsic perspective on the relationship between the asymptotic infinity and bulk geometry.
The structure of the near-horizon solution is exactly such that it realises geometrically the BPS decoupling limit at asymptotic infinity, as the ratio between the radial coordinate $r$ and the characteristic scale $\ell$ becomes infinitely large.
For the D-brane examples, the asymptotic infinity of the near-horizon bulk corresponds to the flat M$p$T background.
For example, the near-horizon AdS${}_5 \times S^5$ geometry can be written as
\begin{align} \label{intro_ads}
    ds^2 = \biggl(\frac{r}{\ell}\biggr)^{\!\!2} \, \dd x^A \, \dd x^B \, \eta^{}_{AB}  + \biggl(\frac{\ell}{r}\biggr)^{\!\!2} \Bigl( \dd r^2 + r^2 \, \dd\Omega_5^2 \Bigr)\,,
\end{align}
with $A=0\,,\,\cdots,\, 3$ and $\ell$ the AdS length.
Comparing with Eq.~\eqref{intro_limit}, and switching to Cartesian coordinates in the transverse space, we find a geometrical version of the M3T decoupling limit with $(r/\ell)^2$ playing the role of the parameter $\omega$.\,\footnote{The promotion of the constant parameter $\omega$ to a background-dependent function works due to an emergent dilatation symmetry, that we discuss around Eq.~\eqref{eq:dilatsmpt}.}

Reversing this logic, the bulk geometry can be viewed as a `deformation' of this non-Lorentzian geometry at asymptotic infinity, by inverting the BPS decoupling limit.
In Section~\ref{sec:TTbar} we study this intrinsic perspective, and discuss the general properties of the field theory deformation that follows from viewing the BPS decoupling limit in reverse.

It turns out that these deformations are related to the $T \bar T$ deformation, which originally appeared in the context of two-dimensional field theories \cite{Zamolodchikov:2004ce,Smirnov:2016lqw}.
This is an irrelevant deformation with many remarkable properties (see~\cite{Jiang:2019epa} for a review).
The most relevant feature for our purposes is the fact that, given some initial Lagrangian $\mathcal{L}(0)$ describing a two-dimensional field theory, the $T \bar T$ deformation introduces a flow parameter $\lambdap$ such that the Lagrangian of the deformed theory obeys 
\be
    \frac{\mathcal{\partial \mathcal{L}(\lambdap)}}{\partial \lambdap} \sim \det T_{\alpha \beta}(\lambdap)\,,
\ee
where $T_{\alpha \beta}(\lambdap)$ denotes the energy-momentum tensor of the theory.
When $\mathcal{L}(0)$ is the theory of $D$ free bosons, the $T \bar T$ deformed Lagrangian is the Nambu-Goto action for a string in ($D$+2)-dimensions, with a $B$-field proportional to $\lambdap^{-1}$~\cite{Cavaglia:2016oda,Bonelli:2018kik}.
On the other hand, starting with this Nambu-Goto action and undoing the deformation, sending $\lambdap\rightarrow 0$ is exactly the decoupling limit adapted to the fundamental string 
that defines \emph{non-relativistic string theory}~\cite{Klebanov:2000pp, Gomis:2000bd, Danielsson:2000gi}. This string theory is unitary and ultra-violet (UV) complete and has a wound string spectrum that is Galilean invariant at fixed winding number. 
This connection between $T\bar T$ deformations and non-relativistic string theory was made in~\cite{Blair:2020ops}.

Here, we exploit the fact that this string decoupling limit is part of the duality web of BPS decoupling limits.
In particular, as we review in Section \ref{sec:matrixstringtheory}, the string decoupling limit leading to non-relativistic string theory is S-dual to the D1-brane decoupling limit leading to M1T~\cite{Ebert:2023hba}.
There, for completeness, we elaborate on how non-relativistic string theory and M1T are related to matrix string theory~\cite{Motl:1997th, Dijkgraaf:1997vv}, which arises from dualising BFSS matrix quantum mechanics on a spatial circle and essentially describes a second quantisation of non-relativistic strings.
Then we explain the link between the $T \bar T$ deformation and the BPS decoupling limit. In Section \ref{sec:TTbarNG} we discuss the relationship in terms of the Nambu-Goto action while in Section \ref{sec:TTbarPoly} we clarify how this can be viewed in terms of the Polyakov action.
We then point out that the D-brane decoupling limits central to this paper should naturally define generalisations of the $T \bar T$ deformation. Finally, in Section~\ref{sec:TTbarBulk} we explain how such a deformation of a flat non-Lorentzian M$p$T spacetime generates the corresponding bulk near-horizon geometry in the context of holography.

\vspace{3mm}

\noindent $\bullet$~\emph{New $p$-brane $T \bar{T}$ deformations.} In Section \ref{sec:pbttfe} we present the flow equations for the $p$-brane generalisations of $T \bar T$ that follow from the logic described above.
These formulae should be of interest beyond the string theory context, as they define potentially interesting field theory deformations, at least classically.
We stress that within the string theory context, they (by definition) are the deformations which induce a flow from SYM to the Dirac-Born-Infeld (DBI) action (at least in the abelian case).
Various higher-dimensional generalisations of $T \bar T$ have been proposed before in the literature \emph{e.g.}~\cite{Bonelli:2018kik,Cardy:2018sdv,Taylor:2018xcy}, and it is especially interesting that there exists a body of work obtaining theories of non-linear electrodynamics from (generalisations of) $T \bar T$ \cite{Conti:2018jho,Brennan:2019azg, Babaei-Aghbolagh:2020kjg, Babaei-Aghbolagh:2022uij,Ferko:2022iru,Conti:2022egv,Ferko:2022cix, Babaei-Aghbolagh:2022leo,Ferko:2023sps,Ferko:2024yua}. None of these cases appear to capture the full DBI brane actions of string theory, though see \cite{Ferko:2023sps, Ferko:2024yua} for recent progress in obtaining flow equations leading from Yang-Mills to abelian and non-abelian Born-Infeld (without scalars).

We firstly obtain a class of flow equations which deform a $(p+1)$-dimensional free scalar field theory to the Dirac-Nambu-Goto action, for any $p$. In flat spacetime, these can be expressed elegantly as:\,\footnote{In the special case where the Lagrangian describes a single scalar and when $p > 0$\,, the flow equation~\eqref{intro_ffe} becomes equivalent to the generalised flow equation obtained in \cite{Ferko:2023sps}, which was used to obtain three-dimensional Born-Infeld from the Maxwell Lagrangian (in three dimensions a vector field is dual to a scalar). However, in the general case, it is the flow equation~\eqref{intro_ffe} that deforms the free multi-scalar field theory to the Dirac-Nambu-Goto action in $p$+1 dimensions. Intriguingly, related formulae also show up in different contexts~\cite{Morone:2024ffm, Tsolakidis:2024wut} (we thank the authors of these references for bringing their works to our attention). See Section~\ref{sec:pbttfe} for more detailed comparisons.}
\be \label{intro_ffe}
    \frac{\p \CL(\lambdap)}{\p \lambdap} = \frac{1}{2 \, \lambdap^2} \, \biggl\{ \tr \, \bigl(\mathbb{1} - \lambdap\, \CT\bigl)  - \bigl(p-1\bigr) \, \Bigl[ \det \bigl(\mathbb{1} - \lambdap\, \CT \bigr)\Bigr]^{\frac{1}{p-1}} - 2 \biggr\}\,,
\ee
where the matrix $\mathcal{T}$ 
has components $\mathcal{T}^\alpha{}_\beta = \eta^{\alpha \gamma} \, T_{\gamma \beta}$\,, with $T_{\alpha \beta}$ denoting the energy-momentum tensor ($\alpha=0\,,\,1\,,\,\dots,\,p$).
For $p=1$, this reduces to the usual $T \bar T$ deformation, and for $p=0$ we obtain the quantum mechanical deformation obtained in \cite{Gross:2019ach}. Our results emphasise that the latter can be viewed simply as arising from the usual non-relativistic limit of a point particle in reverse. 
For $p>1$, we obtain novel higher-dimensional generalisations of $T \bar T$\,.
For instance, the flow equation \eqref{intro_ffe} specialised to $p=2$ is:
\be
\frac{\partial \mathcal{L}(\lambdap)}{\partial \lambdap} = 
\frac{1}{4} \, \Bigl[ \tr\bigl(\mathcal{T}^2\bigl)-(\tr \,\mathcal{T})^2 + 2\, \lambdap \det \bigl( \mathcal{T} \bigr) \Bigr]\,,
\label{intro_membraneTTbar}
\ee
the first two terms of which would be proportional to the determinant of the energy-momentum tensor in a two-dimensional theory, while the third term is a new explicitly $\lambdap$-dependent modification needed in three dimensions. 

We then obtain novel flow equations which, when applied to a $(p+1)$-dimensional field theory of free scalars and an abelian gauge field, produce the D-brane DBI action.
We obtain the full flow equations only 
for $p=1\,,\,2$\,, see Eq.~\eqref{eq:polfd}. 
For $p=1$\,, the flow equation depends explicitly on the field strength of the gauge field as well as the energy-momentum tensor of the theory. 
This equation controls the flow from the (1+1)-dimensional (bosonic) SYM to the D1-brane DBI action, and explains why applying the $T \bar T$ deformation to the same starting point does not lead to the latter~\cite{Conti:2018jho,Brennan:2019azg}.
For $p=2$, the flow equation does \emph{not} depend explicitly on the field strength, and in fact coincides with Eq.~\eqref{intro_membraneTTbar}, reflecting the fact that scalars and one-forms are dual in three dimensions. This is inherent in the relationship between the D2-brane and M2-brane worldvolume theories, which are used to derive the $p=2$ equations. For $p=3$\,, we present the flow equation up to the linear order in $\lambdap$\,.

\section{Matrix Theory: A BPS Perspective}
\label{sec:MxThBPS} 

In this section, we revisit the D$p$-brane decoupling limits that lead to matrix theories.
We start with the $p=0$ case leading to the matrix quantum mechanics on D0-branes \cite{deWit:1988wri, Banks:1996vh,Susskind:1997cw}, which we first motivate as a non-relativistic point particle limit. 
After discussing geometric and algebraic features of the resulting corner of type IIA string theory, which as in \cite{Blair:2023noj} we call matrix 0-brane theory or M0T, we move on via T-duality to the D$p$-brane version. 

\subsection{Decoupling Limit of Charged Particles and BFSS Matrix Theory}
\label{sec:nonrelpp}

The decoupling limits considered throughout this paper can be seen as non-relativistic limits, generalised in a particular way natural to the BPS extended objects of string and M-theory.
To motivate this, let us start with the simple example 
of a charged relativistic particle described by the following action principle:
\be \label{eq:pa}
    S = - m \, c \int \dd\tau \, \sqrt{- \dot{X}^\mu_{\phantom{\dagger}} \, \dot{X}_\mu} + \frac{e}{c} \int \dd\tau \, \dot{X}^\mu \, A_\mu\,,
\ee
where $\mu = 0\,, \, \cdots, \, D$\,, with $D+1$ the spacetime dimension. We have introduced $m$ as the mass of the particle and $e$ the electric charge, and included factors of the speed of light $c$\,, such that  $X^\mu = (c \, t\,, \, X^i)$\,, with $t$ the coordinate time and $i = 1\,, \, \cdots, D$\,. 
In the static gauge, we take the affine parameter $\tau$ to be the coordinate time, \emph{i.e.}~$\tau = t$\,. 
Making this choice, the action can be written as:
\be
    S = - m \, c \int \dd t \, \sqrt{c^2 - \dot{X}^i \, \dot{X}^i} + \frac{e}{c} \int \dd t \, \Bigl( c \, A_0 + \dot{X}^i \, A_i \Bigr)\,.
\ee
In order to take the infinite $c$ limit, we fine tune the gauge potential such that
$A_0 = m \, c^2 / e$ and $A_i = 0$\,. 
The contribution from the gauge potential is a boundary term in the action. 
In the limit where $c \rightarrow \infty$\,, we find the following action describing a non-relativistic particle:
\be \label{eq:nonrelpa}
    S_\text{non-rel.} = \frac{m}{2} \int \dd t \, \dot{X}^i \, \dot{X}^i\,. 
\ee
This limit could be taken for any particle of mass $m$ and charge $e$\,, which would be somewhat \emph{ad hoc}. 
However, a distinguished possibility is to focus on the class of BPS particles whose mass equals their charge, $m=e$\,. 
For such configurations, the electric potential needed is tuned to unit value, meaning $A_0=c^2$.
The infinite speed of light limit with this \emph{critical} choice of gauge potential is then a BPS decoupling limit that decouples relativistic particle states and only leaves us with a non-relativistic spectrum.
Then the action \eqref{eq:nonrelpa} describes the dynamics around the BPS configuration. 
Note that the more general BPS condition is $m=|e|$\,: choosing the electric potential to be $A_0 = -c^2$ then corresponds to keeping the BPS states with $m=-e$ in the limit. This can be viewed as choosing to keep anti-particles rather than particles.

Above we crudely considered taking $c$ to infinity.
However, when taking limits in general, we should really consider an expansion with respect to a control parameter, which must be dimensionless. The speed of light $c$ is dimensionful and therefore does not constitute such a control parameter. This problem can be circumvented by introducing a dimensionless parameter $\omega$ in each place where $c$ shows up. Effectively, in the convention where $c = 1$\,, this can be achieved by performing the following replacements in the particle action~\eqref{eq:pa}:
\be \label{eq:xzxiao}
    m \rightarrow \omega \, m^{}_\text{eff}\,,
        \qquad%
    X^0 \rightarrow \omega \, X^0\,,
        \qquad%
    X^i \rightarrow X^i\,,
        \qquad%
    A^{(1)} \rightarrow \omega^2 \,\frac{m_\text{eff}}{e} \, \dd X^0 \,,
\ee
where $A^{(1)} = A_\mu \, \dd X^\mu$ and $m_\text{eff}$ is the mass of the non-relativistic particle. 

In the context of type IIA superstring theory, the above limiting procedure can be readily applied to the D0-particle. We first focus on the bosonic sector of the D0-particle action, before moving on to discuss the fermionic sector. The D0-particle action is
\be \label{eq:dzpabs}
    S_\text{D0} = - \frac{1}{g_\text{s} \, \sqrt{\alpha'}} \int \dd\tau \, \sqrt{- \dot{X}^\mu_{\phantom{\dagger}} \, \dot{X}_\mu} + \frac{1}{\sqrt{\alpha'}} \int C^{(1)}\,,
        \qquad%
    \mu = 0\,, \, \cdots, \, 9\,.
\ee
Here, $C^{(1)}$ is the RR one-form potential coupled to the D0-particle, which plays the role of the $U(1)$ gauge potential. Moreover, $g_\text{s}$ is the string coupling and $\alpha'$ is the Regge slope, related to the string length $\ell_\text{s}$ by $\alpha' = \ell_\text{s}^2$\,. Effectively, we have the particle mass $m = \bigl(\,g_\text{s} \, \sqrt{\alpha'}\,\bigr)^{-1}$ and the particle charge $e = 1 / \sqrt{\alpha'}$\,. It is useful to recast the reparametrisation~\eqref{eq:xzxiao} equivalently as
\be \label{eq:rpom}
    g^{}_\text{s} \rightarrow \omega^{-3/2} \, g^{}_\text{s}\,,
        \qquad%
    X^0 \rightarrow \sqrt{\omega} \, X^0\,,
        \qquad%
    X^i \rightarrow \frac{X^i}{\sqrt{\omega}}\,,
        \qquad%
    C^{(1)} \rightarrow \omega^2 \, g_\text{s}^{-1} \, \dd X^0\,,
\ee
where $i = 1\,, \, \cdots, \, 9$\,. 
This choice is made such that \emph{no} rescaling of the Regge slope is introduced. This is different from the standard convention in the literature (see \emph{e.g.}~\cite{Gopakumar:2000ep}), but we will see that this parametrisation is convenient for us to gain geometric intuition. 
With $\alpha'$ fixed, it turns out that $X^0$ and $X^i$ have to be rescaled as in Eq.~\eqref{eq:rpom}, even though it looks like one may reshuffle the rescalings of $g^{}_\text{s}$ and $X^\mu$ by an arbitrary factor. A simple way to see why this has to be the case is to consider applying the limit to the fundamental string Nambu-Goto action.
This action with powers of the speed of light $c$ restored is: 
\be
    S_\text{F1} = - \frac{1}{2\pi \alpha' c} \int \dd^2 \sigma \, \sqrt{- \det \Bigl( - c^2 \, \p_\alpha t \, \p_\beta t + \p_\alpha X^i \, \p_\beta X^i \Bigr)}\,.
\ee
Under the assumption that $\alpha'$ is fixed, the rule for the rescaling in terms of $\omega$ follows after absorbing the overall factor of $1/c$ into the determinant, leading to 
\be \label{eq:fsa}
    S_\text{F1} = - \frac{1}{2\pi\alpha'} \int \dd^2 \sigma \, \sqrt{-\det\left(- \p_\alpha \bigl( \sqrt{\omega} \, X^0 \bigr) \, \p_\beta \bigl( \sqrt{\omega} \, X^0 \bigr) + \, \p_\alpha \Bigl( \tfrac{X^i}{\sqrt{\omega}} \Bigr) \, \p_\beta \Bigl( \tfrac{X^i}{\sqrt{\omega}} \Bigr)\right)}\,,
\ee
hence the rescalings in Eq.~\eqref{eq:rpom}. In the $\omega \rightarrow \infty$ limit, the fundamental string action gives rise to the non-vibrating string~\cite{Batlle:2016iel, Gomis:2023eav}. 

We can extend this discussion from particle to superparticle, introducing the superpartners $\Theta$ of the embedding coordinates $X^\mu$\,. These are anticommuting spinor coordinates $\Theta^{a} (\tau)$ with $a = 1\,, \, \cdots, \, 32$\,. The supersymmetric D0-particle action is
\be
     S_\text{D0} = - \frac{1}{g_\text{s} \, \sqrt{\alpha'}} \int \dd\tau \, \Bigl( \sqrt{- \Pi^\mu_{\phantom{\dagger}} \, \Pi_\mu} + \bar{\Theta} \, \gamma_{11} \, \dot{\Theta} \Bigr) +\frac{1}{\sqrt{\alpha'}} \int C^{(1)}\,,
\ee
where $\bar{\Theta} = \Theta^\intercal \, \gamma^0$\,, $\gamma^{}_{11} = \gamma^{}_0 \, \gamma^{}_1 \cdots \gamma^{}_9$\,, with $\gamma^{}_\mu$ the Dirac matrices, and
$\Pi^\mu = \dot{X}^\mu - \bar{\Theta} \, \gamma^\mu \, \dot{\Theta}$\,.
The above superparticle action is invariant under the supersymmetry transformation
\be \label{eq:susytrnsf}
    \delta \Theta = \epsilon\,,
        \qquad%
    \delta X^\mu = \bar{\epsilon} \, \gamma^\mu \, \Theta\,. 
\ee
In order to facilitate the $\omega \rightarrow \infty$ limit, we have to supplement the rules in Eq.~\eqref{eq:rpom} with
\be
    \Theta = \omega^{\frac{1}{4}} \, \Theta_- + \omega^{-\frac{3}{4}} \, \Theta_+\,,
\ee
where $\Theta_\pm$ are the $\pm1$ eigenspinors of $\Gamma = \gamma^0 \, \gamma^{}_{11}$\,. In the $\omega \rightarrow \infty$ limit, we use the kappa symmetry to fix $\Theta_-$\,, such that the resulting non-relativistic superparticle action is
\be\label{eq:NRsuperparticle}
    S = \frac{1}{g_\text{s} \, \sqrt{\alpha'}} \int \dd\tau \, \Bigl( \tfrac{1}{2} \, \dot{X}^i \, \dot{X}^i + 2 \, \psi^\intercal \, \dot{\psi} \Bigr)\,,
\ee
where $\psi = \Theta_+$\,. The supersymmetry transformation is now given by
\be
    \delta \psi = \epsilon_+ + \frac{1}{2} \, \gamma^0 \, \gamma^{}_i \, \dot{X}^i \, \epsilon_-\,,
        \qquad%
    \delta X^i = - 2 \, \bar{\psi} \, \gamma^i \, \epsilon_-\,,
\ee
with $\epsilon_\pm$ the $\Gamma$ eigenspinors and
$\epsilon = \omega^{\frac{1}{4}} \, \epsilon_- + \omega^{-\frac{3}{4}} \, \epsilon_+$\,.
Therefore, $\psi$ is the Goldstone fermion, with $\epsilon_+$ being nonlinearly realised. In contrast, $\epsilon_-$ is linearly realised. See~\cite{Gomis:2004pw} for further details of the above discussion. The nonabelian generalisation of the superparticle action gives the BFSS matrix theory~\cite{baake1985fierz, flume1985quantum, Claudson:1984th, deWit:1988wri, Susskind:1997cw},
\be\label{eq:BFSS}
    S^{}_\text{BFSS} = \frac{1}{R} \int \dd\tau \, \tr\biggl[   \tfrac{1}{2} \dot{X}^i \, \dot{X}^i+ \tfrac{1}{4} \bigl[ X^i\,, \, X^j\bigr] \bigl[ X_i\,, \, X_j\bigr]  + 2 \Bigl( \psi^\intercal \, \dot{\psi} - \psi^\intercal \, \gamma^i \bigl[ \psi\,, \, X^i \bigr] \Bigr) \biggr]\,,
\ee
where now $X^i$ and $\psi$ are $N \times N$ hermitian matrices, we let $R=g_{\text{s}} \sqrt{\alpha'}$ and further fixed string units $2 \pi \alpha' = 1$. For $N, R \rightarrow \infty$ this action was conjectured to describe M-theory in flat spacetime \cite{Banks:1996vh}, and for fixed $N$ to describe the DLCQ of M-theory \cite{Susskind:1997cw}. We will come back to the latter statement in Section \ref{sec:mptmtd}.
The BFSS matrix theory linearly realises the global symmetry transformations associated with 16 supercharges. 

\subsection{Matrix 0-Brane Theory} \label{sec:mzbt}

In the previous subsection, we discussed the BPS decoupling limit of D0-particles in type IIA superstring theory. The definition of this decoupling limit is captured by the rescalings of the spacetime (super-)coordinates together with the appropriate reparametrisations of the string coupling and RR one-form, which we collect below:
\begin{subequations} \label{eq:rpmzt}
\begin{align}
    X^0 & \rightarrow \sqrt{\omega} \, X^0\,, 
        &%
    \Theta & \rightarrow \omega^\frac{1}{4} \, \Theta^{}_- + \omega^{-\frac{3}{4}} \, \Theta^{}_+\,, \\[4pt]
    X^i & \rightarrow \frac{X^i}{\sqrt{\omega}}\,,
        &%
    C^{(1)} & \rightarrow \omega^2 \, g^{-1}_\text{s} \, \dd X^0\,,
        \qquad%
    g^{}_\text{s} \rightarrow \omega^{-\frac{3}{2}} \, g^{}_\text{s}\,.
\end{align}
\end{subequations}
The decoupling limit defined by sending $\omega$ to infinity can be therefore applied to the full-fledged IIA theory, which contains all sorts of extended objects including the fundamental string and various D$p$-branes. In the following, we will focus on the bosonic part of the limit, and generalise the above prescriptions to arbitrary background fields. 

Consider type IIA superstring theory in arbitrary background (string-frame) metric $G_{\mu\nu}$\,, Kalb-Ramond field $B_{\mu\nu}$\,, dilaton $\Phi$\,, and RR $q$-form $C^{(q)}$ with odd $q$\,. 
The essential intuition to bring the reparametrisation~\eqref{eq:rpmzt} to curved spacetime is to replace $\dd X^0$ with a ‘longitudinal’ vielbein, $\tau^0 = \dd x^\mu \, \tau^{}_\mu{}^0$, and to replace $\dd X^i$ with a ‘transverse’ vielbein, $E^i = \dd x^\mu \, E^{}_\mu{}^i$\,. Each of these can be used to define a degenerate ‘metric’, $\tau^{}_{\mu\nu} = - \tau^{}_\mu{}^0 \, \tau^{}_\nu{}^0$ and $E^{}_{\mu\nu} = E^{}_\mu{}^i \, E^{}_\nu{}^i$\,. These have to be orthogonal in the appropriate sense, so that together they define a Newton-Cartan structure, which we will further illustrate in a moment. The reparametrisations of the bosonic fields in Eq.~\eqref{eq:rpmzt} now generalise to arbitrary background fields as
\begin{subequations} \label{eq:rpgcfb}
\begin{align}
    G_{\mu\nu} & = \omega \, \tau^{}_{\mu\nu} + \omega^{-1} \, E^{}_{\mu\nu}\,, 
        &
    \Phi & = \varphi - \tfrac{3}{2} \, \ln \omega\,,
        &
    B^{(2)} &= b^{(2)}\,, \\[4pt]
    C^{(1)} & = \omega^2 \, e^{-\varphi} \, \tau^0 + c^{(1)}\,,
        &
    C^{(q)} & =c^{(q)} \quad\text{ if } q \neq 1\,.
\end{align}
\end{subequations}
Here, in the $\omega \rightarrow \infty$ limit, the background fields are as follows: $\tau_{\mu\nu}$ and $E_{\mu\nu}$ encode the geometry, $\varphi$ denotes the dilaton, $c^{(q)}$ denotes the RR $q$-form potential, and $b^{(2)}$ denotes the Kalb-Ramond field. The original $\mathrm{SO}(1,9)$ local Lorentz transformation is now broken such that,  alongside the $\mathrm{SO}(9)$ transformation acting on $E_\mu{}^i$, we now have the Galilean boost, which acts on the vielbein fields as
\be
    \delta^{}_\text{\scalebox{0.8}{G}} \tau_\mu{}^0 = 0\,,
        \qquad%
    \delta^{}_\text{\scalebox{0.8}{G}} E_\mu{}^i = \Lambda^i \, \tau_\mu{}^0\,.
\ee
Moreover, the RR potential $c^{(1)}$ also transforms nontrivially under the Galilean boost, with
\be \label{eq:gbtc1e}
    \delta^{}_\text{\scalebox{0.8}{G}} c^{(1)} = -e^{-\varphi} \, \dd x^\mu \, E_\mu{}^i \, \Lambda^i \,. 
\ee 
This $\omega \rightarrow \infty$ limit of type IIA superstring theory defines a self-consistent corner of string theory that we refer to as \emph{matrix 0-brane Theory} (M0T), whose spacetime geometry develops a codimension-one foliation structure, described by the Newton-Cartan vielbein fields $\tau^0$ and $E^i$\,. This Newton-Cartan formalism is usually used to covariantise Newtonian gravity. The resulting Newton-Cartan geometry does \emph{not} admit a metric description and is therefore non-Lorentzian. 

It is clear from the defining prescription \eqref{eq:rpgcfb} that a rescaling of the parameter $\omega$, making a replacement   $\omega \rightarrow \omega \, \Delta^{-1}$, does not affect the nature of the limit.
This is true even if $\Delta$ is a (sufficiently well-behaved) function of the coordinates. 
This induces an emergent dilatation symmetry of the M0T target space, acting as 
\begin{align} \label{eq:dilatsm0t}
    \tau^0 \rightarrow \Delta^{\frac{1}{2}} \, \tau^0\,,
        \qquad%
    E^{i} \rightarrow \Delta^{-\frac{1}{2}} \, E^{i},
        \qquad%
    e^\varphi \rightarrow \Delta^{-\frac{3}{2}} \, e^\varphi\,.
\end{align}
Under the reparametrisation~\eqref{eq:rpgcfb}, the curved background generalisation of the single D0-brane action~\eqref{eq:dzpabs},
\be \label{eq:dzpabs0}
    S_\text{\scalebox{0.8}{D0}} = - \frac{1}{\sqrt{\alpha'}} \int \dd \tau \, e^{-\Phi} \sqrt{- \dot{X}^\mu_{\phantom{\dagger}} \, \dot{X}^\nu \, G_{\mu\nu}} + \frac{1}{\sqrt{\alpha'}} \int C^{(1)}\,,
\ee
becomes in the infinite $\omega$ limit,
\be \label{eq:mztdzba}
    S^\text{\scalebox{0.8}{M0T}}_\text{\scalebox{0.8}{D0}} = \frac{1}{2 \, \sqrt{\alpha'}} \int \dd\tau \, e^{-\varphi} \, \frac{\dot{X}^\mu \, \dot{X}^\nu \, E_{\mu\nu}}{\dot{X}^\rho_{\phantom{\dagger}} \, \tau_\rho{}^0} + \frac{1}{\sqrt{\alpha'}} \int c^{(1)}\,.
\ee
Such D0-brane states are the fundamental degrees of freedom in M0T, whose dynamics is captured by the BFSS matrix theory. 
We will derive the non-abelian generalisation of the action \eqref{eq:mztdzba} in Section \ref{sec:mgt} below. 
The action \eqref{eq:mztdzba} is invariant under Galilean boost transformations as well as the local dilatations \eqref{eq:dilatsm0t} of the M0T background.

\subsection{Algebraic Perspective on the BPS Decoupling Limit}
\label{sec:alg} 

In this subsection, we discuss some algebraic aspects of M0T. 
In the BPS decoupling limit, the Poincar\'e symmetry algebra underlying the Lorentzian target space in type IIA superstring theory reduces to the Galilei algebra that underlies M0T.
This algebra consists of the temporal translation generator $H$\,, spatial translation generator $P^{}_i$\,, spatial rotation generator $J^{}_{ij}$\,, and Galilean boost generator $G^{}_i$\,. The non-vanishing commutators are given by
\be \label{eq:ghp}
    \bigl[ G_i\,, \, H \bigr] = P_i\,, 
\ee
together with the expected ones involving $J_{ij}$\,. Note that the commutator $[G_i\,, \, P_j]$ vanishes in the Galilei algebra. M0T also admits a central extension of the Galilei algebra, which enhances the symmetry group to the Bargmann algebra (see \emph{e.g.}~\cite{Bergshoeff:2022eog} for a recent review), with
\be \label{eq:cen}
    \bigl[ G_i \,, \, P_j \bigr] = N \, \delta_{ij}\,,
\ee
where $N$ is the generator associated with particle number conservation. Upon gauging of the Bargmann algebra~\cite{Andringa:2010it}, the temporal vielbein $\tau_\mu{}^0$ introduced in Section~\ref{sec:mzbt} corresponds to the gauge field associated with $H$ and the spatial vielbein $E_\mu{}^i$ to the gauge field associated with $P_i$\,; moreover, the particle number generator $N$ is also associated with a gauge field $m_\mu{}^0$\,, which transforms under the Galilean boost as
\be \label{eq:gel}
    \delta_\text{\scalebox{0.8}{G}} m_\mu{}^0 = E_\mu{}^i \, \Lambda^i\,. 
\ee
The inclusion of $m_\mu{}^0$ allows us to construct the manifestly Galilean boost invariant object,
\be \label{eq:hetm}
    H^{}_{\mu\nu} = E^{}_{\mu\nu} - \tau^{}_\mu{}^0 \, m^{}_\nu{}^0 - \tau^{}_\nu{}^0 \, m^{}_\mu{}^0\,.
\ee
In terms of $H_{\mu\nu}$\,, we write the non-relativistic D0-brane action in M0T as
\be \label{eq:m0td0ba}
    S^\text{\scalebox{0.8}{M0T}}_\text{\scalebox{0.8}{D0}} =  \frac{1}{2 \, \sqrt{\alpha'}} \int \dd\tau \, e^{-\varphi} \, \frac{\dot{X}^\mu \, \dot{X}^\nu \, H_{\mu\nu}}{\dot{X}^\mu \, \tau_\mu{}^0} + \frac{1}{\sqrt{\alpha'}} \int {\tilde{c}}^{\,(1)}\,,
\ee
where ${\tilde{c}}^{\,(1)}$ is invariant under the Galilean boost and $m^{}_\mu{}^0$ transforms nontrivially under the gauge symmetry associated with $N$ as $\delta_\text{\scalebox{0.8}{N}} m_\mu{}^0 = D_\mu \Sigma = (\p^{}_\mu - \p^{}_\mu \varphi) \Sigma$\,. Here, the derivative $D_\mu$ is covariant with respect to the additional dilatation symmetry~\eqref{eq:dilatsm0t}, under which $m^0 \rightarrow \Delta^{-\frac{3}{2}} \, m^0$\,. 
Moreover, Eq.~\eqref{eq:m0td0ba} is invariant under the Stueckelberg  transformation, 
\be
    H^{}_{\mu\nu} \rightarrow H^{}_{\mu\nu} + \tau^{}_\mu{}^0 \, C^{}_\nu + \tau^{}_\nu{}^0 \, C^{}_\mu\,,
        \qquad%
    \tilde{c}\,^{(1)} \rightarrow \tilde{c}\,^{(1)} - e^{-\varphi} \, C^{}_\mu \, \dd x^\mu\,.
\ee
Fixing $C^{}_\mu = m^{}_\mu{}^0$\,, in form, Eq.~\eqref{eq:m0td0ba} becomes Eq.~\eqref{eq:mztdzba}, with $c^{(1)}$ in Eq.~\eqref{eq:mztdzba} replaced with $\tilde{c}\,^{(1)} - e^{-\varphi} \, m^{}_\mu{}^0 \, \dd x^\mu$\,. 

In order to include target space supersymmetry, we need to construct a superalgebra extension of the Bargmann algebra with $32$ supercharges. We will refer to this as the M0T superalgebra.
This can be obtained in two ways.
Firstly, this can be constructed directly as a maximal extension of the super-Bargmann algebra of \cite{Gomis:2004pw}. 
Secondly, it can be derived from the usual IIA superalgebra by an $\dot{\text{I}}$n\"on\"u-Wigner contraction.

We start by describing the first approach where we extend the super-Bargmann algebra.
We introduce supercharges $(Q^{}_+\,, \, Q^{}_-)$ satisfying $\gamma^{}_0 \, \gamma^{}_{11} \, Q^{}_\pm = \pm Q^{}_\pm$\,.\,\footnote{\,The spinors are Majorana with $C^\intercal = -C$ and $\gamma_\mu^\intercal = -C \, \gamma^{}_\mu \, C^{-1}$\,. }
In order to conform with the $[G^{}_i\,, \, G^{}_j]=0$ commutator in the Bargmann algebra, it is necessary that the supercharges form an off-diagonal representation of the boosts
\begin{align}
    [Q^{}_+\,, \, G^{}_i] = -\tfrac{1}{2} \, \gamma^{}_{i0} \, Q^{}_-\,, 
        \qquad%
    [Q^{}_-\,, \, G^{}_i] = 0\,.
\end{align}
The above commutators further restrict the form of the fermionic anticommutators and lead to the super-Bargmann algebra with 
\be
    \{Q_+\,, \, \bar{Q}_+\} = H \, \gamma^{}_{11} \, \pi^{}_-\,, 
        \qquad%
    \{Q_+\,, \, \bar{Q}_-\} = P^i \, \gamma^{}_i \, \pi^{}_+\,,
        \qquad%
    \{Q_-\,, \, \bar{Q}_-\} = 2 \, N \, \gamma^{}_{11} \, \pi^{}_-\,.
\ee
Here, we have defined the projectors $\pi^{}_\pm = \tfrac12\,(\mathbb{1}\pm \gamma^{}_0 \, \gamma^{}_{11})$. This is the algebra realised by the non-relativistic superparticle \eqref{eq:NRsuperparticle} and consequently also the BFSS matrix model \cite{Banks:1996vh, Gomis:2004pw}. 

From this starting point, we can construct the maximally extended version of the super-Bargmann algebra which we call the M0T superalgebra:
\begin{align}
    &\big\{Q_+\,, \, \bar{Q}_+\big\} = \Big(H + H^i \, \gamma_i + \tfrac{1}{4!}\,H^{ijkl} \, \gamma_{ijkl}\Big) \, \gamma_{11} \, \pi_-\,,\\[4pt]
    & \big\{Q_+\,, \, \bar{Q}_-\big\} = \Big(P + P^i \, \gamma_i + \tfrac12\,P^{ij} \, \gamma_{ij} + \tfrac{1}{3!}\,P^{ijk}\,\gamma_{ijk} + \tfrac{1}{4!}\,P^{ijkl}\,\gamma_{ijkl}\Big) \, \pi_+\,,\\[4pt]
    &\big\{Q_-\,, \, \bar{Q}_-\big\} = 2\,\Big( N + N^i \, \gamma_i + \tfrac{1}{4!} \, N^{ijkl}\, \gamma_{ijkl} \Big) \, \gamma_{11} \, \pi_+\,.
\end{align}
This extension introduces extra bosonic charges $H^i$, $H^{ijkl}$, $P$, $P^{ij}$, $P^{ijk}$, $P^{ijkl}$, $N^i$, $N^{ijkl}$ (those with multiple indices are antisymmetric) beyond the Bargmann ones. The interpretation will be clarified below by tracing their origins from the type IIA algebra.
They satisfy non-trivial boost commutators as follows
\begin{align}
    &[H_i\,, \, G_j] = \delta_{ij}P+P_{ij}\,,&&[P\,, \, G_i]=-N_i\,, &&[P_{ij}\,, \, G_k] = 2\,N_{[i} \, \delta_{j]k}\,,&& [N_i\,, \, G_j]=0\,,
\end{align}
and similar for the higher degree charges. To see that the M0T algebra is indeed the maximal extension, we note that $\{Q_\pm\,,\,\bar{Q}_\pm\}$ are symmetric matrices carrying $\mathbf{136}$ components. This anticommutator can be expanded in symmetric gamma matrices that also satisfy $\pi_\pm \, \gamma^{\phantom{\dagger}}_\Box \, \pi_\pm=0$\,, \emph{i.e.}~$(\gamma^{}_{11}\,, \, \gamma^{}_i \, \gamma^{}_{11}\,, \, \gamma^{}_{ijkl} \, \gamma^{}_{11})$\,, with the higher-order elements related by duality. The corresponding charges carry $\mathbf{1}\oplus\mathbf{9}\oplus\mathbf{126}=\mathbf{136}$ components (noting that the order-four charge is self-dual). For the mixed anticommutator $\{Q^{}_+\,, \, \bar{Q}^{}_-\}$ that contains $\mathbf{256}$ components, we have to expand in gamma matrices that satisfy $\pi^{}_- \, \gamma^{\phantom{\dagger}}_\Box \, \pi^{}_+ = 0$\,, leaving $(\mathbb{1}\,, \, \gamma^{}_i\,, \, \gamma^{}_{ij}\,, \, \gamma^{}_{ijk}\,, \, \gamma^{}_{ijkl})$ and their corresponding charges carrying $\mathbf{1}\oplus\mathbf{9}\oplus\mathbf{36}\oplus\mathbf{84}\oplus\mathbf{126}=\mathbf{16}\otimes\mathbf{16}$ components. This proves that the M0T algebra is indeed the maximally extended version of the Bargmann algebra.

Now, we show that the M0T superalgebra is an $\dot{\text{I}}$nönü-Wigner contraction of the relativistic IIA superalgebra. The maximally extended IIA algebra with a 32 component Majorana supercharge $\mathcal{Q}$ can be written as follows \cite{Townsend:1997wg}
\begin{align} \label{eq:IIAsuperalgebra}
    \big\{\mathcal{Q}\,,\,\bar{\mathcal{Q}}\big\} = \mathcal{P}^\mu \, \gamma^{}_\mu + \mathcal{A}^{}_0 \, \gamma^{}_{11} + \mathcal{A}_1^\mu \, \gamma^{}_\mu \gamma^{}_{11} + \tfrac12\,\mathcal{A}_2^{\mu\nu}\, \gamma_{\mu\nu}\,,
\end{align}
where we suppress higher charges for simplicity,
supplemented by the usual Poincar\'e commutators and appropriate ones for the extensions $(\mathcal A_0\,, \, \mathcal A_1\,, \, \mathcal A_2)$\,. The IIA superalgebra allows for an $\dot{\text{I}}$nönü-Wigner contraction that is adapted to the D0-brane states. This contraction is defined by the following prescriptions, where $\mu=(0\,,\,i)$ and a contraction parameter $\epsilon$ is introduced:
\begin{subequations} \label{eq:M0TIW}
\begin{align}
    &Q_- \equiv \tfrac{\epsilon}{2} \, \big(\mathbb{1} -\gamma^{}_0 \gamma^{}_{11}\big) \, \mathcal{Q}\,, && Q_+ \equiv \tfrac{1}{2} \, \big(\mathbb{1} +\gamma^{}_0 \gamma^{}_{11}\big) \, \mathcal{Q}\,, \\[4pt]
    &H^{\phantom i} \equiv \mathcal{P}^0 - \mathcal{A}_0\,,
    && N^{\phantom i} \equiv \tfrac{\epsilon^2}{2}\big(\mathcal{P}^0 + \mathcal{A}_0\big)\,, 
    && P^{\phantom i} \equiv -\epsilon \, \mathcal{A}_1^0\,, 
    &&  P^{ij} \equiv \epsilon \, \mathcal{A}_2^{ij}\,,\\[4pt]
    &H^i = \mathcal{A}_2^{i0} - \mathcal{A}_1^i\,,
    && N^i \equiv \tfrac{\epsilon^2}{2} \, \big(\mathcal{A}_2^{i0} + \mathcal{A}_1^i\big)\,,
    &&P^i \equiv \epsilon\,\mathcal{P}^i \,, 
    && G_i = \epsilon\,J_{0i}\,.  
\end{align}
\end{subequations}
The $\mathfrak{so}(9)$-generators $J_{ij}$ are not rescaled. Higher charges work analogously.
In the $\epsilon\to 0$ limit, the IIA superalgebra contracts to the M0T superalgebra. 

Viewing the M0T algebra as an $\dot{\text{I}}$nönü-Wigner contraction of the IIA superalgebra also enables us to interpret the various bosonic charges that go beyond the Bargmann algebra.
Consider the ground-state configuration with $H=0$\,, where the only nonzero charge is the rest mass $N\neq 0$\,. This M0T state originates from a configuration in the IIA algebra with $\mathcal{P}^0=\mathcal{A}_0$\,, which is the half-supersymmetric D0-brane state in type IIA superstring theory~\cite{Townsend:1997wg}. This shows from the algebraic perspective how the M0T limit is a BPS decoupling limit adapted to the D0-brane states. It is also natural to consider other half-supersymmetric states with other charges turned on. For example, consider an M0T configuration with $\gamma^{}_1 \, (Q_+\,, \, Q_-) = (-Q_+\,, \, Q_-)$\,, as well as $H=-H^1$ and a nonzero $N = N^1$\,. From the contraction prescription~\eqref{eq:M0TIW}, it is clear that this M0T configuration originates from a state with $\mathcal{P}^0=\mathcal{A}_1^1$\,, which is a static fundamental string extending along the $x^1$ direction. For another example, an M0T state with $\gamma^{}_{012}(Q_+,Q_-) = (Q_-,Q_+)$ and $H=2\,N = P^{12}$ corresponds to a D2-brane extending along the $x^1$ and $x^2$ directions. Naturally, one could attempt a more exhaustive analysis of possible supersymmetric states in the M0T algebra following \cite{Townsend:1997wg}. We will not pursue this further here and leave it for future studies.

\subsection{Spatial T-Duality and Matrix \texorpdfstring{$p$}{p}-Brane Theory} \label{sec:stdmpt}

A fundamental feature of string theory is the presence of T-duality when compact isometries are present.
We now discuss how this feature is inherited by the decoupling limits we are going to study, starting with the T-duality transformation along a compact spatial isometry in M0T. We begin with the IIA theory before the decoupling limit is performed. Consider a spacetime Killing vector $k^\mu$ that satisfies $k^\mu \, k^\nu \, G_{\mu\nu} > 0$\,. Choose the coordinate system $x^\mu = (x^m, \, y)$ adapted to $k^\mu$\,, such that $\p_y = k^\mu \, \p_\mu$\,. T-dualising along $y$ gives rise to the dual theory, which is type IIB superstring theory in the following background field configurations~\cite{Buscher:1987sk, Buscher:1987qj}:
\begin{subequations} \label{eq:ubrs}
\begin{align}
    \tilde{G}_{yy} & = \frac{1}{G_{yy}}\,, 
        &
    \tilde{\Phi} & = \Phi - \tfrac{1}{2} \ln G_{yy}\,, \\[4pt]
    \tilde{G}_{my} & = \frac{B_{my}}{G_{yy}}\,, 
        &
    \tilde{B}_{my} & = \frac{G_{my}}{G_{yy}}\,, \\[4pt]
    \tilde{G}_{mn} & = G_{mn} - \frac{G_{my} \, G_{ny} - B_{my} \, B_{ny}}{G_{yy}}\,, 
        &
    \tilde{B}_{mn} & = B_{mn} + \frac{G_{my} \, B_{ny} - G_{ny} \, B_{my}}{G_{yy}}\,,
\end{align}
\end{subequations}
while for the RR sector \cite{Bergshoeff:1995as} (in conventions following \cite{Ebert:2021mfu}),
\begin{subequations}
\begin{align}
    \tilde{C}^{(q+1)}_y & = C^{(q)} - \frac{C^{(q)}_y \wedge G_y}{G_{yy}}\,, \\[4pt]
    \tilde{C}^{(q-1)} & = C^{(q)}_y + C^{(q-2)} \wedge B_y + \frac{C^{(q-2)}_y \wedge B_y \wedge G_y}{G_{yy}}\,. 
\end{align}
\end{subequations}
Here, $G_y = G_{my} \, \dd x^m$, $B_y = B_{my} \, \dd x^m$, and $C^{(q+1)}_y = \frac{1}{q!} \, C^{}_{m_1\cdots m_q \, y} \, \dd x^{m_1} \wedge \cdots \wedge \dd x^{m_q}$\,. 

We would like the compact spatial isometry to reside within the spatial hypersurface in Newton-Cartan geometry after taking the $\omega \rightarrow \infty$ limit. 
We therefore assume that 
\be
    \tau_y{}^0 = 0\,,
        \qquad%
    E_{yy} \neq 0\,,
\ee
which guarantees that $k^\mu \, k^\nu \, G_{\mu\nu} > 0$\,. 
Alongside setting $\tau_y{}^0=0$, we can choose to gauge fix the spatial vielbein such that
\be
    E_y{}^1 \neq 0\,,
        \qquad
    E_y{}^{A'} = 0\,,
        \,\,
    A' = 2\,, \, \cdots, \, 9\,.
\ee
Using this gauge fixing and the T-duality rules~\eqref{eq:ubrs}, it follows that the dual background has the following form:
\begin{subequations} \label{eq:rpgcfbmot}
\begin{align}
    \tilde{G}_{\mu\nu} & = \omega \, \tilde{\tau}^{}_{\mu\nu} + \omega^{-1} \, \tilde{E}^{}_{\mu\nu}\,, 
        &
    \tilde{C}^{(2)} & = \omega^2 \, e^{-\tilde{\varphi}} \, \tau^0 \wedge \tilde{\tau}^1 + \tilde{c}^{\,(2)}\,, \\[4pt]
    \tilde{\Phi} & = \tilde{\varphi} - \ln \omega\,, 
        &
    \tilde{C}^{(q)} & = \tilde{c}^{(q)} \text{ if } q \neq 2\,,
        \qquad%
    \tilde{B}^{(2)} = \tilde{b}^{(2)}\,,
\end{align}
\end{subequations}
where 
\be
    \tilde{\tau}_{\mu\nu} = - \tau_\mu{}^0 \, \tau_\nu{}^0 + \tilde{\tau}_\mu{}^1 \, \tilde{\tau}_\nu{}^1\,, 
        \qquad%
    \tilde{E}_{\mu\nu} = E_\mu{}^{A'} \, E_\nu{}^{A'}\,,
\ee
and
\begin{subequations} \label{eq:mzttmot}
\begin{align}
    \tilde{\tau}_{y}{}^1 &= \frac{1}{E_{y}{}^1}\,, 
        &
    \tilde{b}_{my} & = \frac{E_{my}}{E_{yy}}\,,
    \hskip2.4truecm%
    \tilde{\varphi} = \varphi - \ln \bigl|E_{y}{}^1\bigr|\,,
     \\[4pt]
    \tilde{\tau}^{}_{m}{}^1 &= \frac{b_{my}}{E_{y}{}^1}\,,
        &
    \tilde{b}_{mn} & = b_{mn} + \frac{2 \, E_{y[m} \, b_{n]y}}{E_{yy}}\,,
        &
     \\[4pt]
    \tilde{c}^{}_{my} & = c^{}_m - \frac{c^{}_y \, E_{my}}{E_{yy}}\,, 
        &
    \tilde{c}^{}_{mn} & = c^{}_{mny} + 2 \, c^{}_{[m} \, b^{}_{n]y} - 2 \, \frac{c^{}_y}{E^{}_{yy}} \, E^{}_{y[m} b^{}_{n]y} \,.
\end{align}
\end{subequations}
Dropping the tildes in Eq.~\eqref{eq:rpgcfbmot}, we find the defining prescriptions for a decoupling limit of type IIB superstring theory, where the target space develops a codimension-two foliation structure. In the infinite $\omega$ limit, and after decompactifying the compact isometry direction, we are led to what we will refer to as matrix 1-brane theory (M1T), where a background RR two-form instead of one-form potential is taken to its critical value. The fundamental degrees of freedom in M1T are now the D1-strings that are T-dual to the D0-particles. 
The dynamics of the D1-strings are captured by matrix string theory~\cite{Motl:1997th, Dijkgraaf:1997vv}. 

We now generalise the above T-duality transformation to find the prescriptions for defining \emph{matrix $p$-brane theory} (M$p$T), where the ten-dimensional spacetime develops a codimension-$(p+1)$ foliation structure. The defining prescriptions for M$p$T are~\cite{Blair:2023noj, Gomis:2023eav, Ebert:2021mfu} (see also~\cite{Gopakumar:2000ep, Gomis:2000bd, Gomis:2004pw} for related reparametrisations in flat spacetime)
\begin{subequations} \label{eq:rpgcfbmpt}
\begin{align}
    G_{\mu\nu} & = \omega \, \tau^{}_{\mu\nu} + \omega^{-1} \, E^{}_{\mu\nu}\,, 
        &
    C^{(p+1)} & = \omega^2 \, e^{-\varphi} \, \tau^0 \wedge \cdots \wedge \tau^p + c^{\,(p+1)}\,, \\[4pt]
    \Phi & = \varphi + \tfrac{1}{2} \bigl( p - 3 \bigr) \ln \omega\,, 
        &
    C^{(q)} & = c^{(q)} \quad \text{ if } q \neq p+1\,,
        &
    B^{(2)} &= b^{(2)}\,,
\end{align}
\end{subequations}
where 
\begin{align}
    \tau^{}_{\mu\nu} = \tau^{}_\mu{}^A \, \tau^{}_\nu{}^B \, \eta^{}_{AB}\,,
        \qquad%
    E^{}_{\mu\nu} = E^{}_\mu{}^{A'} \, E^{}_\nu{}^{A'},
\end{align}
with $A = 0\,, \, \cdots, \, p$ and $A' = p+1 \,, \, \cdots, \, 9$\,. 
This defines a decoupling limit of type II string theory.
While here we obtained these definitions via T-duality, in general no isometries are assumed in writing \eqref{eq:rpgcfbmpt}. 
The non-Lorentzian geometry now admits as symmetries local $\mathrm{SO}(1\,,\,p)$ and $\mathrm{SO}(9-p)$ transformations, acting on $\tau_\mu{}^A$ and $E^{}_\mu{}^{A'}$ respectively, as well as $p$-brane Galilean boosts
\be
    \delta^{}_\text{G} \tau_\mu{}^A = 0 \,,
        \qquad%
    \delta^{}_\text{G} E_\mu{}^{A'} = \Lambda^{A'}{}_{\!A} \, \tau_\mu{}^A \,.
\ee
Moreover, as in Eq.~\eqref{eq:gbtc1e} for M0T, the RR $(p + 1)$-form in M$p$T also transforms nontrivially under the $p$-brane Galilean boosts, with
\be
    \delta^{}_\text{G} c^{(p+1)} = \frac{1}{p!} \, e^{-\varphi} \, \Lambda^{A'A} \, E^{A'} \wedge \tau^{A_1} \wedge \cdots \wedge \tau^{A_p} \, \epsilon^{}_{AA_1\cdots A_p}\,. 
\ee
Note that $\Lambda^i$ in Eq.~\eqref{eq:gbtc1e} is identified with $\Lambda^{A'}{}_0 = - \Lambda^{A'0}$ when $p=0$\,. We thus realise a $p$-brane generalisation of the Newton-Cartan geometry that we have discussed in Section~\ref{sec:mzbt} (this is sometimes referred to as `$p$-brane Newton-Cartan geometry'). Furthermore, generalising the case for M0T in Eq.~\eqref{eq:dilatsm0t}, M$p$T also admits an emergent dilatation symmetry arising as an enhancement of the ambiguity in the definition of $\omega$ that controls the limit. Namely, under the gauge transformations,
\begin{align} \label{eq:dilatsmpt}
    \tau^A \rightarrow \Delta^{\frac{1}{2}} \, \tau^A\,,
        \qquad%
    E^{A'} \rightarrow \Delta^{-\frac{1}{2}} \, E^{A'},
        \qquad%
    e^\varphi \rightarrow \Delta^{\frac{p - 3}{2}} \, e^\varphi\,,
\end{align}
M$p$T is mapped to itself. 

The Buscher rules associated with the T-duality transformation from M$p$T to M($p$+1)T can be obtained similarly to those leading from M0T to M1T. We use the untilded notation for the background fields in M$p$T and the tilded notation in M($p$+1)T. We now choose a spacelike Killing vector such that, in the adapted coordinates $x^\mu = (x^m, \, y)$\,, 
\be
\label{eq:gaugefixingMpTBuscher} 
    \tau^{}_y{}^A = 0\,,
        \qquad%
    E^{}_y{}^{p+1} \neq 0\,, 
        \qquad%
    E^{}_y{}^{\CA'}=0, \,\, \CA' = p+2\,, \, \cdots, \, 9\,.
\ee
We then find
\begin{subequations} \label{eq:rpgcfbmppot}
\begin{align}
    \tilde{G}_{\mu\nu} & = \omega \, \tilde{\tau}^{}_{\mu\nu} + \omega^{-1} \, \tilde{E}^{}_{\mu\nu}\,, 
        &
    \tilde{C}^{(p+2)} & = \omega^2 \, e^{-\tilde{\varphi}} \, \tau^0 \wedge \cdots \wedge \tau^p \wedge \tilde{\tau}^{\,p+1} + \tilde{c}^{\,(p+2)}\,, \\[4pt]
    \tilde{\Phi} & = \tilde{\varphi} + \tfrac{1}{2} \bigl( p - 3 \bigr) \ln \omega\,, 
        &
    \tilde{C}^{(q)} & = \tilde{c}^{(q)} \text{ if } q \neq p+2\,,
        \qquad%
    \tilde{B}^{(2)} = \tilde{b}^{(2)}\,,
\end{align}
\end{subequations}
where 
\be
    \tilde{\tau}_{\mu\nu} = - \tau_\mu{}^A \, \tau_\nu{}^B \, \eta^{}_{AB} + \tilde{\tau}_\mu{}^{p+1} \, \tilde{\tau}_\nu{}^{p+1}\,, 
        \qquad%
    \tilde{E}_{\mu\nu} = E_\mu{}^{\CA'} \, E_\nu{}^{\CA'}\,,
\ee
and
\begin{subequations} \label{eq:mzttmpptt}
\begin{align}
    \tilde{\tau}_{y}{}^{p+1} \! &= \frac{1}{E_{y}{}^{p+1}}\,, 
        &
    \tilde{b}_{my} & = \frac{E_{my}}{E_{yy}}\,,
        \qquad\qquad\qquad%
    \tilde{\varphi} = \varphi - \ln \bigl|E_{y}{}^{p+1}\bigr|\,,
     \\[4pt]
    \tilde{\tau}^{}_{m}{}^{p+1} \! & = \frac{b_{my}}{E_{y}{}^{p+1}}\,, 
        &
    \tilde{b}_{mn} & = b_{mn} + \frac{2 \, E_{y[m} \, b_{n]y}}{E_{yy}}\,,
        &
     \\[4pt]
    \tilde{c}^{\,(p+2)}_{y} \! & = c^{(p+1)}_m - \frac{c^{(p+1)}_y \! \wedge E_{y}}{E_{yy}}\,,
        &
    \tilde{c}^{\,(p+2)} \! & = c^{(p+3)}_{y} + c^{(p+1)} \! \wedge b^{(2)}_{y} - \frac{c^{(p+1)}_y \! \wedge E^{}_{y} \wedge b^{}_{y}}{E^{}_{yy}} \,.
\end{align}
\end{subequations}
The Buscher rules in Eq.~\eqref{eq:mzttmpptt} are direction generalisations of the ones in Eq.~\eqref{eq:mzttmot}. Here, $E_y = dx^m \, E_{my}$ and $b_y = dx^m \, b_{my}$\,, and it is understood that the suppressed indices of $\tilde{c}^{\,(q)}$ in Eq.~\eqref{eq:mzttmpptt} are the `$m$' index, excluding the isometry $y$ index. A worldsheet derivation of the same T-duality transformation can be found in~\cite{Gomis:2023eav}.

\subsection{Light D-Branes and Matrix Gauge Theories}
\label{sec:dmgt} 

In this subsection, we discuss some features of D$q$-branes in M$p$T, including demonstrating how to obtain matrix theories on D$p$-branes.

\subsubsection{Generic D\texorpdfstring{$q$}{q}-Branes in Matrix \texorpdfstring{$p$}{p}-Brane Theory} \label{sec:gdbmpt}

We start by considering the M$p$T BPS decoupling limit in a flat background, with trivial gauge fields $c^{(q)} = b^{(2)} = 0$ and constant dilaton with $g^{}_\text{s} \equiv e^{\varphi}$.
We consider the following action that describes a single D$q$-brane action in the original Lorentzian target space:
\be \label{eq:sdpba}
    S^{}_\text{D$q$} = - \frac{T_q}{g_\text{s}} \int \dd^{q+1} \sigma \, \sqrt{-\det \Bigl( \p_{\alpha} X^\mu_{\phantom{\dagger}} \, \p_{\beta} X_\mu \Bigr)} + T_q \int C^{(q+1)}\,,
\ee
where $T_q = (2\pi)^{-q} \, \alpha'{}^{-(q+1)/2}$\,. 
The limit of this action depends on the orientation of the brane. 
We suppose that the brane has $m$ longitudinal directions which are transverse to the M$p$T limit, and $n$ transverse directions which are longitudinal to the M$p$T limit.
We necessarily have $p+m=q+n$.
This configuration can be summarised in the following table:
\begin{center}
\begin{tabular}{c|c|c|c|c}
$\sharp$ of directions & $q-m+1$ & \,$\phantom{q-}m\phantom{-q}$\, & \,$\phantom{q-}n\phantom{-q}$\, & $9 - q - n$ \\[2pt]
\hline 
D$q$-brane & $\times$ & $\times$ & -- & -- \\[2pt]
M$p$T limit & $\times$ & -- & $\times$ & -- 
\end{tabular}
\end{center}

We now consider a static configuration for the D$q$-branes, wrapping the directions illustrated in this table, and at fixed positions in the transverse directions, such that the limiting prescription~\eqref{eq:rpmzt} implies
\be
\label{DqScalingX}
    X^{0\,, \, \cdots, \, q - m} \rightarrow \sqrt{\omega} \, \sigma^{0\,, \, \cdots, \, q - m}\,,
        \qquad%
    X^{q-m+1\,, \, \cdots, \, q} \rightarrow \frac{\sigma^{q-m+1\,, \, \cdots, \, q}}{\sqrt{\omega}}\,,
\ee
along with $g^{}_\text{s} \rightarrow g^{}_\text{s} \, \omega^{(p-3)/2}$\,.
We plug this into \eqref{eq:sdpba}. 
With this static configuration the only case where the Chern-Simons term is relevant is then the special case $p=q$ and $m=n=0$ of a longitudinally aligned D$p$-brane. In this case, $S_{\text{D}p}$ vanishes.
This reflects the fact that these are always present as light excitations in M$p$T. 
Otherwise, the action evaluates to
\be
 S^{}_\text{D$q$} = - \frac{T_q}{g_\text{s}} \, \omega^{\frac{1}{2} (4 - p +q -2m) } \, V_{\text{D}q}\,,
\label{eq:dqmpt}
\ee
where $V_{\text{D}q} = \int \dd^{q+1} \sigma$ is the spacetime volume of the D$q$-brane.
If $4-p+q-2m>0$, the expression \eqref{eq:dqmpt} diverges and we interpret this as signalling that the D$q$-brane configuration has infinite mass in the M$p$T limit.
On the other hand, if $4-p+q-2m=0$ then the configuration has finite mass. 
If $4-p+q-2m < 0$, then it will have vanishing mass. 
We can rewrite the condition for finite/vanishing mass equivalently as $4-n-m \leq 0$.

For example, let us consider the cases where $m=0$, so that the D$q$-brane is longitudinal with respect to the M$p$T limit.
We then see that longitudinal D$(p-4)$-branes have finite mass in M$p$T, while longitudinal D$(p-6)$-branes have vanishing mass.
In particular, the D0-brane in M$4$T has finite mass, and in M$6$T is massless \cite{Sen:1997we}. Indeed, in the former case we can easily write down the abelian D0-brane action after taking the M4T limit in a general background,
\be
S_{\text{D$0$}}= - T_0 \int \dd \tau \,e^{-\varphi} \sqrt{- \dot{X}^\mu \, \dot{X}^\nu \, \tau_{\mu\nu}} + T_0 \int c^{(1)} \,,
\ee
showing that the D0-brane probe only sees the five-dimensional longitudinal part of the M4T geometry.

As a further example, we consider the cases with $n=0$, such that the longitudinal directions of the M$p$T limit are contained within the longitudinal directions of the D$q$-brane.
The condition for finite mass is now $4 + p -q = 0$.
In particular, the D4-brane in M0T has finite mass. 
This is consistent with T-duality between M0T and M4T \cite{Sen:1997we}.
In fact, the D4-brane in M0T plays an important role in the Berkooz-Douglas matrix theory~\cite{Berkooz:1996is}: 
compactifying M5-branes over a lightlike circle in M-theory gives rise to an interacting system of D4-branes and D0-branes, which is described by the Berkooz-Douglas matrix theory. The light modes here are still the D0-branes, but with a modified dynamics due to the D0-D4 bound state.

We will later re-encounter the condition $4-n-m=0$\,, which here singles out the finite mass D$q$-brane configurations in a flat M$p$T background, when we examine the M$p$T limit applied to general curved D$q$-brane supergravity solutions in Section \ref{sec:morebranegeos}. There it will be needed to ensure that the RR gauge potential of the D$q$ solution is finite in the limit.
A complete list of configurations satisfying the condition, with $p\,,\,q \leq 4$, can be found in Table \ref{tab:IIbranes} in that section.

A D$q$-brane in M$p$T with infinite mass according to the above analysis can still have interesting dynamics when more general field configurations are turned on. For example, consider the D2-brane in the M0T limit, \emph{i.e.}~$p=0$, $q=2$, $m=2$ and $n=0$\,. 
In the flat M0T background, the D2-brane action in the static configuration follows from Eq.~\eqref{eq:dqmpt} as $ S_\text{D2} = - g_\text{s}^{-1} \, T_q \, \omega \, V_{\text{D}2}$\,.
However, the outcome of the M$p$T limit of this brane is altered in the presence of a spatial Kalb-Ramond field or equivalently with a non-trivial gauge potential on the brane.
In this case, the limit leads to noncommutative Yang-Mills (NCYM) theory~\cite{Gopakumar:2000na, Gopakumar:2000ep}. In order to understand this subtlety, we consider the conventional worldvolume action describing a single D2-brane in the IIA theory, 
\be \label{eq:sdtbf}
    S^{}_\text{D2} = - \frac{T_2}{g_\text{s}} \int \dd^3 \sigma \, \sqrt{-\det \Bigl( \p_{\alpha} X^\mu_{\phantom{\dagger}} \, \p_{\beta} X_\mu + \CF_{\alpha\beta} \Bigr)} + T_2 \int \Bigl( C^{(3)} + C^{(1)} \wedge \CF \Bigr)\,,
\ee
where $\CF_{\alpha\beta} = B_{\alpha\beta} + F_{\alpha\beta}$\,, with $F = \dd A$ the gauge field strength on the brane. When the brane is static, the M0T version of the reparametrisation~\eqref{DqScalingX} implies
\be
    X^0 \rightarrow \sqrt{\omega} \, \sigma^0\,,
        \qquad%
    X^a \rightarrow \frac{\sigma^a}{\sqrt{\omega}}\,,
        \,\,%
    a = 1\,, \, 2\,,
        \qquad%
    g^{}_\text{s} \rightarrow \omega^{-\frac{3}{2}} \, g^{}_\text{s}\,. 
\ee
Moreover, due to the topological coupling involving $\mathcal{F}$\,, we also have a contribution from the RR one-form (see Eq.~\eqref{eq:rpmzt}),
\be
    C^{(1)} \rightarrow \omega^2 \, g^{-1}_\text{s} \, dX^0\,.
\ee
As a result, at large $\omega$\,, the D2-brane action becomes
\be
    S^{}_\text{D2} = \omega^2 \, \frac{T_2}{g_\text{s}} \, \Bigl( \CF_{12} - |\CF_{12}| \Bigr) - \frac{T_2}{2 \, g_\text{s}} \int \dd^3 \sigma \, \frac{1 - \CF_{0a} \, \CF_{0a}}{|\CF_{12}|} 
    + O \bigl(\omega^{-2}\bigr)
    \,.
\ee
In the case where $\CF_{12} > 0$\,, the $\omega^2$-divergence disappears, and the D2-brane action in M0T is given by
\be
    S^\text{M0T}_\text{D2} = - \frac{T_2}{2 \, g_\text{s}} \int \dd^3 \sigma \, \frac{1 - \CF_{0a} \, \CF_{0a}}{\CF_{12}} \,.
\ee
This is NCYM with its gauge coupling being proportional to $\sqrt{|B_{12}|}$\,. Using the Seiberg-Witten map~\cite{Seiberg:1999vs}, we find that, from the open string perspective, the spatial coordinates on the D2-brane develop a noncommutative behaviour, with
$\bigl[ \sigma^{}_1\,, \, \sigma^{}_2 \bigr] \sim B^{-1}_{12}$\,. See~\cite{Gomis:2023eav, Ebert:2021mfu} for more detailed discussion using the same M$p$T parametrisation. 

The above analysis shows that the physics of D$q$-branes in general M$p$T backgrounds contains rich and interesting physics. We will however set aside the goal of pursuing the worldvolume dynamics of generic D$q$-branes in M$p$T for future work, and instead focus now on the special case which connects directly to the usual framework of matrix gauge theory.

\subsubsection{Matrix Gauge Theory from D\texorpdfstring{$p$}{p}-Branes} \label{sec:mgt}

We now consider the special case with $p = q$\,, \emph{i.e.} D$p$-branes in M$p$T. 
In the case of a single D$p$-brane, if the pullback $\tau^{}_{\alpha\beta}$ is invertible, the $\omega \rightarrow \infty$ limit for such a brane configuration is always finite 
and gives rise to the following action:
\be \label{eq:mptdp}
    S^{\text{\scalebox{0.8}{M$p$T}}}_\text{D$p$} = - \frac{T_p}{2} \int \dd^{p+1} \sigma \, e^{-\varphi} \, \sqrt{-\tau} \, \tau^{\alpha\beta} \, E^{}_{\alpha\beta} + T_p \int c^{(p+1)}\,.
\ee
Importantly, there is an infinite contribution from the pure brane term that is exactly cancelled by an identical divergent piece from the Chern-Simons term. Effectively, the background D$p$-brane charge is fine tuned to cancel its tension. In the special case where $p = 0$\,, this action matches the D0-brane action~\eqref{eq:mztdzba} in M0T.
We now generalise this action -- in curved backgrounds -- to the non-abelian case.

\vspace{3mm}

\noindent $\bullet$~{\it Review of non-abelian D$p$-brane.} 
We will make use of the nonabelian D-brane worldvolume action proposed in~\cite{Myers:1999ps}.
The starting point in \cite{Myers:1999ps} is the extrapolation of the abelian Dirac-Born-Infeld (DBI) action to the nonabelian case for the special case of the D9 brane, which leads to:
\be
    S_\text{D9} = - T_9 \int \dd^{10} \sigma \, \text{STr} \! \left[ e^{-\Phi} \, \sqrt{- \det \bigl( \CE_{\alpha\beta} + F_{\alpha\beta} \bigr)} \, \right] + T_9 \int \text{STr} \biggl(  \sum_q C^{(q)} \wedge e^{B+F} \biggr)\,,
\label{Myers9}
\ee
where $\CE_{\alpha\beta} = G_{\alpha\beta} + B_{\alpha\beta}$\,, and we have introduced the symmetrised trace prescription, denoted by `STr'. 
For D$p$-branes with $p<9$, the same extrapolation turns out to be incompatible with T-duality.
Requiring T-duality covariance leads to the following D$p$-brane non-abelian action~\cite{Myers:1999ps}
\begin{align} \label{Myers} 
S_\text{D$p$} & = - T_p \int \dd^{p+1} \sigma \, \text{STr} \Biggl[ e^{-\Phi} \,
 \sqrt{ - \det\Bigl(\text{P}\bigl[\CE_{\wa \wb} + \CE_{\wa i} \, ( Q^{-1}-\delta)^{i}{}_k \, \CE^{kj} \, \CE_{j \wb}\bigr] + F_{\wa \wb} \Bigr) \, \det \bigl( Q^i{}_j \bigr)} \, \Biggr] \notag \\[4pt]
& \quad + T_p \int \text{STr} \! \left(
\text{P}\Bigl[e^{i \, \iota_X \iota_X} \sum_q C^{(q)} \wedge e^B \Bigr] \wedge e^{F}
\right).
\end{align}
Here we have split the target space coordinates as $X^\mu = (X^\alpha, \, X^i)$ with $i = p+1\,, \, \cdots, \, 9$\,. 
A static gauge choice has been made such that $\sigma^\wa = X^\wa$\,, $\wa=0\,,\,\dots, \, p$\,, while the $X^i$ are (matrix valued) transverse scalars.
We have set $2\pi \alpha'=1$ for simplicity.
There is further a worldvolume gauge field with field strength $F_{\wa \wb}$ and covariant derivative $D_\wa X^i$ acting on the scalars in the adjoint.
The background metric and $B$-field appear via the combination 
\be
    \CE_{\mu\nu} = G_{\mu\nu} + B_{\mu\nu}\,.
\ee
Pullbacks are defined such that given a covector $\omega_\mu$ carrying a spacetime index we have $P[\omega_\wa] \equiv \omega_\wa + \omega_i \, D_\wa X^i$, so explicitly:
\be
\begin{split} 
    \text{P}[\CE_{\wa\wb}]  & \equiv \CE_{\wa\wb} + \CE_{i\wb} \, D_{\wa} X^i 
+ \CE_{\wa i} \, D_{\wb} X^i 
+ \CE_{ij} \, D_\wa X^i D_\wa X^j \,,\\[4pt]
    \text{P}[\CE_{\wa i}] & \equiv \CE_{\wa i} + \CE_{ji} \, D_\wa X^j  \,.
\end{split} 
\ee
In addition, $\CE^{ij}$ denotes the inverse of $\CE_{ij}$, and the matrix $Q$ is defined by
\be
    Q^i{}_j = \delta^i_j + i \, \bigl[X^i, X^k\bigr] \, \CE_{kj} \,.
\ee
The notation $\iota_X$ denotes the interior product, 
\be
    \left(\iota_X C^{(p)} \right){}^{}_{\!\mu_1 \dots \mu_{p-1}} = X^j \, C^{}_{j \mu_1 \dots \mu_{p-1}}\,. 
\ee
The above action is such that if we take the transverse directions to correspond to a torus, then the usual T-duality rules map the action \eqref{Myers} for general $p$ to the action \eqref{Myers9}.
However, more generally the spacetime fields may depend on the non-abelian scalars $X^i$ via their functional dependence on the spacetime coordinates, so implicitly they should be treated as functionals via a matrix Taylor expansion, {\it e.g.}
\be
G_\text{MN}(\sigma^\wa, \, X^i) = 
\exp\!\left( X^i \, \frac{\partial}{\p Y^i} \right) G_\text{MN} \bigl( \sigma^\wa, \, Y^j \bigr) \, \Big|_{Y^i=0} \,.
\ee
See \cite{Myers:1999ps} for further details.
Unlike the abelian DBI action, the non-abelian actions \eqref{Myers9} and \eqref{Myers} are known not to capture all $\alpha'$ orders, and must receive corrections. 
At least for $p\leq 3$, the decoupling limits we consider can be recast as an $\alpha'\rightarrow 0$ limit \cite{Sen:1997we} so this does not affect the conclusions.

\vspace{3mm}

\noindent $\bullet$~{\it Decoupling limits.} It is simplest to first consider the M9T limit of the D$9$-brane action. 
Here there are no transverse scalars, and the non-abelian action for the background defining the M$9$T limit involves:
\be
\begin{split} 
    S_{\text{D9}} & = - T_9 \int \dd^{10} \sigma \,  \text{STr} \left[
\omega^{-3} \, e^{-\varphi} \sqrt{-\det \Bigl( \omega \, \tau_{\wa\wb} + \mathcal{F}_{\wa\wb} \Bigr)} \right] \\[4pt]
    & \quad + T_9 \int \text{STr} \left(
\omega^2 e^{-\varphi} \tau^0 \wedge \dots \wedge \tau^9 + 
P \Bigl[ \sum_n c^{(n)} \wedge e^b \Bigr] \wedge e^{F} \right).
\end{split} 
\ee
Note that
\be
    \sqrt{-\det \Bigl( \omega \, \tau_{\wa\wb} + \mathcal{F}_{\wa\wb} \Bigr)} = \omega^2 \, \sqrt{-\tau} \Bigl[ 1 + \tfrac{1}{4} \, \omega^{-2} \, \tau^{\wa\wb} \, \tau^{\wc\wvd} \, F_{\wa\wc} \, F_{\wa\wvd}  + \mathcal{O} \bigl( \omega^{-4} \bigr) \Bigr]\,.
\ee
Here, $\tau^{\alpha\beta}$ is the inverse of $\tau_{\alpha\beta}$ and $\tau = \det (\tau_{\alpha\beta})$\,. 
In the $\omega \rightarrow \infty$ limit, we find the D9-brane in M9T:
\be \label{eq:mntdn}
\begin{split} 
    S^\text{M9T}_{\text{D9}} 
    & = - \frac{T_9}{4} \int \dd^{10} \sigma \, \text{STr} \left( e^{-\varphi} \, \sqrt{-\tau} \, \tau^{\alpha\beta} \, \tau^{\gamma\delta} \, F^{}_{\alpha\gamma} \, F^{}_{\beta\delta} \right) \\[4pt]
    & \quad + T_9 \int \text{STr} \left( \sum_n c^{(n)} \wedge e^b \wedge e^{F} \right).
\end{split} 
\ee
The first line in Eq.~\eqref{eq:mntdn} reproduces the Maxwell theory in ten-dimensions in the case where $\tau^{}_{\alpha\beta} = \eta^{}_{\alpha\beta}$\,. When a stack of D9-branes in M9T are considered and the fermionic sector is added, this Maxwell theory will be generalised to $\CN = 1$ SYM in ten dimensions.

\vspace{3mm}

\noindent $\bullet$~{\it Longitudinal D$p$-branes in M$p$T.} 
Before diving straight into the M$p$T limit of the D$p$-brane, we stop to think about the consequences of T-duality compatibility in the derivation of the D-brane action. 
Compactifying the D9-brane in M9T over a $(9-p)$-torus, followed by T-dualising in all the toroidal directions, gives rise to the D$p$-brane action \eqref{Myers}. Similarly, starting with M$9$T, the D$p$-brane action in M$p$T should follow from a longitudinal T-duality on $9-p$ directions.

Now, the T-duality transformation between M$p$T and M$q$T in general background fields turns out to contain some subtleties. 
In particular, T-duality on multiple directions is sensitive to the presence of a non-trivial transverse $B$-field, such that the T-duality transformation on $k$ transverse directions may map M$p$T to itself, or to M$(p+r)$T for some $r < k$\,, rather than to M$(p+k)$T.
We will explore this in more detail in another publication \cite{MxReloadedII}.\, One way to understand this subtlety is to realise that we derive the Buscher rules between M$p$T and M$(p+1)$T with a particular gauge fixing of the form~\eqref{eq:gaugefixingMpTBuscher} to ensure that we are performing a longitudinal T-duality.
Iterating this multiple times, it turns out that T-duality between M$p$T and M$q$T with $q > p+1$ on  multiple directions labelled by $i$ requires the conditions
\be \label{eq:zerointernalbfield}
    B^{}_{ij} = 0\,,\quad   
    \tau^{}_{ij} = \tau^{}_{i\alpha} = 0\,.
\ee
To obtain the correct limit of the non-abelian action that describes D$p$-branes in M$p$T, and is compatible with the T-duality between the cases $p=9$ and $p<9$, we have to impose \eqref{eq:zerointernalbfield}.
Pragmatically, these conditions ensure that the limit is indeed finite.
We then find in Eq.~\eqref{Myers} that
\be \label{eq:expqij}
    Q^i{}^{}_j = \delta^i{}^{}_j + i \, \omega^{-1} \, \bigl[ X^i , \, X^k \bigr] \, E^{}_{kj}\,.
\ee
Using the formula
\be
    \det \Bigl( \mathbb{1} + \omega^{-2} \, \CO \Bigr) = 1 + \omega^{-2} \, \tr \, \CO + \frac{1}{2} \, \omega^{-4} \, \Bigl[ \bigl( \tr \, \CO \bigr)^2 - \tr \bigl( \CO^2 \bigr) \Bigr] + O\bigl(\omega^{-6}\bigr)\,.
\ee
we find that, in the M$p$T limit with the above prescriptions, the non-abelian D$p$-brane action~\eqref{Myers} gives
\be \label{eq:mptdpnona}
\begin{split} 
S^\text{M$p$T}_{\text{D}p} & = - \frac{T^{}_p}{2} \int \dd^{p+1} \sigma \, \text{STr} \biggl[
  e^{-\varphi} \sqrt{- \tau} \, \Bigl( 
   \tau^{\wa\wb} \, \hat{\text{P}} \bigl[ E_{\wa\wb} \bigr]
  + \tfrac{1}{2} \, \tau^{\wa\wb} \, \tau^{\wc\wvd} \, \mathscr{F}_{\wa\wc} \, \mathscr{F}_{\wb\wvd} 
   \\ & \qquad\qquad\qquad\qquad\qquad\qquad
   \qquad\qquad\qquad\quad - \tfrac{1}{2} \, \bigl[ X^i, X^k \bigl] \bigl[ X^j, X^l \bigr] \, E^{}_{ij} \, E^{}_{kl}
  \Bigr)
 \biggr] \\[4pt] 
 & \quad + T^{}_p \int \text{STr} \biggl( \text{P} \Bigl[ e^{i \, \iota_X \, \iota_X} \sum_n c^{(n)} \wedge e^{b^{(2)}} \Bigr] \wedge e^{F} \biggr)\,,
\end{split}
\ee
which describes the dynamics of a stack of D$p$-brane in M$p$T. Here,
\be
\begin{split} 
    \mathscr{F}_{\wa\wb}
    & \equiv F_{\wa\wb} + \text{P}[b_{\wa\wb}] +  i \, [ \Phi^i,\Phi^j] \, \text{P}\bigl[b_{\wa i}\bigr] \, \text{P}\bigl[b_{\wb j}\bigr]\,, \\[4pt]
    \hat{\text{P}}[ E_{\wa\wb}]
    & \equiv \text{P}[E_{\wa\wb}] + i \, [\Phi^i, \Phi^j] \, \Bigl( \text{P}[E_{\wa i}] \, \text{P}\bigl[b_{\wb j}\bigr] - \text{P} \bigl[b_{\wa i}\bigr] \, \text{P} \bigl[E_{\wb j}\bigr] \Bigr)\,.
\end{split} 
\ee
Suppose we have $b^{(2)} = c^{(n)} = 0$\,, and that $E_{\mu\nu}$ is restricted so that only its fully transverse components $E_{ij}$ are non zero.
In the flat limit with $\tau_{\wa\wb} = \eta_{\wa\wb}$ and $E_{ij} = \delta_{ij}$\,, the action \eqref{eq:mptdpnona} assumes the form:
\be \label{eq:mgtbs}
\begin{split} 
    S&=  - T_p \int \dd^{p+1} \sigma \, g^{-1}_\text{s} \, \text{Tr} \Big( \tfrac{1}{2} \, D_\wa X^i \, D^\wa X^i + \tfrac{1}{4} \, F_{\wa\wb} \, F^{\wa\wb} - \tfrac{1}{4} \, \big[ X^i, X^j \big]^2
  \Big)\,,
 \end{split}
\ee
which is the bosonic part of the SYM theory in $p+1$ dimensions that arise from compactifying $\CN=1$ SYM in ten dimensions. Moreover, $g^{}_\text{s} = e^{\varphi}$\,.

Allowing the background fields to depend on the spacetime coordinates, the above action \eqref{eq:mptdpnona} gives, in principle, a $p$-brane matrix theory description of curved backgrounds, showing explicitly that this couples to the $p$-brane Newton-Cartan geometry of M$p$T.
This description inherits some of the complexities of the Myers action in~\cite{Myers:1999ps}, for instance one should Taylor expand the non-constant background fields in terms of the non-abelian matrices $X^i$, and apply the symmetrised trace prescription.

\section{Holography: Near Horizon from BPS Decoupling Limit} \label{sec:hnhbpsl}

In this section, we relate the D$p$-brane decoupling limits leading to M$p$T to the near-horizon decoupling limits leading to holography.
It turns out that this follows automatically by taking the M$p$T limit asymptotically in the curved geometries corresponding to D$q$-brane supergravity solutions.
Iterating this procedure will allow us to derive both familiar and non-Lorentzian holographic geometries.

\subsection{Bulk AdS Geometry from Matrix Theory} \label{sec:badsmt}

We start with the classic D$p$-brane holographic correspondence \cite{Maldacena:1997re,Itzhaki:1998dd}, and derive the bulk near-horizon geometry from the BPS decoupling limit that leads to matrix gauge theories at the asymptotic infinity. This examination provides a renewed understanding of the original decoupling limit that leads to the AdS/CFT correspondence, and will act as a foundation for our later discussion on generalised holographic principles.  

\vspace{3mm}

\noindent $\bullet$~\emph{String picture.}
In the terminology of the AdS/CFT correspondence, the previous section on M$p$T essentially deals with the string picture. Here, SYM arises from the M$p$T limit of $N$ coinciding D$p$-branes in type II superstring theory with a flat target space. Denote the target space coordinates in the conventional type II theory as $X^\mu$, $\mu = 0\,, \, \cdots, \, 9$\,, the string coupling $G^{}_\text{s}$\,, and the RR $(p+1)$-form $C^{(p+1)}$\,.
In flat spacetime, the M$p$T prescription~\eqref{eq:rpgcfbmpt} becomes
\begin{subequations} \label{eq:mptp}
\begin{align}
    \bigl( X^0, \, \cdots, \, X^p \bigr) & = \omega^{\frac{1}{2}} \, \bigl( t\,, \, x^1,\, \cdots, \, x^p \bigr)\,, \\[4pt]
    \bigl( X^{p+1}, \, \cdots, \, X^9 \bigr) & = \omega^{- \frac{1}{2}} \, \bigl( x^{p+1}, \, \cdots, \, x^9 \bigr)\,, \\[4pt]
    G^{}_\text{s} & = \omega^{\frac{p-3}{2}} \, g^{}_\text{s}\,
\end{align}
\end{subequations}
The line element of the ten-dimensional target space is
\be
    \dd s^2 = \omega \, \dd x^A \, \dd x^B \, \eta^{}_{AB} 
    + \omega^{-1} \, \Bigl( \, \dd r^2 + r^2 \, \dd \Omega^2_{8-p} \Bigr)\,,
        \qquad%
    r^2 = \bigl( x^{p+1} \bigr)^2 + \cdots + \bigl( x^9 \bigr)^2\,.
\ee
Here, $x^A = (t,x^1,\dots,x^p$)\,. We have introduced spherical coordinates to describe the transverse sector. Additionally, there must be a critical RR ($p$+1)-form, with
\be
    C^{(p+1)} = \omega^2 \, g^{-1}_\text{s} \, \dd t \wedge \cdots \wedge \dd x^{p}\,.
\ee
The $N$ coinciding D$p$-branes extend along $(t\,, \, x^1\,, \, \dots,\, x^p)$ and are transverse to $x^{p+1}, \, \cdots, \, x^9$\,. In the M$p$T limit, $\omega$ is sent to infinity, which leads to SYM living on the D$p$-branes. Moreover, the ten-dimensional geometry acquires a codimension-($p+1$) foliation structure and is non-Lorentzian. In general, these D$p$-branes would couple to non-Lorentzian supergravity in ten-dimensions, as indicated in Section~\ref{sec:mgt}.

The above string picture is valid in the regime where $G_\text{s} \, N \ll 1$\,, so that the backreaction of the D$p$-branes is very small. Under the reparametrisation in Eq.~\eqref{eq:mptp}, na\"{i}vely, the would-be 't Hooft coupling $\lambda^{}_\text{t} = G^{}_\text{s} \, N$ seems to go to either zero for $p > 3$ or infinity for $p < 3$\,, when the $\omega \rightarrow \infty$ limit is performed. However, note that we are also measuring the excitations in a rather different energy regime now, as the time direction is also rescaled in Eq.~\eqref{eq:mptp}. Therefore, the associated energy is scaled up by $\omega^{1/2}$\,, and we are now focusing on the low-energy near-BPS excitations. Ultimately, we end up with a finite effective string coupling $g^{}_\text{s}$ in the resulting M$p$T, which we still require to be sufficiently small such that $g^{}_\text{s} \, N \ll 1$ continues to hold. This is essential for the string picture to hold after the M$p$T limit is performed. We will see at the end of this subsection that the finiteness of the effective string coupling becomes more manifest when we recast the decoupling limit in the more standard form as the Maldacena limit, where the Regge slope or string length is rescaled. 

\vspace{3mm}

\noindent $\bullet$~\emph{Supergravity picture.} In the supergravity picture, where $G_\text{s} N \gg 1$\,, the D$p$-branes backreact strongly and create the following brane geometry as a solution to type II supergravity:
\begin{subequations} \label{eq:dpbg}
\begin{align}
    \dd s^2 & = \frac{1}{\sqrt{H}} \, \dd X^A \, \dd X^B \, \eta^{}_{AB}
    + \sqrt{H} \, \Bigl( \, \dd R^2 + R^2 \, \dd\Omega^2_{8-p} \, \Bigr)\,, \\[4pt]
    C^{(p+1)} & = \frac{1}{G^{}_\text{s} \, H} \, \dd X^0 \wedge \cdots \wedge \dd X^p\,,
        \qquad%
    e^\Phi = G^{}_\text{s} \, H^{\frac{3-p}{4}}\,,
        \qquad%
    H = 1 + \left( \frac{L}{R} \right)^{\!\!7-p}\!\!, 
\end{align}
\end{subequations}
where here $R^2 = \bigl( X^{p+1} \bigr)^2 + \cdots + \bigl( X^9 \bigr)^2$\,, and $\dd\Omega^2_{8-p}$ denotes the line element on the (8-$p$)-sphere. The above solution is valid for $p<7$\,: we do not consider branes of codimension-2 or below in this paper.
A minor technicality is that for $p=3$\,, there is an additional `magnetic' contribution to the RR 4-form such that the field strength is self-dual:
\be
F^{(5)} 
 =
G_\text{s}^{-1} \, \Bigl[ \dd H^{-1} \! \wedge \dd X^0\wedge \cdots \wedge \dd X^3 
+  4 \, L^4 \, \mathrm{Vol}\bigl(S^5\bigr) \Bigr]\,.
\label{F5general}
\ee
The dimensionful constant $L$ scales as 
\be \label{eq:lsp}
    L^{7-p} \sim N \, G^{}_\text{s} \, \ell_\text{s}^{7-p}\,,
\ee
where we omit a numerical constant, with $\ell^{}_\text{s}$ the string length and $G^{}_\text{s}$ the asymptotic string coupling at $R \rightarrow \infty$\,. 

The set of M$p$T prescriptions in Eq.~\eqref{eq:mptp} apply to the regime at the asymptotic infinity, where $R \rightarrow \infty$ and $H (R) \rightarrow 1$\,. The corresponding bulk geometry is generated by 
plugging~\eqref{eq:mptp} into Eq.~\eqref{eq:dpbg}, including the profile of $H$, where both $L$ and $R$ depend on $\omega$ as
\be
    L^{7-p} = \omega^{\frac{p-3}{2}} \, \ell^{7-p}\,,
        \qquad%
    R = \omega^{-\frac{1}{2}} \, r\,.
\ee
Here, $\ell$ defines the characteristic lengthscale of the geometry after the $\omega \rightarrow \infty$ limit. The harmonic function now becomes
\be
    H = 1 + \omega^2 \left( \frac{\ell}{r} \right)^{\!\!7-p}\!\!.
\ee 
As expected, we find a Lorentzian bulk geometry in the $\omega \rightarrow \infty$ limit,
\begin{subequations} \label{eq:dsco}
\begin{align}
    \dd s^2 & = \mathbb{\Omega}(r) \, \dd x^A \, \dd x^B \, \eta^{}_{AB}
    + \frac{\dd r^2 + r^2 \, \dd\Omega^2_{8-p}}{\mathbb{\Omega}(r)}\,, 
        &%
    \mathbb{\Omega}(r) & = \left( \frac{r}{\ell} \right)^{\frac{7-p}{2}}, \\[4pt]
    C^{(p+1)} & = g^{-1}_\text{s} \, \bigl[ \mathbb{\Omega}(r) \bigr]^2 \, \dd t \wedge \dd x^1 \wedge \cdots \wedge \dd x^p\,, 
       &%
    e^\Phi & = \bigl[ \mathbb{\Omega}(r) \bigr]^{\frac{p-3}{2}} \, g^{}_\text{s}\,.
\end{align}
\end{subequations}
This shows that the MpT limit at the asymptotic infinity, which we will refer to as the \emph{asymptotic M$p$T limit} in the following, generates the near-horizon limit of the bulk D$p$-brane geometry~\eqref{eq:dpbg}. 
When $p=3$ (for which we should also include the additional term in the RR 4-form, as the magnetic part of the field strength~\eqref{F5general} is unchanged in this limit) this geometry is exactly AdS${}_5 \times S^5$, while for general $p \neq 3$ the near-horizon geometry is conformal to an AdS${}_{p+2} \times S^{8-p}$ geometry, except for $p=5$ when it is conformal to a Minkowski geometry times $S^3$~\cite{Boonstra:1997dy,Boonstra:1998mp}. 
\begin{figure}[t!]
\centering
\begin{tikzpicture}[boxes/.style={draw=black, rounded corners=5pt, inner sep=5pt, minimum height=1cm,text width=5.5cm,align=center}]

\draw (-3.2,0) node (closed) [text width=5cm, rounded corners=5pt, fill=vub!20,align=center, inner sep=5pt] {\bf Bulk Geometry};

\draw (3.2,0) node (open)  [text width=5cm, rounded corners=5pt, fill=vub!20,align=center, inner sep=5pt] {\bf Asymptotic Infinity};

\node (D3cl) [below =0.5cm of closed, boxes] {D3-brane supergravity solution};

\node (D3op) [below =0.5cm of open, boxes] {D3-brane worldvolume theory};

\node (D3cl_res) [below =1cm of D3cl, boxes] {IIB string theory on AdS${}_5 \times S^5$};

\node (D3op_res) [below =1cm of D3op, boxes] {$\mathcal{N}=4$ super Yang-Mills};

\draw [thick,->] ([yshift=-0.1cm]D3cl.south) -- ([yshift=0.1cm]D3cl_res.north) node [midway,right] {\emph{near-horizon limit}};
\draw [thick,->] ([yshift=-0.1cm]D3op.south) -- ([yshift=0.1cm]D3op_res.north) node [midway,left] {\emph{M3T limit}};

\draw[thick, dashed,<->] ([yshift=-0.15cm]D3cl_res.south) to [out=350,in=190] node [midway,below] {\emph{holographically dual}} ([yshift=-0.15cm]D3op_res.south);

\end{tikzpicture} 
\caption{The usual AdS/CFT correspondence can be realised by taking the near horizon limit as the asymptotic M3T limit, which corresponds to the near-horizon limit of the bulk.}
\label{fig:holo_usual} 
\end{figure}
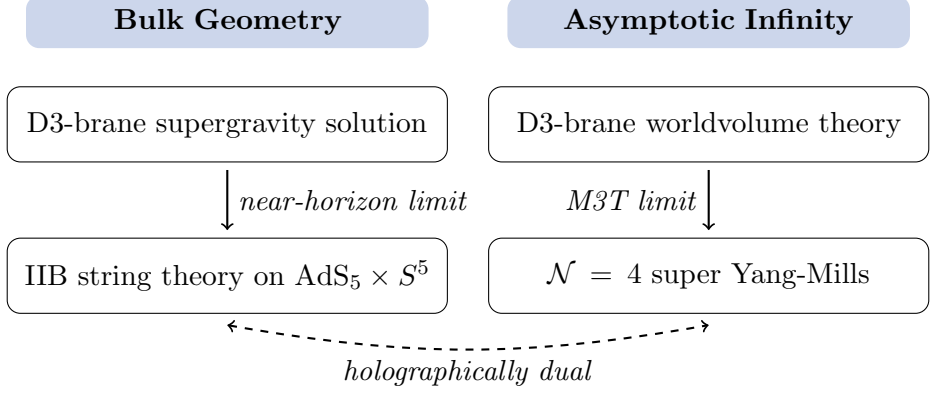

\vspace{3mm}

\noindent $\bullet$~\emph{Relation to the Maldacena limit.}
The limiting prescription discussed above can be equivalently parametrised in a more familiar way. We use a single D$p$-brane to probe the M$p$T geometry at the asymptotic infinity. We align the brane with the longitudinal sector in the M$p$T geometry. Then, under the M$p$T prescription~\eqref{eq:mptp}, the D$p$-brane action in type II string theory can be written as
\begin{align} \label{eq:sdprs}
    S^{}_\text{D$p$} & = - \frac{\omega^{(3-p)/2} \, g^{-1}_\text{s}}{(\alpha' / \omega)^{(p+1)/2}} \int \dd^{p+1} \sigma \, \sqrt{- \det \Bigl[ \p^{}_\alpha x^A \, \p^{}_\beta x^{}_A + \omega^{-2} \, \p^{}_\alpha x^{A'} \, \p^{}_\beta x^{A'} + 2 \pi \bigl( \alpha' / \omega \bigr) \, F^{}_{\alpha\beta} \Bigr]} \notag \\[4pt]
    & \quad + \frac{1}{(\alpha' / \omega)^{(p+1)/2}} \int \omega^{(3-p)/2} \, g^{-1}_\text{s} \, \dd x^0 \wedge \cdots \wedge \dd x^p\,,
\end{align}
where $A=0\,,\,1\,, \, \dots, \, p$ and $A'= p+1\,,\,\dots,\, 9-p$.
Here we have pulled out the factors of $\omega$ to make it clear that this limit can also be viewed as arising from the an alternative set of rescalings, where, in particular the original Regge slope $\alpha'_\text{original}$ is written as 
\be \label{eq:alphapr}
    \alpha'_\text{original} = \frac{\alpha'}{\omega}\,,    
\ee
in terms of the fixed $\alpha'$ that we have been using, together with
\be
    X^A = x^A,
        \quad%
    X^{A'} = \frac{x^{A'}}{\omega}\,,
        \quad%
    G^{}_\text{s} = \omega^{\frac{p-3}{2}} \, g^{}_\text{s}\,,
        \quad%
    C^{(p+1)} = \omega^{\frac{3-p}{2}} \, g^{-1}_\text{s} \, \dd x^0 \wedge \cdots \wedge \dd x^p\,.
\ee
The scaling of the transverse coordinates $X^{A'}$ is such that $R = r / \omega$\,.
This implies that the Yang-Mills coupling $g^{}_\text{YM}$ and the Higgs expectation value $v$ associated with separate D-branes remain constant while bringing the branes together, \emph{i.e.}
\be
    g^2_\text{YM} = (2\pi)^{p-2} \, G^{}_\text{s} \, \bigl( \alpha'_\text{original}
     \bigr)^{\frac{p-3}{2}} = \text{fixed}\,,
        \qquad%
    v = \frac{R}{\alpha'_\text{original}} = \text{fixed}\,.
\ee
Then, $\omega \rightarrow \infty$ precisely gives the Maldacena limit \cite{Maldacena:1997re, Itzhaki:1998dd}. 
Regardless of which prescription is used to define the limit, sending $\omega\to\infty$ in the action \eqref{eq:sdprs} we find that the divergences from the DBI and Wess-Zumino terms cancel. 
Note that the divergent RR gauge potential, which is an essential component of the general M$p$T prescription, would usually not be included. In this case it is a pure gauge contribution and the `constant' divergence that it cancels is often dropped automatically. This is similar to the standard treatment of the rest mass term in the non-relativistic point particle action.

\subsection{Non-Lorentzian Bulk from Double Asymptotic BPS Limits} \label{sec:nlbdal}

We have seen that taking an asymptotic M$p$T limit of the bulk D$p$-brane geometry leads to the near-horizon limit of the latter. This is indicative of a close link between the M$p$T decoupling limits and holography in a more generic setup. It is natural to ask: what happens if one applies an asymptotic M$p$T limit to a bulk D$q$-brane geometry for general $p$ and $q$? The AdS/CFT or IMSY~\cite{Itzhaki:1998dd} correspondence arises as a special case with $p = q$\,. But what happens when $p \neq q$? 

As a warm-up, we now firstly apply the formalism that we have developed in Section~\ref{sec:badsmt} to the case where $p < q$,
and for the moment assume for simplicity that the asymptotic M$p$T longitudinal directions overlap with the longitudinal directions of the bulk D$q$-brane solution.
This choice is purely for pedagogical reasons, and we will consider fully general configurations in the next subsection. 
We will find that this off-aligned case leads to a non-Lorentzian bulk geometry, in which a near-horizon limit is still possible. This near-horizon limit corresponds to a second BPS decoupling limit of the M$q$T form at the asymptotic infinity. 
This points the way to a whole landscape of new holographic duals\,\footnote{We will introduce a series of non-Lorentzian holographic dualities, which can be viewed as limits of known holographic dualities with a Lorentzian bulk geometry. These novel stand-alone examples essentially highlight which subsectors of an SYM could have their own gravity duals.}, as we will see in more detail in the next subsection.

\vspace{3mm}

\noindent $\bullet$~\emph{D$q$-brane geometry in asymptotic matrix $p$-brane limit.} 
The relevant configuration is depicted in the following table, where we have also marked different sectors with the associated indices that we will use later:
\begin{equation*}
\begin{tabular}{c|c|c|c}
 $\sharp$ of directions & $\quad p+1\quad$ & $q-p$ & $9-q$ \\[2pt]
 \hline
 bulk D$q$-brane solution & $\times$ & $\times$ & -- \\[2pt]
asymptotic M$p$T limit & $\times$ & -- & -- \\[2pt]
    \hline
 index & $A$ & $I$ & $i$ 
\end{tabular}
\end{equation*}
Explicitly, the ranges of these indices are
\be
    A = 0\,,\,\dots,\,p\,, 
        \qquad%
    I = p+1\,,\, \dots, \, q\,, 
        \qquad%
    i = q+1\,, \, \dots, \, 9\,.
\ee
We now write down the corresponding bulk D$q$-brane geometry~\eqref{eq:dpbg} reparametrised in terms of the asymptotic M$p$T prescription~\eqref{eq:mptp}:
\begin{subequations} \label{eq:DqSoln_forlimit}
\begin{align}
    \dd s^2 & = H^{-\frac{1}{2}} \biggl\{ \omega \, \Bigl[ - \dd t^2 + \bigl(\dd x^1\bigr)^2 + \dots + \bigl(\dd x^p\bigr)^2 \Bigr] 
    + \omega^{-1} \Bigl[ \bigl( \dd x^{p+1} \bigr)^2+ \dots + \bigl( \dd x^{q} \bigr)^2 \Bigr] \biggr\} \notag\\[4pt]
        &%
    \quad + \omega^{-1} \, H^{\frac{1}{2}} \Bigl[ \bigl( \dd x^{q+1} \bigr)^2 +\dots + \bigl( \dd x^9 \bigr)^2 \Bigr] \,,\\[4pt]
    C^{(q+1)} & = \omega^{\frac{p-q+4}{2}} \, g_\text{s}^{-1} \, H^{-1} \, \dd t \wedge \dd x^1 \wedge \dots \wedge \dd x^q \,,
        \qquad%
    e^\Phi = \omega^{\frac{p-3}{2}} \, g^{}_\text{s} \, H^{\frac{3-q}{4}}\,,
\end{align}
\end{subequations}
where the harmonic function is now given by
\be \label{eq:holr}
    H = 1 +  \omega^{\frac{p-q+4}{2}} \! \left( \frac{\ell}{\| x^i \|} \right)^{\!\!7-q}\!\!,
        \qquad%
    \| x^i \|^2 = \bigl( x^{q+1} \bigr)^2 + \cdots + \bigl( x^9 \bigr)^2\,.
\ee
The metric will be of the correct form to define an M$p$T limit \eqref{eq:rpgcfbmpt} if the harmonic function is finite in the limit. 
From the form of the harmonic function in \eqref{eq:holr}, this immediately singles out the condition $q = p + 4$.
This is the only possibility unless one considers further manipulations such as smearing the brane or introducing a further rescaling of $\ell$, both of which we will discuss later on in Section~\ref{sec:sln}.
The condition  $q=p+4$ also ensures that the RR potential $C^{(q+1)}$ is finite.
To define an M$p$T limit, we also need a critical RR $(p+1)$-form $C^{(p+1)}$, taking the following form: 
\be
    C^{(p+1)} = \omega^2 \, \Bigl( g^{-1}_\text{s} \, H^{\frac{q-3}{4}} \Bigr) \, \tau^0 \wedge \cdots \wedge \tau^p\,,
        \quad%
    \bigl( \tau^0, \, \cdots, \, \tau^p \bigr) = H^{-\frac{1}{4}} \, \bigl( \dd t\,, \, \dd x^1, \, \cdots, \, \dd x^p \bigr)\,.
\ee
In the case where $q = p + 4$\,, we find
\be \label{eq:cppob}
    C^{(p+1)} = \omega^2 \, g^{-1}_\text{s} \, \dd t \wedge \dd x^1 \wedge \cdots \wedge \dd x^p\,,
\ee
which is constant and can be added to the background~\eqref{eq:DqSoln_forlimit}, without affecting its status as a solution of type II supergravity. It then follows that, together with \eqref{eq:cppob}, we find an M$p$T reparametrisation of the desired form \eqref{eq:rpgcfbmpt}, with
\begin{align}
    \dd s^2 = \omega \, \tau^A \, \tau^B \, \eta^{}_{AB} 
    + \omega^{-1} \, \bigl( E^I \, E^I + E^i \, E^i \bigr)\,, 
        \qquad%
    e^\Phi = \omega^{\frac{p-3}{2}} \, e^{\varphi}\,,
        \qquad%
    C^{(q+1)} = c^{(q+1)}\,,
\end{align}
where the M$p$T fields are
\begin{subequations} \label{eq:mptsl0}
\begin{align} 
    \tau^{A} & = H^{-\frac{1}{4}} \, \dd x^A \,,
        &%
    E^{I} & = H^{-\frac{1}{4}} \, \dd x^{I} \,,
        \qquad%
    E^{i} = H^{\frac{1}{4}} \, \dd x^{i} \,,\\[4pt]
    e^{\varphi} & = g^{}_\text{s} \, H^{\frac{3-q}{4}} \,,
        &%
    c^{(q+1)} & = g_\text{s}^{-1} \, H^{-1} \, \dd t \wedge \dd x^1 \wedge \dots \wedge \dd x^{q} \,,
\end{align}
\end{subequations}
with the harmonic function given by 
\be \label{eq:hfmpt}
    H = 1 + \frac{\ell^{7-q}}{\| x^i \|^{7-q}}\,.
\ee
In the $\omega \rightarrow \infty$ limit, we find a non-trivial M$p$T geometry in the bulk described by the data in Eq.~\eqref{eq:mptsl0}. Since the harmonic function $H$ is unaffected under the $\omega \rightarrow \infty$ limit, both the bulk and asymptotic infinity acquire an M$p$T geometry. The simplest example is when $p = 0$ and $q = 4$\,, where the above describes a non-Lorentzian 4-brane geometry of the M0T type. 

\vspace{3mm}

\noindent $\bullet$~\emph{Near horizon from a 2${}^\text{nd}$ asymptotic BPS limit.} 
The fact that we still have a well-defined harmonic function in Eq.~\eqref{eq:hfmpt} is indicative that a standard near-horizon limit can be taken, even though the bulk geometry~\eqref{eq:mptsl0} is now non-Lorentzian. Following the discussion in Section~\ref{sec:badsmt}, we expect that this near-horizon limit is generated by an extra BPS decoupling limit at the asymptotic infinity, which must be the asymptotic M$q$T limit aligned with the bulk D$q$-brane.  
This asymptotic M$q$T limit is in addition to the asymptotic M$p$T limit that we have already taken above. We therefore introduce the asymptotic M$q$T prescription with a second parameter $\tilde{\omega}$\,, with %
\be
    x^A = \tilde{\omega}^{\frac{1}{2}} \, \mathbb{x}^A,
        \qquad%
    x^I = \tilde{\omega}^{\frac{1}{2}} \, \mathbb{x}^I,
        \qquad%
    x^i = \tilde{\omega}^{-\frac{1}{2}} \, \mathbb{x}^i,
        \qquad%
    g^{}_\text{s} = \tilde{\omega}^{\frac{q -3}{2}} \, \mathbb{g}^{}_\text{s}\,.
\ee
Correspondingly, mapping this reparametrisation in the asymptotic infinite regime to the bulk using $\ell^{7 - q} \propto g^{}_\text{s}$\,, we find
\be
    H = 1 + \tilde{\omega}^2 \left( \frac{\lbb}{\|\mathbb{x}^i \|} \right)^{\!7-q}\!\!,
\ee
where $\lbb$ measures the effective scale after the $\tilde{\omega} \rightarrow \infty$ limit. 
Taking the $\tilde{\omega} \rightarrow \infty$ limit of the bulk M$p$T geometry~\eqref{eq:mptsl0}, we find the near-horizon geometry
\begin{subequations} \label{eq:mptsl0NH}
\begin{align} 
    \tau^{A} & = \bigl[ \tilde{\mathbb{\Omega}} (\mathbb{r}) \bigr]^\frac{1}{2} \, \dd \mathbb{x}^A \,,
        &%
    E^{I} & = \bigl[\tilde{\mathbb{\Omega}} (\mathbb{r})\bigr]^{\frac{1}{2}} \, \dd \mathbb{x}^{I} \,,
        \qquad%
    E^{i} = \bigl[ \tilde{\mathbb{\Omega}} (\mathbb{r}) \bigr]^{-\frac{1}{2}} \, \dd \mathbb{x}^{i} \,,\\[4pt]
    e^{\varphi} & = \bigl[ \tilde{\mathbb{\Omega}} (\mathbb{r}) \bigr]^{\frac{q-3}{2}} \, \mathbb{g}_\text{s} \,,
        &%
    c^{(q+1)} & = \bigl[ \tilde{\mathbb{\Omega}}(\mathbb{r}) \bigr]^2 \, \mathbb{g}_\text{s}^{-1} \, \dd \mathbb{t} \wedge \dd \mathbb{x}^1 \dots \wedge \dd \mathbb{x}^{q} \,,
\end{align}
\end{subequations}
where $\mathbb{r}^2 = \mathbb{x}^i \, \mathbb{x}^i$ and
\be
    \tilde{\mathbb{\Omega}} (\mathbb{r}) = \left( \frac{\mathbb{r}}{\lbb} \right)^{\!\!\frac{7-q}{2}}\!\!.
\ee
This resulting bulk geometry should be viewed as leading to an M$p$T analogue of a D-brane near-horizon geometry. 

\begin{figure}[t!]
\centering
\begin{tikzpicture}[boxes/.style={draw=black, rounded corners=5pt, inner sep=5pt, minimum height=1.2cm,text width=5.5cm,align=center}]

\draw (-3.2,0) node (closed) [text width=5cm, rounded corners=5pt, fill=vub!20,align=center, inner sep=5pt] {\bf Bulk Geometry};

\draw (3.2,0) node (open)  [text width=5cm, rounded corners=5pt, fill=vub!20,align=center, inner sep=5pt] {\bf Asymptotic Infinity};

\node (D3cl) [below =0.5cm of closed, boxes] {D4-brane supergravity solution};

\node (D3op) [below =0.5cm of open, boxes ] {D4-brane worldvolume theory};

\node (D3cl_res) [below =1cm of D3cl,boxes ] {IIA string theory on near-horizon geometry
};

\node (D3cl_res2) [below =1cm of D3cl_res, boxes ] {\scalebox{0.92}{M$0$T on background~\eqref{eq:mptsl0NH}}};

\node (D3op_res) [below =1cm of D3op, boxes] {5D super Yang-Mills};
\node (D3op_res2) [below =1cm of D3op_res, boxes ] {QM on instanton moduli space};

\draw [thick,->] ([yshift=-0.1cm]D3cl.south) -- ([yshift=0.1cm]D3cl_res.north) node [midway,right] {\emph{near-horizon limit}};
\draw [thick,->] ([yshift=-0.1cm]D3op.south) -- ([yshift=0.1cm]D3op_res.north) node [midway,left] {\emph{M4T limit}};

\draw [thick,->] ([yshift=-0.1cm]D3cl_res.south) -- ([yshift=0.1cm]D3cl_res2.north) node [midway,right] {\emph{M0T limit}};
\draw [thick,->] ([yshift=-0.1cm]D3op_res.south) -- ([yshift=0.1cm]D3op_res2.north) node [midway,left] {\emph{M0T limit}};

\draw[thick, dashed,<->] ([yshift=-0.15cm]D3cl_res2.south) to [out=350,in=190] node [midway,below] {\emph{holographically dual}} ([yshift=-0.15cm]D3op_res2.south);

\end{tikzpicture} 
\caption{An example of a proposed non-Lorentzian holography \cite{Lambert:2024yjk}, realised by taking both the M0T limit and the near-horizon limit of the bulk D4-brane geometry. 
The order of limits does not matter on the bulk side, which should also be true at the asymptotic infinity. The resulting non-Lorentzian M0T geometry is described by Eq.~\eqref{eq:mptsl0NH} with $p = 0$\,.}
\label{fig:holo_unusual} 
\end{figure}
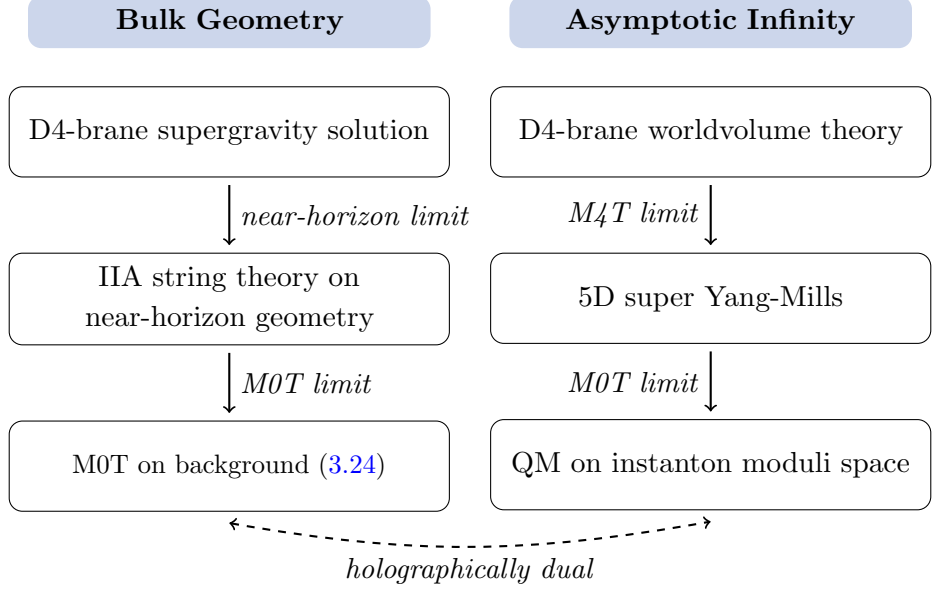

The simplest example is  $p=0$\,,\,\footnote{The case $p=-1$ corresponds to an M$(-1)$T limit of the D3-brane solution. M$(-1)$T follows from a timelike T-duality of M0T \cite{Blair:2023noj,Gomis:2023eav} and it would be interesting to explore how this example relates to the de Sitter/Euclidean YM duality of \cite{Hull:1998vg}.} for which the procedure described here leads to an M0T limit of the near-horizon D4 brane geometry.
In this case, the M$p$T geometry encoded by the vielbein fields in \eqref{eq:mptsl0NH} can be written as:
\begin{subequations}
\begin{align}
\tau_{\mu\nu} \, \dd \mathbb{x}^\mu \, \dd \mathbb{x}^\nu \equiv \tau^A \, \tau^B \, \eta_{AB} 
& = - \left( \frac{\mathbb{r}}{\lbb} \right)^{\!\!\frac{3}{2}} \dd \mathbb{x}^0 \, \dd \mathbb{x}^0,\\[4pt]
E_{\mu\nu} \, \dd \mathbb{x}^\mu \, \dd \mathbb{x}^\nu \equiv
E^{A'} E^{A'} & = 
\left( \frac{\lbb}{\mathbb{r}} \right)^{\!\!\frac{3}{2}} 
\left[
\left( \frac{\mathbb{r}}{\lbb} \right)^{\!\!3} \dd \mathbb{x}^I \dd \mathbb{x}^I 
+ \dd \mathbb{r}^2 + \mathbb{r}^2 \, \dd \Omega_4^2
\right].
\end{align}
\end{subequations}
This can be recognised as being dilatation equivalent to an M0T limit of AdS${}_6 \times S^4$.
It has been proposed in \cite{Lambert:2024yjk} 
that this M0T geometry is the background dual to a field theory obtained by taking a non-relativistic limit of the five-dimensional SYM (or, equivalently, a null reduction of the six-dimensional $(2,0)$ theory). See Figure~\ref{fig:holo_unusual} for a summary. In this field theory, the dynamics reduces to quantum mechanics on instanton moduli space. From our perspective, the boundary theory therefore follows from applying both asymptotic M$4$T and M0T limits to the D4-brane worldvolume theory. 
This example is just one of a number of proposals of non-Lorentzian holographic correspondences contained in \cite{Lambert:2024yjk}. 
We will see below how our limiting procedure systematically generates the others.

Finally, note that while here we have taken the $\omega \rightarrow \infty$ asymptotic M$p$T limit followed by the $\tilde \omega \rightarrow \infty$ asymptotic M$q$T limit, we could equally well have taken them in the opposite order, leading to the same near-horizon geometry \eqref{eq:mptsl0NH}.
In the terminology of our previous paper \cite{Blair:2023noj}, we would refer to this double BPS decoupling limit as a \emph{multicritical} limit. 
We will describe features of these multicritical limits in Section \ref{sec:furtherdlcqs}.

\subsection{More Brane Geometries in Matrix \texorpdfstring{$p$}{p}-Brane Theory} \label{sec:morebranegeos}

In Section~\ref{sec:nlbdal}, we have realised a non-Lorentzian D$q$-brane solution in M$p$T, and then considered its near-horizon limit in connection with non-Lorentzian holography. 
We will argue that the successive application of various asymptotic BPS decoupling limits that we have used above presents a general organising principle for classifying holographic correspondences. Such correspondences in general will involve non-Lorentzian bulk geometry and non-Lorentzian field theory at the asymptotic infinity. To motivate this viewpoint, which we will set out in more detail in Section~\ref{sec:chdlcq}, in this subsection we first investigate the outcomes of taking more general M$p$T limits in curved D$q$-brane geometry. 

\subsubsection{General Brane Configurations} \label{sec:gbc}

In Section~\ref{sec:nlbdal}, we considered a warm-up example where the longitudinal components associated with the asymptotic M$p$T limit are aligned within only the longitudinal directions in the bulk D$q$-brane geometry. Now, we consider more general configurations where the longitudinal directions of the asymptotic M$p$T limit correspond to both transverse and longitudinal directions of the bulk D$q$-brane solution.  
As for the D$q$-brane worldvolume configurations considered in Section~\ref{sec:gdbmpt}, we now let the bulk D$q$-brane have $m$ spatial longitudinal ($n$ transverse) directions overlapping with the transverse (longitudinal) directions in the asymptotic M$p$T limit, corresponding to the following setup:
\begin{center}
\begin{tabular}{c|c|c|c|c}
$\sharp$ of directions & $q-m+1$ & \,$\phantom{q-}m\phantom{-q}$\, & \,$\phantom{q-}n\phantom{-q}$\, & $9 - q - n$ \\[2pt]
\hline 
bulk D$q$-brane solution & $\times$ & $\times$ & -- & -- \\[2pt]
asymptotic M$p$T limit & $\times$ & -- & $\times$ & -- \\[2pt]
\hline
index & $a$ & $I$ & $u$ & $i$
\end{tabular}
\end{center}
Here, $p + m = q + n$ and we have introduced indices,
\begin{subequations} \label{eq:indices}
\begin{align}
    a & = 0\,, \, 1\,, \, \cdots, \, q - m\,, 
        &%
    I & = q - m + 1\,, \, \cdots, \, q\,, \\[4pt]
    u & = q+1\,, \, \cdots, \, q + n\,, 
        &%
    i & = q + n + 1\,, \cdots, \, 9\,.
\end{align}
\end{subequations}
The asymptotic M$p$T prescription~\eqref{eq:mptp} induces the following reparametrised bulk D$q$-brane geometry:
\begin{subequations} \label{DqSoln_forlimit_mixed}
\begin{align} 
    \dd s^2 & = \omega \, \Bigl( H^{-\frac{1}{2}} \, \dd x^a \, \dd x^{}_a + H^{\frac{1}{2}} \, \dd x^u \, \dd x^u \Bigr)  
    + \omega^{-1} \, \Bigl( H^{-\frac{1}{2}} \, \dd x^I \, \dd x^I + H^{\frac{1}{2}} \, \dd x^i \, \dd x^i \Bigr)\,, \\[4pt]
    C^{(q+1)} & = \omega^{\frac{4 - m - n}{2}} \, g_\text{s}^{-1} \, H^{-1} \, \dd t \wedge \dd x^1 \wedge \dots \wedge \dd x^q \,,\
        \qquad%
    e^\Phi = \omega^{\frac{p-3}{2}} \, g_\text{s} \, H^{\frac{3-q}{4}}\,.
\end{align}
\end{subequations}
The harmonic function is now given by
\be \label{eq:hmc}
    H = 1 +  \omega^{\frac{p+q-10}{2}} \left( \frac{\ell}{\sqrt{x^u \, x^u + \omega^{-2} \, x^i \, x^i}} \right)^{\!\!7-q}\!\!.
\ee
Our previous example in Section \ref{sec:nlbdal} corresponded to the case $n=0$. 
When $n > 0$, which we now assume, the harmonic function depends also on the longitudinal coordinates $x^u$ of the asymptotic M$p$T limit rather than purely on transverse coordinates.  
As a result, $H$ has a very different $\omega$-dependence compared to Eq.~\eqref{eq:holr}.

Requiring now that the background  \eqref{DqSoln_forlimit_mixed} defines an M$p$T limit in the bulk, we find two conditions.
Firstly, the metric will lead to a bulk M$p$T limit if the harmonic function is finite in the limit.
In analogy with Section~\ref{sec:nlbdal}, 
and without performing any further manipulations on the harmonic function, this implies
$p + q = 10$\,.
Secondly, we require that the RR potential $C^{(q+1)}$ in Eq.~\eqref{DqSoln_forlimit_mixed} be finite with respect to the infinite $\omega$ limit. This singles out the cases where
$m + n = 4$\,.
We then find the conditions,
\be \label{eq:pmn}
    p = 10 - q\,,
        \qquad%
    m = q - 3\,,
        \qquad%
    n = 7 - q\,,
        \qquad%
    3 \leq q < 7\,.
\ee
Finally, in order to achieve a bulk M$p$T limit, we also need a critical RR potential $C^{(p+1)}$ in the bulk, which must have the form,
\be
    C^{(p+1)} = \omega^2 \, e^{-\varphi} \, \tau^0 \wedge \cdots \wedge \tau^{q-m} \wedge \tau^{q+1} \wedge \cdots \wedge \tau^{q+n}\,.
\ee
Here, $e^{\varphi} = g^{}_\text{s} \, H^{\frac{p-7}{4}}$, $\tau^a = H^{-\frac{1}{4}} \, \dd x^a$ and $\tau^u = H^{\frac{1}{4}} \, dx^u$\,. It follows that
\be
    C^{(p+1)} = \omega^2 \, g^{-1}_\text{s} \, \dd x^0 \wedge \cdots \wedge \dd x^{q-m} \wedge \dd x^{q+1} \wedge \cdots \wedge \dd x^{q+n}\,.
\ee
This is a constant gauge potential, 
whose addition preserves the geometry as a solution to the type II supergravity equations of motion. In the $\omega \rightarrow \infty$ limit, comparing \eqref{DqSoln_forlimit_mixed} subject to the conditions \eqref{eq:pmn} with the defining M$p$T prescription \eqref{eq:rpgcfbmpt}, we find the following M$p$T geometry in the bulk:
\begin{subequations} \label{eq:gbccf}
\begin{align} 
    \tau^a &= H^{-\frac{1}{4}} \, \dd x^a\,, 
        \qquad%
    \tau^u = H^{\frac{1}{4}} \, \dd x^u\,,
        &%
    E^I &= H^{-\frac{1}{4}} \, \dd x^I\,,
        &%
    E^i &= H^{\frac{1}{4}} \, \dd x^i\,, \label{eq:gbccftau} \\[4pt] 
    c^{(q+1)} & = g_\text{s}^{-1} \, H^{-1} \, \dd t \wedge \dd x^1 \wedge \dots \wedge \dd x^{q} \,,
        &%
    e^\varphi &= g^{}_\text{s} \, H^{\frac{3-q}{4}}\,,
        &%
     H &= 1 + \left( \frac{\ell}{\|x^u\|} \right)^{\!\!7-q}\!\!,
     \label{eq:gbccfh}
\end{align}
\end{subequations}
where, $\| x^u \| \equiv \sqrt{x^u \, x^u}$\,. 
A particularly interesting example is when $p = 7$\,, in which case $q = 3\,, \, m = 0$ and $n = 4$\,. Under these conditions, the background fields in Eq.~\eqref{eq:gbccf} encode a non-Lorentzian geometry describing a 3-brane background in M7T.  
Note for $q=3$, the original D3-brane solution involves an extra gauge potential contribution such that the field strength is self-dual.
It can be checked however that the additional magnetic contribution to the field strength scales like $\omega^{n-6}$ for $n>0$ and is finite for $n=0$.
Therefore, with $n=4$\,, this contribution drops out in the $\omega \rightarrow \infty$ limit.

The near-horizon limit of the bulk M$p$T geometry~\eqref{eq:gbccf} proceeds similarly to before. Again, this bulk near-horizon limit is generated by an asymptotic M$q$T limit, with its longitudinal sector aligned with the bulk D$q$-brane configuration. In terms of the parameter $\tilde{\omega}$\,, and according to Eq.~\eqref{eq:rpgcfbmpt}, the asymptotic M$q$T prescription is
\be \label{eq:xaxixuxi}
    x^a = \tilde{\omega}^{\frac{1}{2}} \, \mathbb{x}^a\,, 
        \qquad%
    x^I = \tilde{\omega}^{\frac{1}{2}} \, \mathbb{x}^I\,, 
        \qquad%
    x^u = \tilde{\omega}^{-\frac{1}{2}} \, \mathbb{x}^u\,, 
        \qquad%
    x^i = \tilde{\omega}^{-\frac{1}{2}} \, \mathbb{x}^i\,, 
        \qquad%
    g^{}_\text{s} = \tilde{\omega}^{\frac{q-3}{2}} \, \mathbb{g}^{}_\text{s}\,.
\ee
The induced bulk geometry in the $\tilde{\omega} \rightarrow \infty$ limit is
\begin{subequations} \label{eq:mptsl02}
\begin{align} 
    \tau^{a} & = \bigl[ \tilde{\mathbb{\Omega}} (\mathbb{r}) \bigr]^\frac{1}{2} \, \dd \mathbb{x}^a\,,
        &%
    E^{I} & = \bigl[ \tilde{\mathbb{\Omega}} (\mathbb{r}) \bigr]^{\frac{1}{2}} \, \dd \mathbb{x}^{I}\,, 
        &%
    e^{\varphi} & = \bigl[ \tilde{\mathbb{\Omega}} (\mathbb{r}) \bigr]^{\!\frac{q-3}{2}} \, \mathbb{g}_\text{s} \,, \\[4pt]
    \tau^u & = \bigl[ \tilde{\mathbb{\Omega}} (\mathbb{r}) \bigr]^{-\frac{1}{2}} \, \dd \mathbb{x}^{u}\,,
        &%
    E^{i} & = \bigl[ \tilde{\mathbb{\Omega}} (\mathbb{r}) \bigr]^{-\frac{1}{2}} \, \dd \mathbb{x}^{i} \,,
        &%
    c^{(q+1)} & = \frac{\bigl[ \tilde{\mathbb{\Omega}}(\mathbb{r}) \bigr]^2}{\mathbb{g}_\text{s}} \, \dd \mathbb{t} \wedge \dd \mathbb{x}^1 \wedge \dots \wedge \dd \mathbb{x}^{q} \,,
\end{align}
\end{subequations}
where $\tilde{\mathbb{\Omega}} (\mathbb{r}) = \bigl(\mathbb{r}/\lbb \bigr)^{(7-q)/2}$ and $\mathbb{r} = \| \mathbb{x}^u \|$\,.
The bulk near-horizon geometry remains of the M$p$T type, induced by the double BPS decoupling limits at the asymptotic infinity, consisting of both the asymptotic M$p$T and M$q$T limits, controlled by $\omega$ and $\tilde{\omega}$\,, respectively. 
For the D3-brane solution mentioned above with $q=3$, $m=0$ and $n=4$\,, this gives
\begin{subequations}
    \begin{align}
    \tau^{}_{\mu\nu} \, \dd \mathbb{x}^\mu \, \dd \mathbb{x}^\nu \equiv  \tau^A \,  \tau^B \, \eta^{}_{AB} &= \bigg(\frac{\mathbb{r}}{\tilde{\ell}}\bigg)^{\!\!2} \dd \mathbb{x}^a \, \dd \mathbb{x}_a + \bigg(\frac{\tilde{\ell}}{\mathbb{r}}\bigg)^{\!\!2} \dd \mathbb{r}^2 + \tilde{\ell}\,{}^2\,\dd\Omega_3^2\,,\\[4pt]
    E_{\mu\nu} \, \dd \mathbb{x}^\mu \, \dd \mathbb{x}^\nu \equiv E^{A'} E^{A'} &= \bigg(\frac{\tilde{\ell}}{\mathbb{r}}\bigg)^{\!\!2}\,\dd \mathbb{x}^i \, \dd \mathbb{x}^i\,,
\end{align}
\end{subequations}
which is an AdS${}_5 \times S^3$ longitudinal M7T geometry with a two-dimensional conformally flat transverse sector.

\subsubsection{Smearing and Large \texorpdfstring{$N$}{N}} \label{sec:sln}

So far, we have considered generic configurations involving an asymptotic M$p$T limit applied to a bulk D$q$-brane solution. This gives rise to the reparametrised geometry~\eqref{DqSoln_forlimit_mixed}. In order to facilitate the asymptotic M$p$T limit, we had to require that $m + n =4$\,, such that the RR potential $C^{(q+1)}$ ($q \neq p$) is finite in $\omega$\,. This condition is difficult to evade within the current framework, and is enforced by the BPS nature on the field theory side. The other requirement is that the $\omega$-dependence in the harmonic function~\eqref{eq:hmc} must be of order one, which is crucial for the near-horizon limit and hence the construction of a holographic dual. As a result, we had to impose the condition $p + q = 10$ in Eq.~\eqref{eq:pmn}. However, below we will show that this condition may be relaxed, either by smearing the D$q$-brane solution, or by introducing an additional large $N$ scaling. 

\vspace{3mm}

\noindent $\bullet$~\emph{Smearing and T-duality.} We now return to the harmonic function~\eqref{eq:hmc}, 
which is that of a D$q$-brane localised in all its transverse directions in $x^u$ and $x^i$\,, with $u = q+1\,, \, \cdots, \, q+n$ and $i = n+1\,, \, \cdots, \, 9$\,.
We apply the smearing procedure (see \emph{e.g.}~\cite{Ortin:2015hya}) to the $n$ transverse directions in $x^u$\,. Namely, we compactify $x^u$ over an $n$-torus, with the radii $R_u$\,. The supergravity solution
remains unchanged except that the harmonic function~\eqref{eq:hmc} has to be modified such that it is periodic in $x^u$.
This can be achieved by placing an infinite number of branes at regular intervals along these periodic directions in $x^u$\,. The modified harmonic function is
\be \label{eq:hosm}
    H = 1 + \sum_{u = q + 1}^{q + n} \sum_{k_u = -\infty}^\infty \frac{\omega^{\frac{p-3}{2}} \, \ell^{7 - q}}{\bigl[ \omega \, \sum_u ( x^u + 2 \, \pi \, k_u \, R_u )^2 + \omega^{-1} \, \sum_i x^i \, x^i \bigr]^{\frac{7 - q}{2}}}\,.
\ee
We are interested in the zero modes from the dimensional reduction in $x^u$\,. In the smearing procedure, the infinite sum can be approximated by an integral, \emph{i.e.},
%
%\be
$\sum_{k_u} \rightarrow \int \dd^n \mathbf{k}$\,.
The resulting harmonic function from evaluating the $k_u$ integrals is (assuming for simplicity that $7-q-n>0$, \emph{i.e.}~that the smeared brane has effective codimension greater than two)
\newcommand{\ellsmear}{\boldsymbol{\ell}}
\be \label{eq:hmc_smeared}
    H = 1 + \left( \frac{\ellsmear}{\| x^i \|} \right)^{\!\!7-q-n}\!\!,
        \qquad%
    \ellsmear^{\,7-q-n} = \frac{\Gamma\bigl(\frac{7-q-n}{2}\bigr)}{(4\pi)^{\frac{n}{2}} \, \Gamma\bigl(\frac{7-q}{2}\bigr)} \frac{\ell^{\,7-q}}{R_{q+1} \cdots R_{q+n}}\,.
\ee
We therefore arrive at a harmonic function where the $\omega$-dependence is exactly cancelled out, and it depends only on the doubly transverse coordinates $x^i$\,, with $\|x^i\| \equiv( x^i \, x^i)^{1/2}$\,.
As the harmonic function is manifestly finite, we can drop the condition $p+q=10$ we had previously.
In the $\omega \rightarrow \infty$ limit, the resulting M$p$T geometry takes the form
\begin{subequations} \label{DqSoln_forlimit_mixed_smeared}
\begin{align}
    \tau^a & = H^{-\frac{1}{4}} \, \dd x^a \,,
        \qquad%
    \tau^u = H^{\frac{1}{4}} \, \dd x^u\,,
        &%
    E^I & = H^{-\frac{1}{4}} \, \dd x^I \,,
        \qquad%
    E^i = H^{\frac{1}{4}} \, \dd x^i\,, \\[4pt]
    c^{(q+1)} & = g_\text{s}^{-1} \, H^{-1} \, \dd t \wedge \dd x^1 \wedge \dots \wedge \dd x^{q} \,,
        &%
    e^{\varphi} &= g^{}_\text{s} \, H^{\frac{3-q}{4}}\,,
\end{align}
\end{subequations}
with the harmonic function given by Eq.~\eqref{eq:hmc_smeared}.
For $q=3$ there will be an additional term in the gauge potential following from fact that the limit of the magnetic term in the field strength \eqref{F5general} is finite. 

These backgrounds can be generated by consecutively applying the T-duality transformations that map from M$p$T to M$(p+1)$T as in Section~\ref{sec:stdmpt}, starting with the geometry of~\eqref{eq:mptsl0} specified to $p=0$\,, \emph{i.e.}~starting with the M0T limit of the bulk D4-brane geometry, which required \emph{no} smearing.
For example, this is T-dual on a longitudinal direction of the D4-brane geometry to an M1T limit of the smeared D3 brane geometry, corresponding to~\eqref{DqSoln_forlimit_mixed_smeared} with $p=1$ and $m=3$. 
More generically, the bulk M$p$T geometry~\eqref{DqSoln_forlimit_mixed_smeared} labeled by $(p\,, \, q\,,\, m)$ is T-dual to the M$p$T geometry~\eqref{eq:mptsl0} labeled by $(\tilde{p}\,, \, \tilde{q}$), with
$\tilde{p} = q - m$ and $\tilde{q} = \tilde{p} + 4$\,.

Next, we consider the near-horizon limit applied to the M$p$T geometry~\eqref{DqSoln_forlimit_mixed_smeared} that arises from smearing. This is again generated from the asymptotic M$q$T limit prescribed as in Eq.~\eqref{eq:xaxixuxi}, which further implies that 
$R_u = \tilde{\omega}^{-\frac{1}{2}} \, \mathbb{R}_u$ and
$\ell^{\,7-q} = \tilde{\omega}^{\frac{q-3}{2}} \, \lbb{}^{\,7-q}$\,.
Hence, we find Eq.~\eqref{eq:hmc_smeared} gives
\be
    \ellsmear^{\,7-q-n} = \tilde{\omega}^{\frac{1}{2} (q-3+n)} \, \tilde{\ellsmear}{}^{\,7-q-n}\,,
\ee
where $\tilde{\ellsmear}$ is defined in terms of the fixed radii and string coupling appearing in the definition of the limits involved.
The smeared harmonic function \eqref{eq:hmc_smeared} then goes like $1+ \tilde \omega^2 ( \tilde{\ellsmear} / \|\mathbb{x}^i \|)^{7-q-n}$ and the induced bulk geometry in the $\tilde{\omega} \rightarrow \infty$ limit then becomes formally identical to the M$p$T geometry in Eq.~\eqref{eq:mptsl02}, except that now $\tilde{\mathbb{\Omega}} = \bigl( \| \mathbb{x}^i \| \, / \, \tilde{\ellsmear} \, \bigr)^{7-q-n}$\,.

\vspace{3mm}

\noindent $\bullet$~\emph{Large $N$.} Finally, we consider further non-Lorentzian brane geometries arising from an additional rescaling of the number $N$ of D$q$-branes. We will see that the ones that are relevant to holography involves sending $N$ to $\infty$ as some power of $\omega$. We divide the following discussions into two parts: we first consider a simple flat brane solution in the M$p$T supergravity, and then move on to more curved brane geometries.  

We start with the simple example with $q = p$\,, such that we are back to the configuration that we have considered in Section~\ref{sec:badsmt} in the context of the conventional near-horizon limit. In addition to the asymptotic M$p$T limit defined by the prescription~\eqref{eq:mptp}, we also introduce a further rescaling of $\ell$\,, with 
\be \label{eq:rsl}
    \ell^{7 - p} \rightarrow \omega^{-2} \, \ell^{7-p}\,,
\ee
such that the reparametrised D$p$-brane geometry~\eqref{eq:dpbg} becomes
\begin{subequations}
\begin{align}
    \dd s^2 & = \frac{\omega}{\sqrt{H}} \, \Bigl[ - \dd t^2 + \bigl( \dd x^1 \bigr)^2 + \cdots + \bigl( \dd x^p \bigr)^2 \Bigr] + \frac{\sqrt{H}}{\omega} \, \Bigl( \dd r^2 + r^2 \, \dd\Omega^2_{8-p} \Bigr)\,, \\[4pt]
    C^{(p+1)} & = \frac{\omega^2}{g^{}_\text{s} \, H} \, \dd x^0 \wedge \cdots \wedge \dd x^p\,,
        \qquad%
    e^\Phi = \omega^{\frac{p-3}{2}} \, g^{}_\text{s} \, H^{\frac{3-p}{4}}\,,
        \qquad%
    H = 1 + \left( \frac{\ell}{r} \right)^{\!\!7-p}\!\!.
\end{align}
\end{subequations}
Now, performing this new $\omega \rightarrow \infty$ limit incorporating the further rescaling of $\ell$ no longer leads to the bulk near-horizon limit as in Section~\ref{sec:badsmt}, but instead an M$p$T geometry described by
\be \label{eq:nlgmpt}
    \tau^A = H^{\frac{1}{2}} \, \dd x^A\,,
        \qquad%
    E^{A'} = H^{-\frac{1}{2}} \, \dd x ^{A'}\,,
        \qquad%
    e^\varphi = H^{\frac{p-3}{2}} \, g^{}_\text{s}\,,
\ee
with $A = 0\,, \, \cdots, \, p$ and $A' = p+1\,, \, \cdots, \, 9$\,. Note that all the other background fields are zero. Due to the dilatation symmetry~\eqref{eq:dilatsmpt}, the seemingly curved geometry~\eqref{eq:nlgmpt} is equivalent to the flat brane solution,
\be
    \tau^{}_\mu{}^A = \delta^A_\mu\,,
        \qquad%
    E^{}_\mu{}^{A'} = \delta^{A'}_\mu\,,
        \qquad%
    e^\varphi = g^{}_\text{s}\,.
\ee
Therefore, the solution that we have found in fact describes a ground state in the M$p$T supergravity. This trivialisation of the bulk geometry is of course due to the rescaling~\eqref{eq:rsl}, which is equivalent to rescaling the number of D$p$-branes as $N \rightarrow \omega^{-2} \, N$\,. As the number of $D$-branes is sent to zero, in the limit there is \emph{no} backreaction generating any non-trivial bulk brane geometry.
A similar rescaling was discussed in the context of the non-relativistic string theory limit of the F1 supergravity solution in \cite{Avila:2023aey}, where the limit sending $N$ to zero was described as a `formal manouevre' due in particular to the fact that having $N < 1$ does not make physical sense. This is entirely in accord with what we see here.

Next, we apply similar considerations to the general brane configuration in Section~\ref{sec:gbc}, with $q \neq p$\,. In order for the $\omega$-dependence in the harmonic function~\eqref{eq:hmc} to be of order one, we introduce an additional rescaling 
\be
    \ell^{7-q} \rightarrow \omega^{\tfrac{10-p-q}{2}} \ell^{7-q}\,.
\ee
Taking into account the condition $m+n=4$ required by the asymptotic M$p$T limit, this is equivalent to rescaling the number of D$q$-branes as 
\be
    N \rightarrow \omega^{3-q+m} \, N.
\ee 
For $q-m<3$\,, \emph{i.e.}~$p + q < 10$\,, this is a large $N$ limit. We will therefore require this condition in the following discussion. 
This mechanism of an additional rescaling is \emph{not} part of our intrinsic definition of an M$p$T limit, and its physical significance is not completely clear. 
This sort of rescaling was used in~\cite{Lambert:2024uue} when studying an M2-decoupling limit of the M2-brane supergravity solution, and also in~\cite{Fontanella:2024rvn} when studying the F1-decoupling limit of the D3-brane supergravity.
For D$q$-branes in M$p$T, the same effective rescaling was arrived at in \cite{Lambert:2024yjk} in terms of viewing the D$q$/M$p$T background as an intersecting brane solution, and then solving the resulting Laplacian equation albeit without sources -- it appears that including a brane source term leads inevitably back to the interpretation in terms of scaling $N$.
We leave aside the issues of the physical origin of this rescaling for now, and simply view it as a trick which engineers backgrounds of potential interest. 

We can then read off the M$p$T geometry describing the limit of the D$q$-brane with this additional rescaling introduced.
It takes the same form as \eqref{DqSoln_forlimit_mixed_smeared}, except with the harmonic function $H$ given by 
\be \label{Hlong}
    H = 1 + \frac{\ell^{7-q}}{\|x^u\|^{7-q}} \,,
\ee
which depends only on the M$p$T longitudinal coordinates $x^u$ rather than the M$p$T transverse coordinates $x^i$\,. Moreover, the near-horizon limit generated by a further asymptotic M$q$T limit also implies the correct scaling at large $\tilde{\omega}$\,, with
\be
    H = \tilde{\omega}^2 \, \lr \frac{\lbb}{\| \mathbb{x}^u \|} \rr^{\!\!7-q} + O\bigl(\tilde{\omega}^0\bigr)\,.
\ee
The near-horizon geometry is in form the same as in Eq.~\eqref{eq:mptsl02}, except that the condition $p + q = 10$ is now relaxed to be $p + q < 10$\,. 

\begin{table}[ht!]
\centering
\addtolength{\tabcolsep}{-2pt}    
\small
\begin{tabular}{ccccccccccc}
\multicolumn{11}{c}{\quad\quad\quad$\boxed{p=0 \quad q=4 \quad m=4 \quad n=0}$}\\[4pt]
& 0 & 1 & 2 & 3 & 4 & 5 & 6 & 7 & 8 & 9 \\
D4& $\times$ & $\times$ & $\times$ & $\times$ & $\times$ & -- & -- & -- & -- & -- \\
M$0$T& $\times$ & -- & -- & -- & -- & -- & -- & -- & -- & -- \\
\end{tabular}\hspace{2em}
\begin{tabular}{ccccccccccc}
\multicolumn{11}{c}{\quad\quad\quad$\boxed{p=2 \quad q=2 \quad m=2 \quad n = 2}$}\\[4pt]
& 0 & 1 & 2 & 3 & 4 & 5 & 6 & 7 & 8 & 9 \\
D2& $\times$ & $\times$ & $\times$ & -- & -- & -- & -- & -- & -- & -- \\
M$2$T& $\times$ & -- & -- & $\times$ & $\times$ & -- & -- & -- & -- & -- \\[4pt]
\end{tabular}

\vspace{1em}
\begin{tabular}{ccccccccccc}
\multicolumn{11}{c}{\quad\quad\quad$\boxed{p=2 \quad q=4\quad m=3 \quad n=1}$}\\[4pt]
& 0 & 1 & 2 & 3 & 4 & 5 & 6 & 7 & 8 & 9 \\
D4 & $\times$ & $\times$ & $\times$ & $\times$ & $\times$ & -- & -- & -- & -- & -- \\
M2T & $\times$ & $\times$ & -- & -- & -- & $\times$ & -- & -- & -- & -- \\
\end{tabular}\hspace{2em}
\begin{tabular}{ccccccccccc}
\multicolumn{11}{c}{\quad\quad\quad$\boxed{p=4 \quad q=0 \quad m=0 \quad n=4}$}\\[2pt]
& 0 & 1 & 2 & 3 & 4 & 5 & 6 & 7 & 8 & 9 \\
D0 & $\times$ & -- & -- & -- & -- & -- & -- & -- & -- & -- \\
M4T & $\times$ & $\times$ & $\times$ & $\times$ & $\times$ & -- & -- & -- & -- & -- \\
\end{tabular}

\vspace{1em}\hspace{1mm}
\begin{tabular}{ccccccccccc}
\multicolumn{11}{c}{\quad\quad\quad$\boxed{p=4\quad q=2\quad m=1 \quad n=3}$}\\[4pt]
& 0 & 1 & 2 & 3 & 4 & 5 & 6 & 7 & 8 & 9 \\
D2& $\times$ & $\times$ & $\times$ & -- & -- & -- & -- & -- & -- & -- \\
M4T& $\times$ & $\times$ & -- & $\times$ & $\times$ & $\times$ & -- & -- & -- & -- \\
\end{tabular}\hspace{2em}
\begin{tabular}{ccccccccccc}
\multicolumn{11}{c}{\quad\quad\quad\!\!$\boxed{p=4 \quad q=4 \quad m=2 \quad n=2}$}\\[2pt]
& 0 & 1 & 2 & 3 & 4 & 5 & 6 & 7 & 8 & 9 \\
D4 & $\times$ & $\times$ & $\times$ & $\times$ & $\times$ & -- & -- & -- & -- & -- \\
M$4$T & $\times$ & $\times$ &$\times$ & -- & -- & $\times$ & $\times$ & -- & -- & --  \\
\end{tabular}

\vspace{1em}\hspace{1mm}
\begin{tabular}{ccccccccccc}
\multicolumn{11}{c}{\quad\quad\quad$\boxed{p=1 \quad q=3 \quad m=3 \quad n=1}$}\\[2pt]
& 0 & 1 & 2 & 3 & 4 & 5 & 6 & 7 & 8 & 9 \\
 D3& $\times$ & $\times$ & $\times$ & $\times$ & -- & -- & -- & -- & -- & -- \\
 M$1$T& $\times$ & --  & -- & -- & $\times$ & -- & -- & -- & -- & -- \\
\end{tabular}\hspace{2em}
\begin{tabular}{ccccccccccc}
\multicolumn{11}{c}{\quad\quad\quad$\boxed{p=3 \quad q=1 \quad m=1 \quad n=3}$}\\[2pt]
& 0 & 1 & 2 & 3 & 4 & 5 & 6 & 7 & 8 & 9 \\
 D1& $\times$ & $\times$ & -- & -- & -- & -- & -- & -- & -- & -- \\
 M3T& $\times$ & -- & $\times$ & $\times$ & $\times$ & --  & -- & -- & -- & -- \\
\end{tabular}

\vspace{1em}
\begin{tabular}{ccccccccccc}
\multicolumn{11}{c}{\!\quad\quad\quad$\boxed{p=3 \quad q=3 \quad m=2 \quad n=2}$}\\[2pt]
& 0 & 1 & 2 & 3 & 4 & 5 & 6 & 7 & 8 & 9 \\
D3& $\times$ & $\times$ & $\times$ & $\times$ & -- & -- & -- & -- & -- & -- \\
M$3$T& $\times$ & $\times$ & -- & -- & $\times$  & $\times$ & -- & -- & -- & -- \\
\end{tabular} \hspace{6.6cm}
\caption{Possible D$q$-branes admitting M$p$T limits with $p\leq 4$ and $q \leq 4$\,, satisfying $m + n = 4$\,. Here, $m$ is the number of spatial directions of the D$q$-brane that are transverse with respect to the M$p$T limit, and $n$ the number of directions transverse to the D$q$-brane that are longitudinal with respect to the M$p$T limit. 
Note that the same condition implies that the analogous D$p$-D$q$ brane intersection is 1/4 BPS.
The case $m=4$ is special as taking the limit does \emph{not} require additional manipulation (smearing or large $N$ rescaling) in the harmonic function. 
In other cases, the D$q$ brane is either localised only on transverse directions of the M$p$T limit (if we smear) or on longitudinal directions (if we use the extra rescaling $N$).
}
\addtolength{\tabcolsep}{2pt}  
\label{tab:IIbranes}
\end{table}

\vspace{3mm}
\noindent $\bullet$~\emph{A catalogue of M$p$T D$q$-brane geometries.}
It is straightforward to solve the condition $m + n = 4$ for the allowed branes.
For $q\,,\,p\leq 4$ the complete set is shown in Table~\ref{tab:IIbranes}.
Let us make two remarks about the cases contained in this table.
The first remark concerns the match with the analysis of D$q$-brane worldvolume actions in section \ref{sec:gdbmpt}, and the second concerns the match with the proposed non-Lorentzian holographic correspondences of \cite{Lambert:2024yjk}.

Firstly, we note that the D$q$-brane configurations in M$p$T with $m + n =4$ were identified in section \ref{sec:gdbmpt} as the D$q$-brane static configurations which have finite mass in the M$p$T limit. 
Here, we see that they correspond to the supergravity solutions whose M$p$T limits can be taken and produce a curved M$p$T configuration (potentially only after smearing or rescaling $N$).
This is reminiscent of observations made in~\cite{Guijosa:2023qym,Avila:2023aey}, where D$q$-branes in non-relativistic string theory were examined. 
It was argued that depending on the orientation with respect to the longitudinal direction of the non-relativistic string limit, the D$q$-brane in non-relativistic string theory either gave rise to a geometry which was necessarily \emph{Lorentzian} if the brane wrapped the longitudinal direction or else a genuine \emph{non-Lorentzian} string Newton-Cartan background if the brane was transverse.
It seems likely there is a similar interpretation here: if we insert a D$q$-brane whose mass diverges in the M$p$T limit, it will backreact sufficiently strongly so as to source a relativistic supergravity geometry. On the other hand, a D$q$-brane whose mass is finite can exist as a source for curved M$p$T geometries.

Secondly, the possible solutions encapsulated in Table~\ref{tab:IIbranes} include all the limits discussed as candidate geometries for new non-Lorentzian holographic duals in \cite{Lambert:2024yjk} (see their Figure~1).
In this paper, the extra rescaling of $\ell$ was (effectively) incorporated into the limit such that they obtained the longitudinal-coordinate dependent solutions characterised by the harmonic function $H$ of \eqref{Hlong}.
We will take these holographic proposals as input into our more general treatment of holography in the next section (see in particular Section \ref{sec:furtherdlcqs}).
Note that our analysis shows that there exist two seemingly different versions of these D$q$-brane solutions: one localised only in transverse directions of the M$p$T limit and the other localised only in longitudinal directions.
It would be interesting to understand if these are simply smeared cases of a more general allowed D$q$-brane geometry in M$p$T, and to clarify which backgrounds are actually relevant for holography.

\section{A Conjecture: Holography as \texorpdfstring{DLCQ${}^n$/DLCQ${}^{m}$}{DLCQn/DLCQm} Correspondence} \label{sec:chdlcq}

In Section~\ref{sec:hnhbpsl}, we discussed further BPS decoupling limits of the bulk AdS geometries and saw that the standard AdS/CFT correspondence can be reincarnated in the language of an asymptotic M$p$T limit. We have also shown that the geometry at the asymptotic infinity is intrinsically non-Lorentzian.
Furthermore, we applied multiple BPS decoupling limits to generate non-Lorentzian versions of holography. 
This hints at a general principle for classifying holographic duals using the duality web of BPS decoupling limits. 

More precisely, what we have seen in the previous section is the following. 
We apply the D$p$-brane version of the BPS decoupling limit at asymptotic infinity. 
In the bulk geometry, this induces the near-horizon limit. This does not change the character of the bulk geometry itself, \emph{i.e.}~if it is Lorentzian then it leads to a Lorentzian geometry, while if it is a non-Lorentzian M$q$T geometry (obtained by applying some separate D$q$-brane decoupling limit) it stays an M$q$T geometry.
At the asymptotic infinity, the limit does lead to a different geometry, and furthermore acts non-trivially when applied to the brane worldvolume field theory living at infinity.
Note that if the bulk-geometry is M$q$T, then applying a further M$p$T limit leads to a novel, more involved non-Lorentzian geometry at infinity. 
In the terminology of \cite{Blair:2023noj}, this arises from a \emph{multicritical} BPS decoupling limit.

In this section, we will find that it is convenient to organise the BPS decoupling limits in terms of (U-)duality orbits of M-theory in the Discrete Light Cone Quantisation (DLCQ)~\cite{Blair:2023noj, Gomis:2023eav, bpslimits}. We will see that the results in Section~\ref{sec:hnhbpsl} lead us to a natural conjecture that the AdS/CFT correspondence generalises to correspondences between: (1) a bulk geometry in the DLCQ${}^n$ orbit, and (2) a field theory in the DLCQ${}^{n+1}$ orbit. The discrepancy in the number of DLCQs is accounted for by the fact that the near-horizon decoupling limit does \emph{not} alter the nature of the geometry in the bulk. 

Furthermore, in Section~\ref{sec:dlcqmn}, we generalise the DLCQ${}^n$/DLCQ${}^{n+1}$ correspondence to the DLCQ${}^n$/DLCQ${}^m$ correspondence with $m > n$\,. The extra $m - n$ DLCQs on the field theory side correspond to performing multiple `near-horizon' limits on the bulk side: namely, we consider a bound D-brane
bulk geometry with multiple harmonic functions of the form $H = 1 + \cdots$, and each of the `near-horizon' limits gets rid of the first term `1' in one of the harmonic functions. Each of these `near-horizon' limits corresponds to a BPS decoupling limit at the asymptotic infinity. We will discuss a concrete example of DLCQ${}^0$/DLCQ${}^2$ based on the AdS${}_3$/CFT${}_2$ correspondence from the D1-D5 system. 

\subsection{Matrix \texorpdfstring{$p$}{p}-Brane Theory from M-Theory in the DLCQ} \label{sec:mptmtd}

As a preparation for our discussion, we first illustrate the connection between M$p$T and M-theory in the DLCQ. In particular, in this subsection, we describe how such a connection works in detail for M$0$T.

The DLCQ of M-theory can be viewed as a null compactification of M-theory. 
It is commonly defined as a decoupling limit of M-theory, and intuitively interpreted as an infinite boost limit of M-theory along a compactified spatial direction \cite{Seiberg:1997ad}. Even though the spatial compactification breaks the boost symmetry, it is still convenient to use the terminology of `infinite boost' in a heuristic way.
We will show that this infinite boost followed by a simple change of coordinates \cite{Bilal:1998vq} leads automatically to the M0T limit.

We begin with M-theory on a spatial circle of radius $R_{\text{s}}$\,.
Denote the eleven-dimensional spacetime coordinates by $(X^0, \, X^{10}, \, x^i)$\,, $i=1\,, \, \dots, \, 9$\,, with the periodic boundary condition $X^{10} \sim X^{10} + 2 \pi R_{\text{s}}$\,. The line element is
\be
\dd s^2 = - \bigl( \dd X^0 \bigr)^2 + \bigl( \dd X^{10} \bigr)^2 + \dd x^i \, \dd x^i \,.
\label{linelt}
\ee
Now, we apply a boost transformation to define M-theory on an almost lightlike circle.
The boost defines new coordinates by
\be
    x^0 = \gamma \, \bigl( X^0 - v \, X^{10} \bigr)\,,
        \qquad%
    x^{10} = \gamma \, \bigl( X^{10} - v \, X^0 \bigr)\,,
        \qquad%
    \gamma = \frac{1}{\sqrt{1-v^2}}\,,
\ee
with $v$ the boost velocity and $\gamma$ the Lorentz factor. 
We then define the almost `lightcone' coordinates 
\be 
    x^- = \frac{x^{10} - x^0}{\sqrt{2}}\,,
        \qquad%
    x^+ = \frac{x^0 + x^{10}}{\sqrt{2}} - \frac{x^-}{2 \, \omega^2}\,, 
        \qquad%
    \omega = \frac{1}{\sqrt{2}} \sqrt{\frac{1 + v}{1 - v}}\,,
\label{almostlccoords}
\ee
which become truly lightlike in the infinite boost limit $v \rightarrow 1$\,, \emph{i.e.}~$\omega \rightarrow \infty$\,. 
This is exactly the same parameter $\omega$ that we have been using throughout the paper, and it corresponds to the Lorentz factor $\gamma$ associated with the infinite boost in eleven dimensions, as can be seen by noting $\gamma \sim \omega / \sqrt{2}$ at large $\omega$\,. The $O(\omega^{-2})$ term in the definition of $x^+$ is introduced such that only $x^-$ is periodic, with
\be \label{eq:pllc}
    x^+ \sim x^+\,,
        \qquad%
    x^- \sim x^- \!+ 2 \pi R\,,
        \qquad%
   R \equiv \omega \, R_{\text{s}}\,.
\ee
A well-defined infinite boost limit then also requires that the effective radius $R$ of the null circle in $x^-$ is fixed in the $\omega \rightarrow \infty$ limit. In terms of the coordinates \eqref{almostlccoords}, the line element \eqref{linelt} becomes
\be \label{eq:edle}
    \dd s^2_{11} = \frac{1}{\omega^2} \, \bigl( \dd x^- + \omega^2 \, \dd x^+ \bigr)^{2} - \omega^2 \, \bigl( \dd x^+ \bigr)^2 + \dd x^i \, \dd x^i\,.
\ee
For finite $\omega$\,, the $x^-$ direction remains a spacelike circle. In the $\omega \rightarrow \infty$ limit, the line element~\eqref{eq:edle} becomes
\be
    \dd s^2_{11} = 2\, \dd x^- \, \dd x^+ + \dd x^i \, \dd x^i\,,
\ee
which means that we are led to M-theory with a lightlike compactification in $x^-$\,, \emph{i.e.}~the DLCQ of M-theory. Note that, as indicated by Eq.~\eqref{eq:pllc}, the lightlike circle has a finite effective radius $R$\,.

We now switch to the perspective of IIA superstring theory by viewing the $x^-$ compactification as the M-theory circle, and revisit the limiting procedure above. When we do so, the spacetime coordinates that we are initially using are the ones appropriate in M-theory, where the length scale is the eleven-dimensional Planck scale. In order to measure in terms of the string scale $\ell^{}_\text{s}$ in ten dimensions, we have to perform the following coordinate rescalings:
\be
    x^- \rightarrow \frac{R}{\ell^{}_\text{P}} \, x^-\,,
        \qquad%
    x^\mu \rightarrow \frac{\ell^{}_\text{s}}{\ell^{}_\text{P}} \, x^\mu\,.
\ee
The rescaling of $x^-$ is introduced such that the dependence on the radius of the null compactification is relocated to be in the background fields. 
Here, we have introduced the ten-dimensional coordinates $x^\mu = (t = x^+, \, x^i)$\,, with the lightcone coordinate $x^+$ in eleven dimensions now playing the role of time in the ten-dimensional IIA theory. Using the standard relations $R = g^{}_\text{s} \, \ell^{}_\text{s}$\,, with $g^{}_\text{s}$ the effective string coupling, and $\ell^{}_\text{P} = g^{1/3}_\text{s} \, \ell^{}_\text{s}$\,, we find that the eleven-dimensional line element then decomposes into
\begin{align} \label{eq:edlet}
\begin{split}
    \dd s^2_{11} & = \frac{g^{4/3}_\text{s}}{\omega^{2}} \, \Bigl( \dd x^- + \omega^2 \, g^{-1}_\text{s} \, \dd t \Bigr)^2 + g^{-2/3}_\text{s} \Bigl( - \omega^2 \, \dd t^2 + \dd x^i \, \dd x^i \Bigr) \\[4pt]
    & \equiv e^{4\Phi/3} \, \bigl( \dd x^- + C^{}_\mu \, \dd x^\mu \bigr)^2 + e^{-2\Phi/3} \, g^{}_{\mu\nu} \, \dd x^\mu \, \dd x^\nu\,,
\end{split}
\end{align}
where $g_{\mu\nu}$ is the metric in ten dimensions, with $C^{(1)} = C_\mu \, dx^\mu$ the RR one-form potential and $\Phi$ the dilaton field. From Eq.~\eqref{eq:edlet}, we find
\be \label{eq:sten}
    \dd s^2_{10} = g^{}_{\mu\nu} \, \dd x^\mu \, \dd x^\nu = - \omega \, \dd t^2 + \frac{1}{\omega} \, \dd x^i \, \dd x^i\,,
        \qquad%
    C^{(1)} = \omega^2 \, g^{-1}_\text{s} \, \dd t\,,
        \qquad%
    e^\Phi = \omega^{-\frac{3}{2}} \, g^{}_\text{s}\,,
\ee
which matches the M0T prescription~\eqref{eq:rpom} that we derived from the non-relativistic particle limit applied to D0-branes. Therefore, in the $\omega \rightarrow \infty$ limit, M0T in flat spacetime is recovered.  

As we mentioned after Eq.~\eqref{eq:rpom}, this limit keeps $\alpha'$ and hence the string length fixed.
In the literature on matrix theory and DLCQ M-theory it is more common to phrase the limit in terms of a rescaling of $\ell_\text{s}$.
In particular, we could adopt an alternative string length  $\tilde{\ell}^{}_\text{s} = \omega^{-\frac{1}{2}} \, \ell^{}_\text{s}$, which also induces a new set of spacetime coordinates $\tilde{x}^\mu$ and string coupling $\tilde{g}^{}_\text{s}$ that satisfy 
$\dd\tilde{s}^2_{10} / \, \tilde{\ell}^2_\text{s} = \dd s^2_{10} / \ell^2_\text{s}$ and
$\tilde{\ell}^{}_\text{P} = \ell^{}_\text{P}$\,.
It then follows that 
\be
    \dd \tilde{s}^2_{10} = - \dd t^2 + \omega^{-2} \, \dd x^i \, \dd x^i\,,
        \qquad%
    C^{(1)} = \sqrt{\omega} \, \tilde{g}^{-1}_\text{s} \, \dd t\,,
        \qquad%
    e^\Phi = \tilde{g}^{}_\text{s}\,.
\ee
This essentially recovers the scaling limit arrived at in \cite{Seiberg:1997ad}.

From our discussion in Section~\ref{sec:stdmpt}, the other M$p$T limits then follow by T-duality of \eqref{eq:sten}. 
It is in this sense that they can all be viewed as lying in the duality orbit of the M-theory DLCQ which leads to M0T.
We will shortly recast this relationship in terms of U-duality in M-theory itself.

\subsection{AdS/CFT from M-Theory in the DLCQ} \label{sec:adsdlcq}

So far, we have reviewed the DLCQ of M-theory, recast in our framework of M0T. It is well known that, after performing the infinite boost limit in M-theory, all excitations except the Kaluza-Klein modes in the lightlike circle are decoupled. From the IIA perspective, these Kaluza-Klein modes correspond to the D0-branes in M0T, whose dynamics is described by BFSS matrix theory~\cite{Susskind:1997cw,Seiberg:1997ad,Sen:1997we}. At large $N$, the null circle in DLCQ M-theory decompactifies and BFSS matrix theory is believed \cite{Banks:1996vh} to encode the physics of M-theory in eleven-dimensional Minkowski spacetime, with the eleven-dimensional Lorentz symmetry restored (see \cite{Tropper:2023fjr, Herderschee:2023bnc} for recent progress on understanding this particular issue).

In this subsection, we will discuss examples of holographic dualities that arise from the duality orbit related to the DLCQ of M-theory, which include the standard AdS/CFT correspondence that we will recast as a correspondence between sectors of M-theory with and without the DLCQ. We will then discuss further the M-theory uplifts of the various M$p$Ts associated with IMSY holography~\cite{Itzhaki:1998dd}, which leads us to the so-called \emph{non-relativistic M-theory} that is U-dual to DLCQ M-theory~\cite{Ebert:2023hba, Blair:2023noj}. 

\vspace{3mm}

\noindent $\bullet$~\emph{AdS/CFT as a DLCQ$\,{}^0$/DLCQ$\,{}^1$ correspondence.}
As we reviewed in Section \ref{sec:badsmt}, the IMSY holographic duality applied to general D$p$-branes follows by taking the M$p$T decoupling limit both in the open string picture, where it leads to matrix theory on $p$-branes, and in the supergravity (closed string) picture, where it induces the near-horizon limit of the D$p$-brane geometry.  
Let us focus on $p=0$ case relevant for BFSS matrix theory.\footnote{We can view this as a refinement of the flat spacetime limit of Section \ref{sec:mptmtd} taking into account the backreaction of the D0-branes \cite{Balasubramanian:1997kd, Polchinski:1999br}.}
From Eq.~\eqref{eq:dsco}, we read off the corresponding bulk conformal AdS${}_2$ geometry generated by the asymptotic M0T limit. In particular, the dilaton develops a profile given by,
\be
e^\Phi = \lr \frac{\ell}{r} \rr^{\!\!\frac{21}{4}} g^{}_\text{s}\,.
\ee
Deep in the bulk where $r$ is small, the effective string coupling becomes large and the M-theory description is needed, which naturally corresponds to the large $N$ limit of BFSS matrix theory. Instead, when one approaches the asymptotic infinity, the $\alpha'$ corrections cannot be ignored.
Therefore, it is only the intermediate bulk region ($0 \ll r \ll \infty$) where the physics is described by IIA supergravity. The same behaviour also apply to general holographic duals involving D$p$-branes with $p < 3$~\cite{Itzhaki:1998dd}. For the $p=3$ case of AdS${}_5$/CFT${}_4$\,, the dilaton remains constant throughout the bulk, and therefore leads to a simpler duality relation between $\CN = 4$ SYM and the bulk AdS${}_5$ geometry. 

In the framework of M$p$T, the conventional D$p$-brane holographic correspondences are formally related to each other via T-duality transformations, as M$p$Ts are T-dual to each other (see Section~\ref{sec:stdmpt}). In this sense, this whole zoo of holographic duals may be thought of as being generated by the DLCQ limit of M-theory: the infinite-boost limit in eleven dimensions corresponds to the BPS decoupling limit of type IIA superstring theory, which leads to M0T whose dynamics is described by BFSS matrix theory. This corner is then dualised to other M$p$Ts, and thus generates a set of BPS decoupling limits in type II superstring theory. Such BPS decoupling limits are then used to induce the bulk near-horizon limits as in Section~\ref{sec:badsmt}. In this regard, the DLCQ limit in M-theory provides us with a `parent' operation which then cascades to generate various decoupling limits in string theory that are associated with holographic dualities.

In this above sense the field theories living at the asymptotic infinity are all related to the DLCQ of M-theory. More loosely, we refer to the full duality orbit of such BPS decoupling limits as belonging to the DLCQ (or `the 1${}^\text{st}$ DLCQ') of M-theory. On the bulk side, however, the asymptotic M$p$T limit corresponds to the near-horizon limit, and the bulk geometry is Lorentzian. In this sense, the IMSY holographic duals can be viewed as examples of a duality between the DLCQ${}^0$ of closed strings in the bulk and the DLCQ${}^1$ of open strings at the asymptotic infinity. This observation shows that the IMSY correspondence is a DLCQ${}^0$/DLCQ${}^1$ correspondence. At this point, all we have done is a rephrasing of the conventional story of holographic duals, but we will see in the next subsection how this idea generalises to the DLCQ${}^n$/DLCQ${}^{n+1}$ correspondence. 

\vspace{3mm}

\noindent $\bullet$~\emph{M-theory uplifts from U-duality.} Before we proceed, let us first clarify how to view M$p$T, $p>0$, directly in terms of M-theory uplifts. 
We will use this excursion to illustrate a generic lesson that also applies to the later discussions in this paper: the null compactification in DLCQ M-theory can be mapped to a spatial compactification via a duality transformation, where the latter dual picture makes the BPS nature of the DLCQ limit manifest. This is the key observation that underlies the validity of the DLCQ in string and M-theory as well as matrix gauge theories at finite $N$\,.

\begin{figure}[t!]
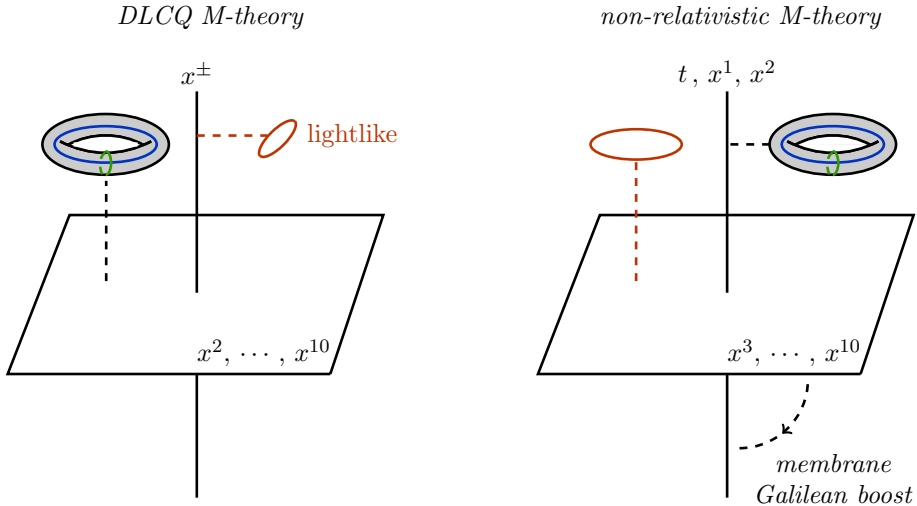

\begin{center}

\tikzpicture[scale=0.47, line width=0.35mm]

\draw (3.375+1,-4.5) -- (-3.625-0.125-1,-4.5) -- (-2-1,0) -- (5-0.125+1,0) -- (3.375+1,-4.5);

\draw (0.6,-8) -- (0.6,-4.5);
\draw (0.6,-2.2)--(0.6,3.5);
\draw (1+15,5.6) node{\scalebox{0.9}{\emph{non-relativistic M-theory}}};
\draw (0.6+15,4) node{\scalebox{0.9}{$t\,, \, x^1, \, x^2$}};
\draw (4.9-2.4+15,-3.9) node{\scalebox{0.9}{$x^3, \, \cdots, \, x^{10}$}};
\draw (20-15,2.25) node{\scalebox{0.9}{\textcolor{DarkRed}{lightlike}}};
\draw [DarkRed,dashed] (.6,2+.25)--(.6+2,2+.25);
%\draw [DarkRed] (0.6cm,2+.25) node{$\boldsymbol{\times}$};
\draw [DarkRed, postaction={decorate},cm={cos(45) ,-sin(45) ,sin(45) ,cos(45) ,(-0.4 cm, 2.7 cm)}] (2.4+0.3,1.25) arc (-90:90:0.35cm*0.7 and 0.99*0.7cm);
\draw [DarkRed, cm={cos(45) ,-sin(45) ,sin(45) ,cos(45) ,(-0.4 cm, 2.7 cm)}] (2.4+0.3,1.25+1.9*0.7+0.05) arc (90:270:0.35cm*0.7 and 0.99*0.7cm);

\draw (15+3.375+1,-4.5) -- (15-3.625-0.125-1,-4.5) -- (15-2-1,0) -- (15+5-0.125+1,0) -- (15+3.375+1,-4.5);

\draw (15+0.6,-8) -- (15+0.6,-4.5);
\draw (15+0.6,-2.2)--(15+0.6,3.5);

\draw (1,5.6) node{\scalebox{0.9}{\emph{DLCQ M-theory}}};
\draw (0.6,4) node{\scalebox{0.9}{$x^\pm$}};
\draw (4.9-2.4,-3.9) node{\scalebox{0.9}{$x^2, \, \cdots, \, x^{10}$}};
\scope[xshift=13 cm,yshift=2cm,scale=.43]

\fill[LightGray] (13,0) ellipse (5.2*0.8 and 2.5*0.8);

\begin{scope}
  \clip (13,0) ellipse (5.2*0.8 cm and 2.5*0.8 cm);
  \path[draw,name path= A](-3+13,0.2) .. controls (-1+13,-0.8) and (1+13,-0.8) .. (3+13,0.2);
  \clip (13,0) ellipse (5.2*0.8 cm and 1.5*0.8 cm);
  \path[draw,name path=B] (-2.5+13,-0.05) .. controls (-1+13,0.8) and (1+13,0.8) .. (2.5+13,-0.05);
\fill [white, intersection segments={of=A and B, sequence={A1--B1}}];
\end{scope}

\draw[] (13,0) ellipse (5.2*0.8 cm and 2.5*0.8 cm);
\draw[] (-3+13,0.2) .. controls (-1+13,-0.8) and (1+13,-0.8) .. (3+13,0.2);
\draw[] (-2.5+13,-0.05) .. controls (-1+13,0.8) and (1+13,0.8) .. (2.5+13,-0.05);

\draw[DarkBlue, postaction={decorate}] (13,0) ellipse  (4.2*0.8 cm and 1.5*0.8 cm);

\draw[olivegreen] (13,-2) arc (-90:90:0.45*0.8cm and 0.97*0.8cm);
\draw[olivegreen,dashed] (13,-0.575) arc (90:270:0.45*0.8cm and 0.97*0.8cm);

\draw[ultra thick, white] (-2.1,-.0035) -- (2.1,-.0035);

%\draw[black] (-34.8,-9) node{$\boldsymbol{\times}$};
\draw [black,dashed] (-34.8,-9)--(-34.8,-2.4);
\endscope

%%%%%%%%%%%% SECOND TORUS

\scope[xshift=3.61cm,yshift=2cm,scale=.43]

\fill[LightGray] (0-13,0) ellipse (5.2*0.8 and 2.5*0.8);

\begin{scope}
  \clip (0-13,0) ellipse (5.2*0.8 cm and 2.5*0.8 cm);
  \path[draw,name path= A](-3-13,0.2) .. controls (-1-13,-0.8) and (1-13,-0.8) .. (3-13,0.2);
  \clip (0-13,0) ellipse (5.2*0.8 cm and 1.5*0.8 cm);
  \path[draw,name path=B] (-2.5-13,-0.05) .. controls (-1-13,0.8) and (1-13,0.8) .. (2.5-13,-0.05);
\fill [white, intersection segments={of=A and B, sequence={A1--B1}}];
\end{scope}

\draw[] (0-13,0) ellipse (5.2*0.8 cm and 2.5*0.8 cm);
\draw[] (-3-13,0.2) .. controls (-1-13,-0.8) and (1-13,-0.8) .. (3-13,0.2);
\draw[] (-2.5-13,-0.05) .. controls (-1-13,0.8) and (1-13,0.8) .. (2.5-13,-0.05);

\draw[DarkBlue, postaction={decorate}](0-13,0) ellipse  (4.2*0.8 cm and 1.5*0.8 cm);

\draw[olivegreen, postaction={decorate}] (0-13,-2) arc (-90:90:0.45*0.8cm and 0.97*0.8cm);
\draw[olivegreen,dashed] (0-13,-0.575) arc (90:270:0.45*0.8cm and 0.97*0.8cm);

\draw[ultra thick, white] (-2.1-13,-.0035) -- (2.1-13,-.0035);

%\draw[black] (-7+34.9,0) node{$\boldsymbol{\times}$};
\draw [black,dashed] (-6.8+34.9,0)--(-4.2+34.9,0);
\draw[DarkRed] (-13+34.9,0) ellipse  (3 cm and 1cm);
%\draw[DarkRed] (-13+34.9,-9) node{$\boldsymbol{\times}$};
\draw [DarkRed,dashed] (-13+34.9,-9)--(-13+34.9,-1);
\draw[decoration={markings, mark=at position 0.5 with {\arrow{<}}},
        postaction={decorate}, dashed] (-6.2+34.9,-20) arc (-90:0:4.5cm and 4.5cm);
\draw (0+34.9,-21) node{\scalebox{0.9}{\emph{membrane}}};
\draw (0+34.9,-23) node{\scalebox{0.9}{\emph{Galilean boost}}};
\endscope

\endtikzpicture
\caption{The U-duality relation between DLCQ and non-relativistic M-theory.
The torus (shaded grey) transverse to the DLCQ direction is exchanged for a torus in the longitudinal non-relativistic M-theory directions. Similarly the red lightlike circle in the DLCQ M-theory is mapped to the red spacelike circle in the transverse sector of non-relativistic M-theory. The figure is adapted from \cite{Ebert:2023hba}.}   
\label{fig:DLCQnrmt}
\end{center}
\end{figure}

In view of the T-dual relations between the M$p$Ts, we are required to find the U-dual frame of DLCQ M-theory in order to identify their M-theory uplifts. This requires compactifying DLCQ M-theory over an extra two-torus.
U-duality then acts on this two-torus together with the direction that becomes null in the DLCQ: collectively these three directions can be thought of as a three-torus with one of the cycles being null. Starting with the metric~\eqref{eq:edle} whose $\omega \rightarrow \infty$ limit defines M-theory in the DLCQ, applying the standard U-duality transformation~\cite{Obers:1998fb} (see Figure.~\ref{fig:DLCQnrmt} for a pictorial illustration) leads to M-theory in the following background metric and three-form fields:
\begin{subequations} \label{eq:nrmtdef}
\begin{align}
	\dd s^2_{11} & = \omega^{\frac{4}{3}} \, \Bigl[ - \dd t^2 + \bigl( \dd x^1 \bigr)^2 + \bigl( \dd x^2 \bigr)^2 \Bigr] + \omega^{- \frac{2}{3}} \, \Bigl[ \bigl( \dd x^3 \bigr)^2 + \cdots + \bigl( \dd x^{10} \bigr)^2 \Bigr]\,, \\[4pt]
	A^{(3)} & = \omega^2 \, \dd t \wedge \dd x^1 \wedge \dd x^2\,.  
\end{align}
\end{subequations}
Here, $\omega$ is related to the Lorentz factor associated with the DLCQ limit in the original M-theory in the DLCQ. The $\omega \rightarrow \infty$ limit defines a BPS decoupling limit that zooms in on a background M2-brane. Applying this limit to M-theory leads to the so-called non-relativistic M-theory (this is called the `Galilean Membrane' theory in \cite{Gomis:2000bd}, and `Wrapped M2-brane' theory in  \cite{Danielsson:2000gi, Garcia:2002fa}). In the decompactification limit, the eleven-dimensional spacetime geometry admits a codimension-three foliation structure, whose curved version is referred to as membrane Newton-Cartan geometry~\cite{Blair:2021waq}.\,\footnote{See also~\cite{Ebert:2021mfu} for an intrinsic derivation of how the membrane Newton-Cartan geometry arises starting from a string theory perspective using probe branes.} Importantly, non-relativistic M-theory arises from a standard BPS decoupling limit where \emph{no} null compactification is involved. The wrapped membranes on the longitudinal torus in non-relativistic M-theory are dual to the Kaluza-Klein excitations in the null circle of DLCQ M-theory. In this sense, DLCQ M-theory on a two-torus receives a solid definition as a U-dual of non-relativistic M-theory, and the genuine DLCQ M-theory without any further toroidal compactification arises in a decompactification limit~\cite{Ebert:2023hba, Blair:2023noj}.

We discussed in Section \ref{sec:badsmt} how the M$p$T decoupling limit was related to the near-horizon D$p$-brane decoupling limit and holography. 
The BPS decoupling limit defined by the prescription in Eq.~\eqref{eq:nrmtdef} can similarly be related to the holographic prescription for M2-branes and the AdS${}_4$/CFT${}_3$ correspondence~\cite{Maldacena:1997re}. 
Let us illustrate this on the geometric side. Consider the bulk M2-brane geometry, 
\begin{subequations} \label{eq:mtbg}
\begin{align}
    \dd s^2_{11} &= H^{- \frac{2}{3}} \, \Bigl[ - \bigl( \dd X^0 \bigr) + \bigl( \dd X^1 \bigr)^2 + \bigl( \dd X^2 \bigr)^2 \Bigr] + H^{\frac{1}{3}} \, \Bigl( \dd R^2 + R^2 \, \dd \Omega^2_7 \Bigr)\,, \\[4pt]
    A^{(3)} &= \frac{1}{H} \, \dd X^0 \wedge \dd X^1 \wedge \dd X^2\,, 
        \qquad%
    H = 1 + \left( \frac{L}{R} \right)^{\!\!6}\,,
\end{align}
\end{subequations}
where $R^2 = ( X^3 )^2 + \cdots + ( X^{10} )^2$\,. Applying the prescription~\eqref{eq:nrmtdef} at the asymptotic infinity with $R \rightarrow \infty$ implies $X^A = \omega^{2/3} \, x^A$ with $A = 0\,, \, 1\,, \, 2$ and $R = \omega^{-1/3} \, r$ with $r^2 = (x^3)^2 + \cdots + (x^{10})^2$\,. We therefore find at large $\omega$ that
$H \sim \omega^2 \, (L / r)^6$\,.
In the $\omega \rightarrow \infty$ limit, the M2-brane geometry~\eqref{eq:mtbg} gives rise to the AdS${}_4 \times S^7$ geometry:
\begin{subequations} \label{eq:adsf}
\begin{align}
    \dd s^2_{11} &= \frac{r^4}{L^4} \, \Big[ - \dd t^2 + \bigl( \dd x^1 \bigr)^2 + \bigl( \dd x^2 \bigr)^2 \Bigr] + \frac{L^2}{r^2} \Bigl( \dd r^2 + r^2 \, \dd\Omega^2_7 \Bigr)\,, \\[4pt]
    A^{(3)} &= \frac{r^6}{L^6} \, \dd t \wedge \dd x^1 \wedge \dd x^2\,. 
\end{align}
\end{subequations}
At the asymptotic infinity, as discussed earlier, we are led to the regime of non-relativistic M-theory in membrane Newton-Cartan geometry, which should have direct connections to the BLG and ABJM multiple membrane field theories \cite{Bagger:2007jr, Gustavsson:2007vu, Aharony:2008ug}. Taking a further asymptotic non-relativistic M-theory limit that is not aligned with the near-horizon limit leads to the non-Lorentzian version of the AdS${}_4$/CFT${}_3$ correspondence proposed in \cite{Lambert:2024uue}.

Depending on which spatial directions in non-relativistic M-theory are compactified, we will be led to different BPS decoupling limits of 10-dimensional type II string theory (using additional T-dualities to arrive at type IIB).
We can classify various scenarios of this sort below:
\begin{equation*}
\begin{tabular}{c|c|ccc|ccccc}
\textbf{non-relativistic M-theory} & \emph{codim.} & $t$ & $x^1$ & $x^2$ & $x^3$ & $x^4$ & $x^5$ & $\cdots$ & $x^{10}$ \\
    \hline%
matrix 1-brane theory & \emph{2} & -- & -- & $\times$ & $\otimes$ & -- & -- & $\cdots$ & -- \\
matrix 2-brane theory & \emph{3} & -- & -- & -- & $\otimes$ & -- & -- & $\cdots$ & -- \\
matrix 3-brane theory & \emph{4} & -- & -- & -- & $\otimes$ & $\times$ & -- & $\cdots$ & -- \\
    \hline%
IIB \,DLCQ string theory & \emph{0} & -- & $\times$ & $\otimes$ & -- & -- & -- & $\cdots$ & -- \\
    \hline%
IIA non-relativistic string & \emph{2} & -- & -- & $\otimes$ & -- & -- & -- & $\cdots$ & -- \\
IIB non-relativistic string & \emph{2} & -- & -- & $\otimes$ & $\times$ & -- & -- & $\cdots$ & --
\end{tabular}
\end{equation*}
Here, `$\otimes$' means that the relevant compactification is treated as the M-theory circle, while `$\times$' refers to the direction that one T-dualises to arrive at a type IIB theory, and `\emph{codim.}' signifies the codimension of the associated Newton-Cartan foliation. 
We will discuss non-relativistic string theory later in Section~\ref{sec:matrixstringtheory}, as well as its relationship to matrix string theory (or matrix 1-brane theory).

We can apply a further U-duality acting on three transverse directions of non-relativistic M-theory to obtain its magnetic dual, which is a BPS decoupling limit associated with the M5-brane.
This is defined by:
\begin{subequations} \label{eq:nrmtFivedef}
\begin{align}
	\dd s^2_{11} & = \omega^{\frac{2}{3}} \, \Bigl[ - \dd t^2 + \bigl( \dd x^1 \bigr)^2 + \cdots \bigl( \dd x^5 \bigr)^2 \Bigr] + \omega^{- \frac{4}{3}} \, \Bigl[ \bigl( \dd x^6 \bigr)^2 + \cdots + \bigl( \dd x^{10} \bigr)^2 \Bigr]\,, \\[4pt]
	A^{(6)} & =  \omega^2 \, \dd t \wedge \dd x^1 \wedge \cdots \wedge \dd x^5\,,
\end{align}
\end{subequations}
using $A^{(6)}$ the electromagnetic dual of the eleven-dimensional three-form.
An analogous (and again, non-exhaustive) table of reductions can be made for this limit as below: 
\begin{equation*}
\begin{tabular}{c|c|cccc|ccc}
\textbf{non-rel.~M-theory (mag.~dual)} & \emph{codim.} & $t$ & $x^1 \cdots x^3$ & $x^4$ & $x^5$ & $x^6$ & $x^7$ & $x^8 \cdots x^{10}$ \\
    \hline%
matrix 3-brane theory & \emph{4} & -- & --\,\,$\cdots$\,\,-- & $\times$ & $\otimes$ & -- & -- & --\,\,$\cdots$\,\,--  \\
matrix 4-brane theory & \emph{5} & -- & --\,\,$\cdots$\,\,-- & -- & $\otimes$ & -- & -- & --\,\,$\cdots$\,\,-- \\
matrix 5-brane theory & \emph{6} & -- & --\,\,$\cdots$\,\,-- & -- & $\otimes$ & $\times$ & -- & --\,\,$\cdots$\,\,-- \\
    \hline%
S-dual of M$5$T & \emph{6} & -- & --\,\,$\cdots$\,\,-- & -- & $\times$ & $\otimes$ & -- & --\,\,$\cdots$\,\,-- 
\end{tabular}
\end{equation*}
Here, in the S-dual of M5T, a BPS decoupling limit is taken associated with a background NS5-brane. This is the magnetic dual of non-relativistic string theory, which we will not discuss in this paper. 
Again, the limiting prescription~\eqref{eq:nrmtFivedef} can be applied to the M5-brane geometry to generate the near-horizon limit, now leading to the M5-brane AdS${}_7$/CFT${}_6$ correspondence~\cite{Maldacena:1997re}.

Finally, for completeness, we add a summary for the string theory limits obtained by direct reduction of DLCQ M-theory:\footnote{Matrix $(-1)$-brane theory (M$(-1)$T) is formally T-dual to M0T in the presence of a timelike compactification. The fundamental string in M$(-1)$T is identical to the one in tensionless string theory. Further spacelike T-duality of M$(-1)$T leads to various Carrollian string theories. See~\cite{Blair:2023noj, Gomis:2023eav}.}
\begin{equation*}
\begin{tabular}{c|c|cc|ccccc}
\textbf{DLCQ M-theory} & \emph{codim.} & $x^+$ & $x^-$ & $x^2$ & $x^3$ & $x^4$ & $\cdots$ & $x^{10}$ \\
    \hline%
matrix 0-brane theory & \emph{1} & -- & $\otimes$ & -- & -- & -- & $\cdots$ & -- \\
matrix 1-brane theory & \emph{2} & -- & $\otimes$ & $\times$ & -- & -- & $\cdots$ & -- \\
    \hline%
matrix (-1)-brane theory & \emph{0} & $\times$ & $\otimes$ & -- & -- & -- & $\cdots$ & -- \\
    \hline%
IIA \,DLCQ string theory & \emph{0} & -- & -- & $\otimes$ & -- & -- & $\cdots$ & -- \\
IIB \,DLCQ string theory & \emph{0} & -- & -- & $\otimes$ & $\times$ & -- & $\cdots$ & -- \\
    \hline%
IIB non-relativistic string & \emph{2} & -- & $\times$ & $\otimes$ & -- & -- & $\cdots$ & -- 
\end{tabular}
\end{equation*}

\subsection{Non-Lorentzian Holography from Further DLCQs}
\label{sec:furtherdlcqs}

\noindent $\bullet$~\emph{The second DLCQ.} We are now ready to move on to the next DLCQ orbit. This second DLCQ is made possible as non-relativistic M-theory contains a three-dimensional longitudinal sector that is Minkowskian. Within this longitudinal sector, a second null circle can be formed, which leads to non-relativistic M-theory in the DLCQ. 
Similarly to the DLCQ of relativistic M-theory seen in Section~\ref{sec:mptmtd}, the null circle in the DLCQ of non-relativistic M-theory can be thought of as an infinite-boost limit of a longitudinal spatial circle. Denote the Lorentz factor associated with this infinite boost as $\tilde{\omega}$\,, then the line element in the defining prescription~\eqref{eq:nrmtdef} for non-relativistic M-theory becomes (see also Eq.~\eqref{eq:edle})
\begin{align} \label{eq:dlcqnrmtdef}
\begin{split}
	\dd s^2_{11} = \omega^{\frac{4}{3}} \, \Bigl[ \tilde{\omega}^{-2} \, \bigl( \dd x^- + \tilde{\omega}^2 \, \dd x^+ \bigr)^2 - \tilde{\omega}^2 \, \bigl( \dd x^+ \bigr)^2 + \bigl( \dd x^2 \bigr)^2 \Bigr] & \\[4pt]
    + \omega^{- \frac{2}{3}} \, \Bigl[ \bigl( \dd x^3 \bigr)^2 + \cdots + \bigl( \dd x^{10} \bigr)^2 \Bigr] & \,,
\end{split}
\end{align}
for the `almost' lightlike coordinates $x^\pm$ defined analogously as in Eq.~\eqref{almostlccoords}. Just as the null circle in DLCQ M-theory is mapped to a spatial circle in non-relativistic M-theory via a U-duality transformation, the null circle in the DLCQ of non-relativistic M-theory also maps to a spatial circle after a U-duality transformation is applied. 
To see this, we further compactify two of the transverse directions, say $x^3$ and $x^4$\,. U-dualising in $x^-, \, x^3$ and $x^4$\,, we find that, measured in the new Planck length, 
\begin{subequations} \label{eq:MMTdef}
\begin{align} 
	\dd s^2_{11} & = - \bigl( \omega \, \tilde{\omega} \bigr)^{\frac{4}{3}} \, \dd t^2 + \lr \frac{\omega^2}{\tilde{\omega}} \rr^{\!\!\frac{2}{3}} \, \Bigl[ \bigl( \dd x^1 \bigr)^2 + \bigl( \dd x^2 \bigr)^2 \Bigr] \notag \\[4pt]
        & \hspace{2.55cm} + \lr \frac{\tilde{\omega}^2}{\omega} \rr^{\!\!\frac{2}{3}} \, \Bigl[ \bigl( \dd x^3 \bigr)^2 + \bigl( \dd x^4 \bigr)^2 \Bigr]
    + \frac{\bigl( \dd x^5 \bigr)^2 + \cdots + \bigl( \dd x^{10} \bigr)^2}{\bigl( \omega \, \tilde{\omega} \bigr)^{2/3}}\,, \\[4pt] 
    A^{(3)} & = \omega^2 \, \dd t \wedge \dd x^1 \wedge \dd x^2 +\tilde{\omega}^2 \, \dd t \wedge \dd x^3 \wedge \dd x^4\,,
\end{align}
\end{subequations}
where $t = x^+$\,.
The U-dual of the DLCQ of non-relativistic M-theory then arises from the double-scaling limit where $\omega$ and $\tilde{\omega}$ are sent to infinity simultaneously. This is a double BPS decoupling limit that zooms in on two orthogonal background M2-branes, both of which are coupled to critical divergent gauge fields. One of the M2-branes is associated with the parameter $\omega$ and longitudinal to $t\,, \, x^1\,, \, x^2$\,, while the other with parameter $\tilde{\omega}$ and longitudinal to $t\,, \, x^3\,, \, x^4$\,. The resulting M-theory is referred to as \emph{multicritical M-theory} in~\cite{Blair:2023noj}, and the light excitations are the orthogonal M2-brane states, which are $\frac{1}{4}$-BPS. 

\vspace{3mm}

\noindent $\bullet$~\emph{Multicritical M-theory.} In order to make connection with the double BPS decoupling limits considered in Sections~\ref{sec:nlbdal} and \ref{sec:morebranegeos}, we compactify the multicritical M-theory prescription~\eqref{eq:MMTdef} along $x^{10}$ to find a multicritical IIA string theory. Denote the radius of the compactification along $x^{10}$ as $R$\,, then we have $R = g^{}_\text{s} \, \ell^{}_\text{s}$\,, with $g^{}_\text{s}$ the string coupling and $\ell^{}_\text{s}$ the string length. Dimensionally reducing gives the following ten-dimensional IIA background: \begin{subequations} \label{eq:mdtdt}
\begin{align} 
	\dd s^2_{10} & = - \omega \, \tilde{\omega} \, \dd t^2 + \frac{\omega}{\tilde{\omega}} \, \Bigl[ \bigl( \dd x^1 \bigr)^2 + \bigl( \dd x^2 \bigr)^2 \Bigr] \notag \\[4pt]
    & \hspace{1.98cm} + \frac{\tilde{\omega}}{\omega} \, \Bigl[ \bigl( \dd x^3 \bigr)^2 + \bigl( \dd x^4 \bigr)^2 \Bigr]  + \bigl( \omega \, \tilde{\omega} \bigr)^{-1} \, \Bigl[ \bigl( \dd x^5 \bigr)^2 + \cdots + \bigl( \dd x^{9} \bigr)^2 \Bigr]\,, \\[4pt] 
    C^{(3)} & = g^{-1}_\text{s} \, \dd t \wedge \Bigl( \omega^2 \, \dd x^1 \wedge \dd x^2 + \tilde{\omega}^2 \, \dd x^3 \wedge \dd x^4 \Bigr)\,,
        \qquad%
    e^\Phi = \bigl( \omega \, \tilde{\omega} \bigr)^{- \frac{1}{2}} \, g^{}_\text{s}\,.
\end{align}
\end{subequations}
This is precisely applying the M2T limiting prescription~\eqref{eq:rpgcfbmpt} (with $p=2$) twice, once with the parameter $\omega$ and once with $\tilde{\omega}$\,. Physically, this is the double BPS decoupling limit where we zoom in on two orthogonal background D2-branes in type IIA. We depict this background D2-brane configuration below:
\begin{center}
\begin{tabular}{c|c|c|c|c}
& $t$ & \,$x^1\,, \, x^2$\, & \,$x^3\,, \, x^4$\, & $x^5\,, \, \cdots, \, x^9$ \\[2pt]
\hline 
D2 & $\times$ & $\times$ & -- & -- \\[2pt]
D2 & $\times$ & -- & $\times$ & -- 
\end{tabular}
\end{center}
T-dualising this multicritical IIA theory generates all the corners associated with various double BPS decoupling limits at the asymptotic infinity as discussed in Sections~\ref{sec:nlbdal} and~\ref{sec:morebranegeos}.

If one instead compactifies multicritical M-theory along a direction that is longitudinal to one of the background M2-branes, it then leads to the multicritical matrix 2-brane theory (MM2T) introduced in~\cite{Blair:2023noj,Gomis:2023eav}, where one zooms in on a background F1-D2 configuration. T-dualising MM2T gives rise to more general MM$p$Ts as discussed in~\cite{Blair:2023noj,Gomis:2023eav}, which arise from the double BPS decoupling limits zooming on the F1-D$p$ configurations. For the IIB cases where $p$ is odd, an S-duality maps an MM$p$T to the multicritical string theory where D1-D$p$ is critical. Finally, T-duals of such a critical D1-D$p$ scenario leads to the critical D$p$-D$q$ theories considered in Sections~\ref{sec:nlbdal} and~\ref{sec:morebranegeos}. We refer to the duality orbit of such limits as being the DLCQ${}^2$ of string/M-theory (see also~\cite{bpslimits}).

Now, we have reached an intriguing observation: in the context of holography, the double BPS decoupling limits at the asymptotic infinity that we discussed in Section~\ref{sec:hnhbpsl} are associated with multicritical M-theory, which lives in the DLCQ${}^2$ orbit of the full M-theory; on the other hand, the dual bulk M$p$T geometry induced by an asymptotic double BPS decoupling limit lives in the DLCQ${}^1$ orbit, as we have explained in Section~\ref{sec:adsdlcq}. The proposed holographic duals relevant to the non-Lorentzian bulk geometries introduced in Section~\ref{sec:hnhbpsl} are examples of the DLCQ${}^1$/DLCQ${}^2$ correspondence, where the extra DLCQ limit at the asymptotic infinity corresponds to the near-horizon limit in the bulk. 

\vspace{3mm}

\noindent $\bullet$~\emph{Field theoretical aspects of DLCQ${}^{\,1}$/DLCQ${}^{\,2}$.} 
We can interpret recent proposals for non-Lorentzian holography in both string and M-theory \cite{Lambert:2024uue, Fontanella:2024rvn, Lambert:2024yjk, Fontanella:2024kyl} as fully non-Lorentzian realisations of a DLCQ${}^{\,1}$/DLCQ${}^{\,2}$ correspondence.
Let us summarise in more detail the examples studied in \cite{Lambert:2024yjk} which involve a web of non-relativistic field theories related to D-branes \cite{Lambert:2018lgt,Lambert:2019nti}. As we noted in Section~\ref{sec:sln}, the possible D$q$-brane solutions in M$p$T encapsulated in Table~\ref{tab:IIbranes} correspond to the T-duality web highlighted in~\cite{Lambert:2024yjk} (see their Figure~1).
The perspective advocated there is that the near-horizon limits of the resulting M$p$T geometries are dual to non-relativistic limits of the low-energy effective action of the D$q$-brane worldvolume theory.
The resulting field theory dynamics depends on the coordinates corresponding to the intersection of the longitudinal directions of the M$p$T limit and the longitudinal directions of the D$q$-brane worldvolume, and localises on the moduli space of BPS solutions in the directions orthogonal to this intersection. The non-relativistic limit applied to the worldvolume is inherited from the M$p$T limit in spacetime restricted to the worldvolume directions (up to conformal rescaling).

From our perspective, the M$p$T holographic correspondences of~\cite{Lambert:2024yjk} are naturally viewed to arise from two decoupling limits.
We have to take the near-horizon limit of the D$q$-brane as well as the asymptotic M$p$T limit.
The former can also be viewed as a longitudinal M$q$T decoupling limit: its effect is to reduce the D$q$-worldvolume theory to its low energy SYM form, to which the M$p$T limit is then further applied to obtain a non-relativistic field theory. In the unified language of DLCQ, this realises the DLCQ${}^2$ at the asymptotic infinity and the DLCQ${}^1$ in the bulk. It has been argued in \cite{Fontanella:2024rvn} that the decoupling and near-horizon limits commute for the case where the former is the non-relativistic string limit (see Section~\ref{sec:matrixstringtheory}) and the latter (in our parlance) is the M3T limit. It would be nice to verify this more generally. 

Given the logic that string theory on the near-horizon of the D3-brane is dual to the wordvolume $\CN = 4$ SYM theory, it is natural to suggest that the bulk theory which is dual to the non-relativistic field theories appearing in~\cite{Lambert:2024yjk} is the M$p$T decoupling limit of string theory.
Moreover, in view of the connection to matrix theory, the bulk closed string theory associated with DLCQ${}^2$ should be described \emph{not} using strings but D$p$-branes, whose dynamics are given by the (supersymmetric extension) of the action \eqref{eq:mptdpnona} specified to the relevant non-Lorentzian near-horizon geometries.

The proposed DLCQ${}^1$/DLCQ${}^2$ correspondence is also reminiscent of the classic example of the BMN limit \cite{Berenstein:2002jq} of $\CN = 4$ SYM, where the bulk geometry is related to the Penrose limit of the AdS geometry, \emph{i.e.} the pp-wave geometry with a null isometry. It would be interesting to explore this intuition further.

\vspace{3mm}

\noindent $\bullet$~\emph{A DLCQ${}^{\,2}$/DLCQ${}^{\,3}$ correspondence?} 
Having established how our perspective classifies Lorentzian and non-Lorentzian versions of AdS/CFT, it is natural to then attempt to continue the  DLCQ${}^{\,n}$/DLCQ${}^{\,n+1}$ pattern to $n>1$. 

For $n=2$, a possible example is provided by Spin Matrix Theory (SMT)~\cite{Harmark:2014mpa,Harmark:2019zkn,Baiguera:2022pll} (see \cite{Harmark:2017rpg,Harmark:2018cdl, Harmark:2019upf, Harmark:2020vll,Roychowdhury:2020yun,Roychowdhury:2021wte,Bidussi:2021ujm, Bidussi:2023rfs} for a string worldsheet perspective on SMT).
This is obtained by taking further limits of $\mathcal{N}=4$ SYM dual to type IIB on AdS${}_5 \times S^5$. SMT arises from a class of special BPS decoupling limits of $\CN = 4$ SYM that involve making  R-charges critical. Concretely, the most interesting SMT limit involves first performing a null reduction of $\CN = 4$ SYM followed by identifying the (fixed) null momentum with an R-charge. This leads to the PSU($1,2|3$)-symmetric SMT and it can be argued that this theory lives in the DLCQ${}^3$ orbit: first, we have discussed that $\CN = 4$ SYM lives in the dual orbit of a single DLCQ of M-theory; second, the null reduction is clearly related to a second DLCQ; finally, the unconventional further procedure of making the R-charge critical is may be associated with a third DLCQ. This third DLCQ can be performed by first taking a magnetic dual of multicritical M-theory defined by the prescription~\eqref{eq:mdtdt} followed by an infinite boost limiting that leads to a lightlike compactification. On the bulk side, Spin Matrix Theory (SMT) corresponds to closed strings on a bulk geometry that arises from the DLCQ${}^2$ orbit, which is demonstrated using a probe fundamental string in~\cite{Blair:2023noj, Gomis:2023eav}. 

For $n=3$ and beyond, it is an open question whether there are further examples. 

\subsection{Generalisation to \texorpdfstring{DLCQ${}^n$/DLCQ${}^m$}{DLCQn/DLCQm} Correspondence} \label{sec:dlcqmn}

Through Sections~\ref{sec:hnhbpsl} and~\ref{sec:chdlcq}, we have focused on bulk geometries sourced by a single stack of branes, involving a single harmonic function to which the near-horizon limit is applied. This is the configuration that underlies the DLCQ${}^n$/DLCQ${}^{n+1}$ correspondence. However, it is also natural to consider intersecting brane geometries, which in fact lead us to a more general DLCQ${}^n$/DLCQ${}^{m}$ correspondence, where $m > n$\,. 

As a proof of concept, we focus here on the example of the D1-D5 geometry and its associated AdS${}_3$/CFT${}_2$ correspondence~\cite{Maldacena:1997re}. Based on the brane configuration,
\begin{center}
\begin{tabular}{c|c|c|c}
& $X^0\,,\, X^1$  & $X^2 \cdots X^{5}$ & $X^6 \cdots X^9$ \\[2pt]
\hline 
D1 & $\times$ & -- & -- \\[2pt]
D5 & $\times$ & $\times$ & -- \\[2pt]
\hline
index & $a$ & $I$ & $i$
\end{tabular}
\end{center}
we consider the following D1-D5 intersecting brane solution to IIB supergravity:
\begin{subequations} \label{eq:d1d5}
\begin{align}
    \dd s^2 & = \frac{\dd X^a \, \dd X_a}{\sqrt{H_1 \, H_5}} + \sqrt{\frac{H_1}{H_5}} \, \dd X^I \, \dd X^I + \sqrt{H_1 \, H_5} \, \dd X^i \, \dd X^i \,,
        \qquad%
    e^\Phi = \sqrt{\frac{H_1}{H_5}} \, G^{}_\text{s}\,, \\[4pt]
    C^{(2)} & = \frac{1}{G^{}_\text{s} \, H_1} \, \dd X^0 \wedge \dd X^1 \,,
        \hspace{3cm}%
    C^{(6)} = \frac{1}{G^{}_\text{s} \, H_5} \, \dd X^0 \wedge \cdots \wedge \dd X^5 \,,
\end{align}
\end{subequations}
where $a=0\,,\,1$\,, $I=2\,,\,3\,,\,4\,,\,5$ and $i=6\,,\,7\,,\,8\,,\,9$\,. The harmonic functions take the form
\begin{subequations}
\begin{align}
    H_1 &= 1 + \frac{L_1^2}{\|X^i\|^2} \,,  
        &%
    L_1^2 &= \frac{(2\pi)^4 \, N^{}_1 \, G^{}_\text{s} \, \ell_\text{s}^{\,6}}{V^{}_4}\,, \\[4pt]
    H_5 &= 1 + \frac{L_5^2}{\|X^i\|^2} \,,
        &%
    L_5^2 &= N_5 \, G_\text{s} \, \ell_\text{s}^2\,.
\end{align}
\end{subequations}
Here, $N_1$ and $N_5$ are the numbers of D1- and D5-branes. Moreover, $V_4$ is the volume of the four-dimensional internal compact manifold which is wrapped by the D5-branes, which here for simplicity we take to be a torus with coordinates $X^I$\,.
It can easily be checked that taking the asymptotic M$1$T limit replaces $H_1$ with its near-horizon form (\emph{i.e.}~dropping 1) and leaves $H_5$ unchanged, while taking the asymptotic M5T limit has the opposite effect. 

We now consider a multicritical D1-D5 limit at the asymptotic infinity, akin to the multicritical D2-D2 limit defined via Eq.~\eqref{eq:mdtdt}. The defining prescriptions for the D1-D5 limit are
\be \label{eq:d1d5p}
    X^a = \sqrt{\omega \, \tilde{\omega}} \, x^a \,,
        \qquad%
    X^I = \sqrt{\frac{\tilde{\omega}}{\omega}} \, x^I \,,
        \qquad%
    X^i = \frac{x^i}{\sqrt{\omega \, \tilde{\omega}}}\,, 
        \qquad
    G_\text{s} = \frac{\tilde{\omega}}{\omega} \, g_\text{s} \,,
\ee
together with the critical RR $2$-form and $6$-form.
Here, $\omega$ parametrises the asymptotic M1T limit longitudinal to the bulk D1-brane and $\tilde{\omega}$ parametrises the asymptotic M5T limit longitudinal to the bulk D5-brane. It then follows that the characteristic lengths $L_1$ and $L_5$ scale as $L_1^2 = \omega \, \tilde{\omega}^{-1} \, \ell_1^2$ and $L_5^2 = \omega^{-1} \, \tilde{\omega} \, \ell_5^2$\,. Plugging Eq.~\eqref{eq:d1d5p} into the original D1-D5 geometry~\eqref{eq:d1d5} and then sending both $\omega$ and $\tilde{\omega}$ to infinity, we are led to the AdS${}_3\times S^3 \times M_4$ geometry
\begin{subequations} \label{eq:adstg}
\begin{align}
    \dd s^2 & = \frac{r^2}{\ell^2} \, \dd x^a \, \dd x_a + \frac{\ell^2}{r^2} \bigl( \dd r^2 + r^2 \, \dd \Omega_3^2 \bigr) + \frac{\ell_1}{\ell_5} \, \dd x^I \, \dd x^I\,,
        \qquad%
    e^\Phi = \frac{\ell_1}{\ell_5} \, g^{}_\text{s}\,, \\
    C^{(2)} & = \frac{r^2}{g^{}_\text{s} \, \ell_1^2} \, \dd x^0 \wedge \dd x^1 \,,
        \hspace{2.85cm}%
    C^{(6)} = \frac{r^2}{g^{}_\text{s} \, \ell_5^2} \, \dd x^0 \wedge \cdots \wedge \dd x^5 \,.
\end{align}
\end{subequations}
where $r \equiv \|x^i\|$ and we defined the AdS${}_3$ length $\ell = \sqrt{\ell_1 \, \ell_5}$\,. In the associated AdS${}_3$/CFT${}_2$ correspondence, the AdS${}_3\times S^3 \times M_4$ geometry~\eqref{eq:adstg} is dual to a two-dimensional CFT with (4\,,\,4) supersymmetry living at the D1-D5 intersection, which describes the light excitations of this multicritical D1-D5 limit of type IIB superstring theory. Note that, when $\tilde{\omega} = \omega$\,, our multicritical limit can be identified with the original Maldacena limit in~\cite{Maldacena:1997re} up to rescalings of the coordinates and the Regge slope $\alpha'$\,.  

Note that if we considered a configuration in which one brane is totally smeared, so that its corresponding harmonic function is just $H=1$, then the above procedure reduces to an M$p$T limit of D$q$-brane of the sort we considered before.
This totally smeared brane intersection interpretation of the limit is used in \cite{Lambert:2024yjk}. 

In the above example of the AdS${}_3$/CFT${}_2$ correspondence, the bulk geometry is Lorentzian and therefore lives in the DLCQ${}^0$ orbit. However, since the bulk AdS${}_3$ geometry arises from a double near-horizon limit, we are led to a double BPS decoupling limit at the asymptotic infinity. As we have discussed earlier, each of these BPS decoupling limits is associated with a DLCQ of M-theory. Therefore, the non-Lorentzian geometry at the asymptotic infinity lives in the DLCQ${}^2$ duality orbit, ensuing that AdS${}_3$/CFT${}_2$ should be viewed as a DLCQ${}^0$/DLCQ${}^2$ correspondence. 

It is also possible to consider more involved asymptotic BPS decoupling limits as in Sections~\ref{sec:nlbdal} and \ref{sec:morebranegeos}, which will lead to non-Lorentzian bulk geometries associated with M-theory in a single or multiple DLCQs. 
We could further extend the discussion here to situations with three or more brane charges.
These possibilities lead us to conjecture a DLCQ${}^n$/DLCQ${}^m$ correspondence with $m > n$\,. The extra $m-n$ DLCQs are performed at the asymptotic infinity, which correspond to a multiple `near-horizon' limit in the bulk. Each near-horizon limit essentially drops a 1 in one of the $m-n$ harmonic functions that appear in a supergravity solution describing multiple intersecting branes.

\section{\texorpdfstring{$T\bar{T}$}{TTbar} Deformation: Generating the Bulk Geometry} 
\label{sec:TTbar}

In Sections~\ref{sec:hnhbpsl} and~\ref{sec:chdlcq}, we have shown that the near-horizon limits can be generated by using asymptotic BPS decoupling limits. 
In this section, we focus on these near-horizon geometries, and discuss how they can be viewed \emph{intrinsically} as a particular geometrically-realised deformation of an asymptotic non-Lorentzian geometry.

Recall that, in Section~\ref{sec:badsmt}, we discussed the usual near-horizon limits of D$p$-brane solutions, realising these limits via an asymptotic M$p$T prescription.
It is manifest that the geometry at the asymptotic infinity of the near-horizon bulk is of the M$p$T type, which is non-Lorentzian. 
We can see this from the expression \eqref{eq:dsco} for the geometry in the near-horizon limit, which involved the metric
\begin{align}
\label{eq:dscoagain}
    \dd s^2 & = \mathbb{\Omega}(r) \, \Bigl( - \dd t^2 + \dd x^i \, \dd x^i \Bigr) + \frac{\dd r^2 + r^2 \, \dd\Omega^2_{8-p}}{\mathbb{\Omega}(r)}\,, 
        &%
    \mathbb{\Omega}(r) & = \left( \frac{r}{\ell} \right)^{\frac{7-p}{2}}\,.
\end{align} 
In the asymptotic infinite regime where the scale $\ell$ is small compared to the radial direction $r$\,, the factor $\mathbb{\Omega} (r)$ in the near-horizon geometry~\eqref{eq:dscoagain} is large. 
Using the dilatation symmetry~\eqref{eq:dilatsmpt}, we can redefine the constant parameter $\omega$ in the M$p$T prescription~\eqref{eq:rpgcfbmpt} to be $\mathbb{\Omega}(r)$\,, such that the infinite $\mathbb{\Omega}$ limit of the near-horizon geometry is of the M$p$T type. 
As M$p$T arises from a BPS decoupling limit, one expects that the bulk geometry can be generated by deforming the asymptotic geometry by the `inverse' of this limit procedure, controlled by the radial-dependent parameter $\mathbb{\Omega}(r)$\,. 

In this section, we will demonstrate that such an inverse BPS decoupling limit is the $p$-brane version of the $T\bar{T}$ deformation, generalised to arbitrary dimension.

To show this, we need to build on an observation of \cite{Blair:2020ops}, which showed that the BPS decoupling limit defining non-relativistic string theory could be viewed in reverse as the $T \bar T$ deformation of a two-dimensional field theory. 
We will review and motivate this statement below, in the case of a constant deformation parameter. 
The geometric extension to the coordinate-dependent $\mathbb{\Omega}(r)$ then follows.

To establish the link to non-relativistic string theory in more detail, we will first focus on the S-dual picture of M1T. 
There, if the gauge potentials are set to zero, the $T\bar{T}$ deformation will be the standard 
one in two dimensions. The dynamics of M1T is, on the S-dual side, described by matrix string theory.
We will review this and clarify the relation to type IIB non-relativistic string theory: the latter is a perturbative string theory, and the deformation of its string worldsheet action toward the usual string sigma model is precisely the $T\bar{T}$ deformation~\cite{Blair:2020ops}. After reviewing these foundational aspects, and giving an improved understanding of their relationships in our current framework, we will finally discuss how the bulk geometry is generated via a $p$-brane $T\bar{T}$ deformation.
In this section we will define the latter using formal duality arguments, as we know how to map between non-relativistic string theory and other BPS decoupling limits.
In the following Section \ref{sec:pbttfe} we will explicitly derive and present the relevant generalised $T \bar T$ deformations.

\subsection{Non-Relativistic and Matrix String Theory}
\label{sec:matrixstringtheory}

We now start with M1T and a review of matrix and non-relativistic string theory, which plays a fundamental role in the $T\bar{T}$ story. 

In M1T, the light excitations are the D1-strings, whose dynamics is described by $\CN = 4$ SYM on a torus, which can be equivalently viewed as BFSS matrix theory on a circle transverse to the D0-branes, followed by a T-duality transformation along the circle. Matrix theory now becomes a two-dimensional $\CN = 8$ SYM with $U(N)$ gauge symmetry, whose bosonic sector can be obtained from Eq.~\eqref{eq:mptdpnona} with $p = 1$\,,
\be \label{eq:mptdpnonat}
\begin{split} 
S^\text{M$p$T}_{\text{D}1} = - \frac{1}{2} \int \dd^2 \sigma \, \text{STr} \biggl[
  e^{-\varphi} \sqrt{- \tau} \, \Bigl( 
   \tau^{\wa\wb} \, {\text{P}} \bigl[ E_{\wa\wb} \bigr]
  & + \tfrac{1}{2} \, \tau^{\wa\wb} \, \tau^{\wc\wvd} \, F_{\wa\wc} \, F_{\wb\wvd} 
   \\ & - \tfrac{1}{2} \, \bigl[ X^i, X^k \bigl] \bigl[ X^j, X^l \bigr] \, E^{}_{ij} \, E^{}_{kl}
  \Bigr)
 \biggr]\,. 
\end{split}
\ee
For simplicity, we have set the $B$-field and RR potentials to zero, and let $2\pi\alpha' = 1$\,. S-dualising the $N$ D1-strings in M1T leads to \emph{matrix string theory}, whose large $N$ limit describes the second quantised IIA strings~\cite{Motl:1997th, Dijkgraaf:1997vv}.

Under S-duality, M1T is mapped to a perturbative string theory that arises from a decoupling limit zooming in on a background fundamental string in type IIB. This decoupling limit leads to what is known in the literature as \emph{non-relativistic string theory}~\cite{Klebanov:2000pp, Gomis:2000bd, Danielsson:2000gi, Danielsson:2000mu}. See~\cite{Blair:2023noj, Ebert:2021mfu, Ebert:2023hba} for recent studies of this S-duality between M1T and non-relativistic string theory and \cite{Bergshoeff:2022iss,Bergshoeff:2023ogz} for its SL(2\,, $\mathbb{Z}$) generalisation.
We will now explain how matrix string theory provides a second-quantisation of non-relativistic string theory~\cite{Danielsson:2000gi}. We elaborate on these ingredients below, starting with non-relativistic string theory. 

\vspace{3mm}

\noindent $\bullet$~\emph{First quantisation: non-relativistic string theory.}
The S-duality of M1T is inherited from the S-duality in type IIB superstring theory.
This is a discrete symmetry acting on the metric $G^{}_{\mu\nu}$\,, dilaton field $\Phi$\,, and RR potential $C^{(2)}$ in the string frame as
\be \label{eq:sdual}
    \tilde{G}^{}_{\mu\nu} = e^{-\Phi} \, G^{}_{\mu\nu}\,,
        \qquad%
    e^{\tilde{\Phi}} = e^{-\Phi}\,,
        \qquad%
    \tilde{B}^{(2)} = - C^{(2)}\,.
\ee
Here, $\tilde{G}^{}_{\mu\nu}$ is the metric field, $\tilde{\Phi}$ the dilaton, and $\tilde{B}^{(2)}$ the Kalb-Ramond field in the S-dual frame. We have not included any Kalb-Ramond field in the original theory as the M1T limit does not involve any divergence in $B^{(2)}$. However, it is necessary to introduce $C^{(2)}$, which appears as the critical gauge potential in M1T. We have also set all the remaining RR potentials to zero for simplicity. Using the M1T prescription~\eqref{eq:rpgcfbmppot} with $p = 1$\,, we write
\be
    G^{}_{\mu\nu} = \omega \, \tau^{}_{\mu\nu} + \omega^{-1} \, E^{}_{\mu\nu}\,,
        \qquad%
    e^\Phi = \omega^{-1} \, e^\varphi\,,
        \qquad%
    C^{(2)} = \omega^2 \, e^{-\varphi} \, \tau^0 \wedge \tau^1\,.
\ee
Then, the S-dual map~\eqref{eq:sdual} implies that
\be \label{eq:tgtptb}
    \tilde{G}^{}_{\mu\nu} = \omega^2 \, \tilde{\tau}^{}_{\mu\nu} + \tilde{E}^{}_{\mu\nu}\,,
        \qquad%
    e^{\tilde{\Phi}} = \omega \, e^{\tilde{\varphi}}\,,
        \qquad%
    \tilde{B}^{(2)} = - \omega^2 \, \tilde{\tau}^0 \wedge \tilde{\tau}^1\,,
\ee
where
\be \label{eq:tttesd}
    \tilde{\tau}^A = e^{-\frac{\varphi}{2}} \, \tau^A,
        \qquad%
    \tilde{E}^{A'} = e^{-\frac{\varphi}{2}} \, E^{A'},
        \qquad%
    e^{\tilde{\varphi}} = e^{-\varphi}\,,
\ee
with $A = 0\,, \, 1$ and $A' = 2\,, \, \cdots, 9$\,. In order to understand the dynamics in the S-dual frame of M1T, we apply the prescription~\eqref{eq:tgtptb} to the fundamental string action in the Nambu-Goto formulation,
\be \label{eq:ngfos}
    S^{}_\text{F1} = - \frac{1}{2\pi\alpha'} \int \dd^2 \sigma \, \sqrt{- \det \Bigl[ \p_\alpha X^\mu_{\phantom{\dagger}} \, \p_\beta X^\nu_{\phantom{\dagger}} \, \tilde{G}^{}_{\mu\nu} (X) \Bigr]} - \frac{1}{2\pi\alpha'} \int \tilde{B}^{(2)}\,,
\ee
where $\sigma^\alpha = (\tau\,, \, \sigma)$ are the worldsheet coordinates. 
In the $\omega \rightarrow \infty$ limit, we find~\cite{Andringa:2012uz} 
\be \label{eq:nrstng}
    S^\text{NRST}_\text{F1} = - \frac{1}{4\pi\alpha'} \int \dd^2 \sigma \, \sqrt{-\tilde{\tau}} \, \tilde{\tau}^{\alpha\beta} \, \p_\alpha X^\mu \, \p_\beta X^\nu \, \tilde{E}^{}_{\mu\nu} (X)\,,
        \qquad%
    \tilde{\tau}^{}_{\alpha\beta} = \p_\alpha X^\mu \, \p_\beta X^\nu \, \tilde{\tau}^{}_{\mu\nu}\,.
\ee
Here, $\tilde{\tau}^{\alpha\beta}$ is the inverse of $\tilde{\tau}_{\alpha\beta}$ and $\tilde{\tau} = \det \bigl( \tilde{\tau}_{\alpha\beta} \bigr)$\,. This worldsheet action describes non-relativistic string theory. In flat target space and in static gauge, with $\tilde{\tau}^{}_{\alpha\beta} = \eta^{}_{\alpha\beta}$ and $\tilde{E}^{}_{\mu\nu} = \delta^{A'}_{\mu} \, \delta^{A'}_\nu$\,, this action reduces to
\be \label{eq:nrstngf}
    S^\text{NRST}_\text{F1} = - \frac{1}{4\pi\alpha'} \int \dd^2 \sigma \, \p_\alpha X^{A'} \, \p^\alpha X^{A'},
        \qquad%
    A' = 2\,, \, \cdots, \, 9\,.
\ee
The closed string spectrum is only nontrivial when the longitudinal spatial direction $X^1$ is compactified over a circle of radius $R$\, \cite{Gomis:2000bd, Danielsson:2000gi}. For a string state with winding number $w \in \mathbb{Z}$ and momentum number $n \in \mathbb{Z}$ in $X^1$, with $w\,, \, n \in \mathbb{Z}$\,, the dispersion relation is
\be \label{eq:nrstdr}
    \varepsilon = \frac{1}{2\,wR} \, \Bigl( \tfrac{1}{2} \, \alpha' \, k^{}_{A'} \, k^{}_{A'} +  N + \bar{N} - 2 \Bigr)\,,
        \qquad%
    N - \bar{N} = n \, w\,.
\ee
where $\varepsilon$ is the energy, $k_i$ the transverse momentum, and $(N, \bar{N})$ the string excitation numbers. This dispersion relation is only valid in the case where $w \neq 0$\,. The theory is referred to as `non-relativistic' as the dispersion relation is Galilean invariant at a fixed $w$\,. Moreover, the target space geometry is equipped with a codimension-two foliation structure via string Newton-Cartan geometry \cite{Andringa:2012uz}. There are \emph{no} massless gravitons in this theory; instead, instantaneous gravitational forces are experienced by wound strings \cite{Gomis:2000bd,Danielsson:2000gi,Danielsson:2000mu}, which provide a dual non-relativistic string description of the Newton-like gravitational force between the D0-branes in M0T.  

\vspace{3mm}

\noindent $\bullet$~\emph{Second quantisation: matrix string theory.} 
Next, we move on to matrix string theory~\cite{Motl:1997th, Dijkgraaf:1997vv}. 
The two-dimensional SYM action~\eqref{eq:mptdpnonat} describes a stack of coinciding D1-strings in M1T. Matrix string theory in the S-dual frame then arises from applying the duality transformation~\eqref{eq:tttesd} to Eq.~\eqref{eq:mptdpnonat}. 
In the flat limit where $\tau^{}_\mu{}^A = \delta_\mu^A$ and $E^{}_\mu{}^{A'} = \delta_\mu^{A'}$\,, we also take the string coupling $g^{}_\text{s} = e^\varphi$ to be constant. We take the convention $2\pi\alpha'=1$ here. The bosonic sector of matrix string theory is then described by the action~\cite{Dijkgraaf:1997vv}
\be \label{eq:matrixstring}
\begin{split} 
S^{}_{\text{MST}} & = - \int \dd^2 \sigma \, \tr \Bigl(
  \tfrac{1}{2} \, \p_\alpha X^{i} \, \p^\alpha X^{i}
  + \tfrac{1}{4} \, g^2_\text{s} \, F_{\wa\wb} \, F^{\wa\wb} 
   - \tfrac{1}{4} \, g^{-2}_\text{s} \, \bigl[ X^{i}, X^{j} \bigl]^2
 \Bigr)\,. 
\end{split}
\ee
The Yang-Mills coupling is $g^{}_\text{YM} = g^{-1}_\text{s}$\,. In the deep infrared, the two-dimensional SYM becomes strongly coupled with $g^{}_\text{YM} \rightarrow \infty$\,, which is a relevant coupling with a positive scaling dimension in mass. Correspondingly, the string coupling $g^{}_\text{s}$ goes to zero, which implies that we have free strings. Performing this $g^{}_\text{s} \rightarrow 0$ limit at the level of the matrix string action~\eqref{eq:matrixstring} imposes that $[X^i, X^j] = 0$\,, \emph{i.e.}~$X^i$ 
is restricted to be in the Cartan subalgebra. This results in the $g^{}_\text{s} \rightarrow 0$ limit in a free superconformal field theory, whose bosonic sector is described by $N$ copies of non-relativistic strings in Eq.~\eqref{eq:nrstngf}. 
Subtleties of this somewhat na\"{i}ve argument have been discussed in~\cite{Harvey:1995tg}. For example, it is possible that there is \emph{no} mass gap between the ground and excited states, in which case it is not completely clear how the fields orthogonal to the Cartan subalgebra decouple. However, it is believed that the non-relativistic string sigma model captures at least part of the story. 

The requirement that $X^i$ must be in the Cartan subalgebra is subject to gauge transformations. In general, we write $X^i = V \, x^i \, V^{-1}$\,, where $V \in U(N)$ and $x^i$ is a diagonal matrix. Recall that matrix sting theory ultimately comes from compactifying BFSS matrix theory over a spatial circle. Traversing along this spatial circle may interchange the eigenvalues of $X^i$, which are gauge invariant. Therefore, the eigenvalues $x^i_r (\sigma)\,, \, r = 1\,, \, \cdots, \, N$ are in general multivalued. This leads to twisted sectors where $x^i$ transforms adjointly under the Weyl group. In the case of $U(N)$ gauge group, the associated Weyl group is the symmetric group $S_N$\,. Therefore, the CFT in the deep IR is an $S_N$ orbifold field theory, whose Hilbert space is decomposed into twisted sectors labeled by the conjugacy classes of the orbifold group $S_N$~\cite{Harvey:1995tg, Bershadsky:1995vm}. A particular twisted sector corresponds to a string with certain length. 

From the perspective of non-relativistic string theory, the length $n_i$ of the string in a twisted sector consisting of a finite number of cycles corresponds to the winding number of a non-relativistic string. On the other hand, this $n_i$ corresponds to the number of D0-branes within a particular bound state in M0T. The $S_N$ orbifold CFT at $g^{}_\text{s} = 0$ describes the multi-string states that form a non-interacting superselection sector, where the number of the strings is conserved. 

Interactions can be introduced in matrix string theory by going slightly away from the $g^{}_\text{s} = 0$ Gaussian fixed point. This essentially means that we introduce twist field operators that join or split the strings, at the positions where two eigenvalues of $X^i$ coincide. Therefore, matrix string theory at a finite $g^{}_\text{s}$ describes a second quantisation of non-relativistic string theory. In the large $N$ limit, the total winding number of all the interacting non-relativistic strings under consideration becomes infinite, and the physical meaning of non-relativistic string theory in this limit becomes murky. This difficulty is circumvented by passing to a T-dual frame. It is known that T-dualising the IIB version of non-relativistic superstring theory along the longitudinal $X^1$ circle before performing the $N \rightarrow \infty$ limit defines the DLCQ of the conventional IIA theory, where the winding number is now mapped to the Kaluza-Klein number of a lightlike momentum. The $N \rightarrow \infty$ limit of the IIA theory in the DLCQ gives rise to the complete IIA theory with a decompactified lightlike isometry. Therefore, the large $N$ limit of matrix string theory describes second-quantised strings.

\subsection{Undo the BPS Decoupling Limits: \texorpdfstring{$T\bar{T}$}{TTbar} Deformation}
\label{sec:TTbarNG} 

We have seen in Section~\ref{sec:matrixstringtheory} that non-relativistic string theory~\eqref{eq:nrstngf} in flat spacetime arises from an $\omega \rightarrow \infty$ limit of the conventional string action~\eqref{eq:ngfos}, which is a BPS decoupling limit that zooms in on a background fundamental string. On the other hand, deforming the non-relativistic string action~\eqref{eq:nrstng} by reintroducing the $\omega$-dependence leads us back to the conventional string sigma model. It was shown in~\cite{Blair:2020ops} that this deformation can be viewed as the $T\bar{T}$ deformation.

In the simplest setup, we consider the Nambu-Goto action~\eqref{eq:nrstngf} describing the non-relativistic string flat target space and in a static gauge, with the Regge slope $\alpha'$ set to $(2\pi)^{-1}$\,. Even though we have in mind the superstrings (see \emph{e.g.}~\cite{Gomis:2005pg, Kim:2007pc, Park:2016sbw, Blair:2019qwi}), we will only focus on the bosonic sector as a proof of concept in this section.
Where possible, we keep the spacetime dimension generic, and use the transverse index $A' = 2\,, \, \cdots, \, D-1$\,. The standard $T\bar{T}$ deformation deforms the Lagrangian of \eqref{eq:nrstngf} by an irrelevant operator that is the determinant of the energy-momentum tensor $T_{\alpha\beta}$\,. As a result, we follow a trajectory in the space of field theories parametrised by $\lambdap$.
Each point of this trajectory is associated with a two-dimensional sigma model, whose Lagrangian we denote as $\CL(\lambdap)$\,.
Such a trajectory is defined (in our conventions) via the following flow equation:
\be \label{eq:ttbarfe} 
   \frac{\p \CL(\lambdap)}{\p \lambdap} = \frac{1}{2} \, \det \bigl[ T^{}_{\alpha\beta} (\lambdap) \bigr]\,.
\ee
Solving this flow equation for the non-relativistic string action~\eqref{eq:nrstngf} of $D-2$ free bosons, one is led to the standard result in the $T\bar{T}$ literature ~\cite{Cavaglia:2016oda,Bonelli:2018kik}
\be
    \CL (\lambdap) = \frac{1}{\lambdap} \, \biggl[ 1-\sqrt{1 + \lambdap\, \p^{}_{\alpha} X^{A'} \p^{\alpha} X^{A'} - \lambdap^2 \, \det \bigl( \p^{}_\alpha X^{A'} \p^{}_\beta X^{A'} \bigr)}  \biggr]\,.
\ee
This is the known result that the $T\bar{T}$-deformed theory of $D-2$ free bosons is the Nambu-Goto string action in a static gauge for a $D$-dimensional target space, in the presence of a $B$-field proportional to $\lambdap^{-1}$\,. Note the original papers~\cite{Cavaglia:2016oda,Bonelli:2018kik} used Euclidean signature in two dimensions; here we adhere to Lorentzian signature (loosely) following ~\cite{Blair:2020ops}.

One remarkable property of the $T\bar{T}$ deformation is that  the deformed spectrum can be derived exactly from the undeformed one, if the latter is known. In two dimensions, a CFT with conformal dimension $\Delta$ and spin $s$ on a cylinder of radius $\CR$ has energy
\be \label{eq:cftsp}
    \varepsilon \bigl( 0\,, \, \CR \bigr) = \frac{2\pi}{\CR} \left( \Delta - \frac{c}{12} \right)\,,
\ee
and momentum $p = s / \CR$\,. 
In particular, for the bosonic string sigma model~\eqref{eq:nrstngf} of $D=24$ free bosons, we have
\be \label{eq:mfsnrs}
    \Delta = \pi \, k^2 + N + \bar{N}\,,
        \qquad%
    s = N - \bar{N}\,,
        \qquad%
    c = 24\,,
        \qquad%
    \CR = w \, R\,,
        \qquad%
    p = \frac{n}{R}\,.
\ee
The resulting spectrum corresponds to free strings with winding number one. Furthermore, $w$ is the winding number, $n$ the Kaluza-Klein momentum number, and $R$ the radius of the compact circle. After plugging~\eqref{eq:mfsnrs} into the CFT spectrum, the Galilean-invariant dispersion relation~\eqref{eq:nrstdr} is recovered. Note that $k^2 = k_{A'} k_{A'}$ with $k_{A'}$ the transverse momentum. 
The $T\bar{T}$ deformation of the CFT spectrum~\eqref{eq:cftsp} can be derived as 
\be
    \varepsilon \bigl( \lambdap\,, \CR \bigr) = \frac{2 \pi \, \CR}{\lambdap} \ls \sqrt{1 + \frac{\Delta - \frac{c}{12}}{\pi \, \CR^2} \, \lambdap + \lr \frac{s \, \lambdap}{2 \pi \, \CR^2} \rr^{\!\!2}} - 1 \rs.
\ee
Using $\lambdap = 1$ and Eq.~\eqref{eq:mfsnrs}, we recover the spectrum for conventional free bosonic strings winding $w$ times around a circle of radius $R$\,, in the presence of a constant $B$-field. 

Next, we consider the non-relativistic string theory in general background fields, for which the associated sigma model extending~\eqref{eq:nrstng} is:
\be \label{eq:nganrst}
    S^\text{NRST}_\text{F1} = \int \dd^2 \sigma \, \CL(0)\,,
\ee
where (restoring the factors of $\alpha'$)
\begin{align} \label{eq:lag0} 
    \CL(0) & = - \frac{\sqrt{-\det \tau}}{4\pi\alpha'} \, \p_\alpha X^\mu \, \p_\beta X^\nu \, \Bigl[ \tau^{\alpha\beta} \, 
    E^{}_{\mu\nu} (X) + \varepsilon^{\alpha\beta} \, b^{}_{\mu\nu} (X) \Bigr]\,. 
\end{align}
Here, $\tau^{}_{\alpha\beta} = \p^{}_\alpha X^\mu \, \p^{}_\beta X^\nu  \tau^{}_{\mu\nu} (X)$, and $\tau^{\alpha\beta}$ is the inverse of $\tau^{}_{\alpha\beta}$\,. Moreover, $\varepsilon^{\alpha\beta}$ is the Levi-Civita tensor.  In order to make connection
with the $T\bar{T}$ deformation, it is sufficient to take a static gauge where $\tau^{}_{\alpha\beta}$ is constant instead of the Minkowski metric. 
The $T\bar{T}$ deformation that we have reviewed above can be readily generalised to the Lagrangian~\eqref{eq:lag0} with a constant $\tau^{}_{\alpha\beta}$\,, and the solution of the flow equation now takes the form: 
\begin{align}
\begin{split}
    \CL (\lambdap) = & - \frac{\sqrt{-\det \tau}}{\lambdap} \, \left[ \sqrt{1 + \tau^{\alpha \beta} E^{}_{\alpha \beta} \left(\frac{\lambdap}{2\pi\alpha'} \right) + \frac{\det \bigl( E_{\alpha \beta} \bigr)}{\tau} \left(\frac{\lambdap}{2\pi\alpha'}\right)^{\!\!2}} - 1 \right] \\[4pt]
    & - \frac{\sqrt{-\det\tau} \, \varepsilon^{\alpha\beta} \, b^{}_{\alpha\beta}}{4\pi\alpha'}\,,
\label{eq:F1FlowSoln}
\end{split}
\end{align}
where $E_{\alpha\beta}$ and $b_{\alpha\beta}$ are pullbacks to the worldsheet.
Defining $\lambdap = 2 \, \pi \, \alpha' \, \omega^{-2}$\,, we find that the $T\bar{T}$-deformed action can be written as 
\be \label{eq:odnga}
    S = - \frac{1}{2\pi\alpha'} \int \dd^2 \sigma \, \sqrt{-\det \Bigl( \omega^2 \, \tau^{}_{\alpha\beta} + E^{}_{\alpha\beta} \Bigr)} - \frac{1}{2\pi\alpha'} \int \Bigl( - \omega^2 \, \tau^0 \wedge \tau^1 + b^{(2)} \Bigr)\,.
\ee
This is the conventional Nambu-Goto action reparametrised in terms of the non-relativistic string prescription~\eqref{eq:tgtptb}. In this sense, the $T\bar{T}$ deformation essentially undoes the BPS decoupling limit that sends $\omega$ to infinity. It is then natural to conjecture that such a deformation of the second quantised non-relativistic strings, \emph{i.e.}~matrix string theory, gives rise to the second quantisation of type IIB superstring theory. This is in spirit similar to the field-theoretical discussion in~\cite{Benjamin:2023nts}.

\subsection{Polyakov Formulation: Mapping \texorpdfstring{$T\bar{T}$}{TTbar} to a Marginal Deformation}
\label{sec:TTbarPoly}

In the above presentation using the Nambu-Goto formulation, the $T\bar{T}$ deformation is generated by turning on an irrelevant operator in the theory. This is rather unconventional when it comes to the renormalisation group (RG) flow, which is a semi-group. Therefore, it only makes sense to think about RG as a flow from the UV to the IR, not \emph{vice versa}. Nevertheless, as we have already observed, the $T\bar{T}$ deformation precisely introduces the irrelevant operator such that non-relativistic string theory with non-Lorentzian target space is deformed back to the conventional string sigma model with a Lorentzian target space. One would expect that there must be an RG way to understand the $T\bar{T}$ deformation in this stringy context. 

The difficulty of having a standard RG interpretation of the $T\bar{T}$ deformation in the context of string theory is related to the usual difficulties of quantising the Nambu-Goto action, except in the lightcone gauge. Instead, in the Polyakov formulation, the $T\bar{T}$ deformation is mapped to a current-current deformation with a marginal coupling in a two-dimensional worldsheet field theory, without the necessity of committing to a static gauge.
The Polyakov formulation of non-relativistic string theory in flat spacetime is known as the Gomis-Ooguri theory~\cite{Gomis:2000bd} in the literature, which in curved backgrounds (and on a curved worldsheet) takes the form~\cite{Bergshoeff:2018yvt} 
\begin{align} \label{eq:pfnrst}
\begin{split}
    S^{}_\text{P} & = - \frac{1}{4\pi\alpha'} \int \dd^2\sigma \, \sqrt{-h} \, \Bigl( h^{\alpha\beta} \, \p^{}_\alpha X^\mu \, \p^{}_\beta X^\nu \, E^{}_{\mu\nu} + \lambda \, \bar{e}^\alpha \, \p^{}_\alpha X^\mu \, \tau^{}_\mu + \bar{\lambda} \, e^\alpha \, \p^{}_\alpha X^\mu \, \bar{\tau}^{}_\mu \Bigr) \\[4pt]
    & \quad - \frac{1}{4\pi\alpha'} \int \dd^2 \sigma \, \sqrt{-h} \, \varepsilon^{\alpha\beta} \, \p_\alpha X^\mu \, \p_\beta X^\nu \, b^{}_{\mu\nu} + \frac{1}{4\pi} \int \dd^2 \sigma \, \sqrt{-h} \, R(h) \, \varphi\,,
\end{split}
\end{align}
where $\tau_\mu = \tau_\mu{}^0 + \tau_\mu{}^1$ and $\bar{\tau}_\mu = \tau_\mu{}^0 - \tau_\mu{}^1$\,, $h^{}_{\alpha\beta} = e^{}_\alpha{}^a \, e^{}_\beta{}^b \, \eta_{ab}$ is the worldsheet metric, $e^{}_\alpha{}^a$ is the worldsheet zweibein field and $R(h)$ is the worldsheet Ricci scalar. Moreover, $h = \det (h_{\alpha\beta})$\,, $h^{\alpha\beta}$ is the inverse of $h^{}_{\alpha\beta}$ and $e^\alpha{}_a$ the inverse of $e^{}_\alpha{}^a$\,, and we have also introduced the lightcone notation $e^\alpha = e^\alpha{}^{}_0 + e^\alpha{}^{}_1$ and $\bar{e}^\alpha = e^\alpha{}^{}_0 - e^\alpha{}^{}_1$\,. 
Finally, $\lambda$ and $\bar{\lambda}$ are one-form fields that play the role of Lagrange multipliers imposing the constraints 
\be
    \bar{e}^\alpha \, \p^{}_\alpha X^\mu \, \tau^{}_\mu = e^\alpha \, \p^{}_\alpha X^\mu \, \bar{\tau}^{}_\mu = 0\,. 
\ee
These constraints are solved by $e^{}_\alpha \propto \p_\alpha X^\mu \, \tau^{}_\mu$ and $\bar{e}^{}_\alpha \propto \p_\alpha X^\mu \, \bar{\tau}^{}_\mu$\,. Plugging these solutions back into the Polyakov action~\eqref{eq:pfnrst} recovers the Nambu-Goto formulation~\eqref{eq:nganrst}.

In this Polyakov formulation, the deformation parametrised by $\omega$ in Eq.~\eqref{eq:odnga} is reincarnated into the following marginal operator:
\be \label{eq:llbd}
    - \frac{1}{4\pi\alpha'} \int \dd^2 \sigma \, \sqrt{-h} \, \omega^{-2} \, \lambda \, \bar{\lambda}\,. 
\ee
This is the operator that deforms non-relativistic string theory back to the conventional string theory.  The full Polyakov action combining Eq.~\eqref{eq:pfnrst} and \eqref{eq:llbd} is equivalent to the Nambu-Goto action~\eqref{eq:odnga}, except that now the original irrelevant $T\bar{T}$ deformation in the gauge-fixed Nambu-Goto action has been recast as a marginal deformation in the Polyakov formulation. 
As discussed in \cite{Blair:2020ops}, this can also be viewed as the deformation arising from a TsT transformation which acts on the string Newton-Cartan geometry and maps it (via null duality) to a Lorentzian one.

This mapping of the $T\bar{T}$ deformation to a marginal coupling is partly possible because the undeformed theory in the covariant quantisation is described by a different CFT. We further elaborate on this in flat target spacetime. In conformal gauge and after integrating out the Lagrange multipliers $\lambda$ and $\bar{\lambda}$\,, the matter sector of~\eqref{eq:pfnrst} is the same as the Nambu-Goto action~\eqref{eq:nganrst} describing the non-relativistic strings when a static gauge is chosen. This Nambu-Goto action in flat spacetime reduces to the free theory ~\eqref{eq:nrstngf} of 24 scalar fields, upon the choice of the static gauge. However, in the Polyakov formulation, the theory can now be quantised in a covariant way without committing ourselves to a lightcone or static gauge. In the presence of the longitudinal embedding coordinates, a $bc$-ghost sector needs to be introduced in the BRST quantisation, due to the gauge fixing of the worldsheet metric. Now, the current-current deformation interpolates from the undeformed non-relativistic string sigma model
\be
    S_0 = \frac{1}{2} \int \dd^2\sigma \, \Bigl( \p_\alpha X^{A'} \, \p^\alpha X^{A'} + \lambda \, \bar{\p} X + \bar{\lambda} \, \p \bar{X} + b \, \bar{\p} c + \bar{b} \, \p \bar{c} \Bigr)\,, 
\ee
to the relativistic string sigma model
\begin{align} \label{eq:csm}
    S = \frac{1}{2} \int \dd^2\sigma \, \Bigl( \p_\alpha X^{A'} \, \p^\alpha X^{A'} + \lambda \, \bar{\p} X + \bar{\lambda} \, \p \bar{X} + 2 \, \omega^{-2} \, \lambda \, \bar{\lambda} + b \, \bar{\p} c + \bar{b} \, \p \, \bar{c} \Bigr)\,.
\end{align}
The contribution from the marginal current-current term $\lambda\bar{\lambda}$ is equal to the contribution from summing over all the irrelevant operators
in the Nambu-Goto action (\emph{i.e.} in that case we obtain $\mathcal{L}(\lambdap)$ by solving \eqref{eq:ttbarfe}, meaning integrating up the contributions of the deforming operator at each value of the deformation parameter from $0$ to finite $\lambdap$).
Integrating out $\lambda$ and $\bar{\lambda}$ in the relativistic string sigma model~\eqref{eq:csm} leads to the conventional Polyakov string action in conformal gauge.

In general, the marginal $\lambda\bar{\lambda}$ term is generated by quantum corrections in the non-relativistic string sigma model~\eqref{eq:pfnrst}, which then drives a RG flow toward the full string theory~\cite{Gomis:2019zyu, Gallegos:2019icg, Yan:2019xsf, Bergshoeff:2021bmc, Yan:2021lbe}.
However, a more detailed analysis of the underlying symmetry principles reveal that, at least within the bosonic sector, there exists various extensions of the global symmetries used to define the string sigma model such that the $\lambda\bar{\lambda}$ operator is not generated at all loops~\cite{Yan:2021lbe}. As a result the target space gauge group is also enlarged, such that certain geometric constraints are imposed on the longitudinal vielbein $\tau^{}_\mu{}^A$~\cite{Andringa:2012uz, Bergshoeff:2018yvt, Bergshoeff:2018vfn}. These geometric constraints set restrictions on the torsion of $\tau^{}_\mu{}^A$\,. Intriguingly, similar geometric constraints also arise in the study of non-Lorentzian supergravity in ten- and eleven-dimensions \cite{Bergshoeff:2021tfn, Bergshoeff:2024nin}, where they are required for supersymmetry.
This suggests that such torsional constraints are indispensable for the self-consistency of non-relativistic string theory (and its D-brane and M-theory cousins). For the above reason, the $\lambda\bar{\lambda}$ deformation in Eq.~\eqref{eq:llbd} is also referred to as the torsional deformation~\cite{Yan:2021lbe}. 

Now, we have shown that, in the S-dual frame, the same $\omega$ parametrises the M1T limit. Via T-dualities, the same $\omega$ translates to the parameter controlling various M$p$T limits. Therefore, the deformations of various M$p$T brane actions toward the DBI action in type II superstring theory are naturally dual to the $T\bar{T}$ deformation of non-relativistic string sigma model. 
This suggests that we can find interesting generalisations of the $T\bar{T}$ deformation by studying the features of these decoupling limits on the worldvolumes of the relevant branes.
In Section~\ref{sec:pbttfe}, we show that this is indeed the case.

\subsection{Bulk Geometry from Dual \texorpdfstring{$T\bar{T}$}{TTbar} Deformation} 
\label{sec:TTbarBulk} 

So far we have discussed how the $T\bar{T}$ deformation in the Polyakov formulation generates a marginal RG flow from the worldsheet theory describing the non-relativistic string to one for the relativistic string. This RG flow is associated with the coupling constant in front of the current-current term~\eqref{eq:llbd}, which we write as $\lambda \, \bar{\lambda}$\,. 

Applying the intuition that we have developed above to superstring theory, it follows from the formal duality relationships between the different decoupling limits that  this $T\bar{T}$ deformation is mapped to a deformation of M$p$T back to the full type II superstring theory, controlled by the parameter $\omega$\,. 
This dual $T\bar{T}$ deformation induces a flow from the M$p$T brane toward the DBI action. We will refer to this natural generalisation of the standard $T\bar{T}$ deformation to the branes as the \emph{$p$-brane $T\bar{T}$ deformation}. We will derive the explicit form of the flow equations generalising \eqref{eq:ttbarfe} in Section \ref{sec:pbttfe}.

From the perspective of the string sigma model, the coupling $\omega^{-2}$ in Eq.~\eqref{eq:llbd} can be background dependent, replacing~\eqref{eq:llbd} with
\be
    - \frac{1}{4\pi\alpha'} \int \dd^2 \sigma \, \sqrt{-h} \, U(X) \, \lambda \, \bar{\lambda}\,. 
    \label{ttbartada}
\ee
Due to the dilatation symmetry~\eqref{eq:dilatsmpt}, $\omega$ can be
promoted to a function of the spacetime coordinate $\mathbb{\Omega} = \mathbb{\Omega}(X)$\,, with $U(X) = \bigl[ \mathbb{\Omega} (X) \bigr]^{-2}$. 

Let us outline a pertinent example.
Applying a non-relativistic string decoupling limit to the F1-string supergravity solution leads to a near-horizon geometry of the following form: \cite{Danielsson:2000mu, Avila:2023aey, soliton}
\be
    \dd s^2 = \mathbb{\Omega}(r)^2 \, \dd x^A \, \dd x^B \, \eta^{}_{AB} + \dd x^{A'} \, \dd x^{A'} \,,
        \quad%
    B = -\mathbb{\Omega(r)}^2 \, \dd x^0 \wedge \dd x^1 \,,
\label{F1nhor}
\ee
with $A=0\,,\,1$ and $A'=2\,,\,\dots,\,9$\,. Moreover, $\mathbb{\Omega}(r) = (r/\ell)^3$\,, where $r\equiv \|x^{A'}\|$\,. At $r \rightarrow \infty$ this gives a flat string Newton-Cartan geometry.
The Polyakov action \eqref{eq:pfnrst} for the non-relativistic string in flat target space, deformed by 
\eqref{ttbartada}, is
\begin{align} \label{eq:pfnrst_example}
    S^{}_\text{P} & = - \frac{1}{4\pi\alpha'} \int \dd^2\sigma \, \sqrt{-h} \, \Bigl[ h^{\alpha\beta} \, \p^{}_\alpha X^{A'} \, \p^{}_\beta X^{A'} \ + \lambda \, \bar{e}^\alpha \, \p^{}_\alpha (X^0+X^1)\,  + \bar{\lambda} \, e^\alpha \, \p^{}_\alpha (X^0-X^1) \,\Bigr] \notag \\[4pt]
    & \qquad - \frac{1}{4\pi\alpha'} \int \dd^2 \sigma \, \sqrt{-h} \, \mathbb{\Omega}^{-2}(X) \, \lambda \, \bar{\lambda}\,. 
\end{align}
Integrating out $\lambda$ and $\bar \lambda$\,,
we obtain the Polyakov action for a string in the Lorentzian target spacetime \eqref{F1nhor}.

Generalising to the $p$-brane case, such a background-dependent $p$-brane $T\bar{T}$ deformation has an immediate application in holography: it allows us to generate the curved geometry in the bulk by starting with the associated asymptotic M$p$T in flat spacetime. This provides us with an opportunity to build an intrinsic relation between the asymptotic and bulk geometries. 

We start by illustrating how the bulk geometry is generated by a $p$-brane $T\bar{T}$ deformation. In the asymptotic limit we have SYM living on the longitudinal sector of the flat spacetime with a codimension-($p+1$) foliation. This is a ten-dimensional M$p$T geometry,
\begin{align} \label{eq:asyMpTgeo}
	\tau^{}_\mu{}^A \, \dd x^\mu & = \dd x^A \,,
        &%
    E^{}_\mu{}^{A'} \dd x^\mu & = \dd x^{A'}\,.
\end{align}
with $A=0,1,\dots,p$, $A' = p + 1\,, \, \cdots, \, 9$\,, and a constant string coupling $g^{}_\text{s}$\,. All the other background fields are set to zero. This geometry admits a Galilean boost-like symmetry,
\be
    \delta^{}_\text{G} x^{A} = 0\,,
        \qquad%
    \delta^{}_\text{G} x^{A'} = \Lambda^{}_A{}^{A'} x^A\,,
\ee
under which $\dd x^A$ and $\p / \p x^{A'}$ are invariant. The $p$-brane $T\bar{T}$ deformed geometry in terms of the background-dependent parameter $\mathbb{\Omega}(r)$ is
\begin{subequations} \label{eq:mptdtg}
\begin{align} 
	\dd s^2 & = \mathbb{\Omega} (r) \, \dd x^A \, \dd x^{B} \, \eta^{}_{AB} + \frac{\dd x^{A'} \, \dd x^{A'}}{\mathbb{\Omega} (r)}\,, \\[4pt] 
    C^{(p+1)} & = \bigl[ \mathbb{\Omega}(r) \bigr]^2 \, g^{-1}_\text{s} \, \dd t \wedge \dd x^1 \! \wedge \cdots \wedge \dd x^{p}\,, 
        \qquad%
    e^\Phi = \bigl[ \mathbb{\Omega}(r) \bigr]^{\!\frac{p-3}{2}} g^{}_\text{s}\,.
\end{align}
\end{subequations}
Upon the identification $\mathbb{\Omega} (r) = (r / \ell)^{\frac{7-p}{2}}$, Eq.~\eqref{eq:mptdtg} describes the geometry~\eqref{eq:dsco} arising in the D$p$-brane near-horizon solution. In the IMSY holographic duality, the closed string mode captured by $\mathbb{\Omega}(r)$\,, which is a background field that can be thought of as a coherent state of closed strings, corresponds to the open string modes that give rise to the SYM at the asymptotic infinity. The flat M$p$T geometry at the asymptotic infinity is recovered by sending $\mathbb{\Omega}$ to infinity, \emph{i.e.}~the radial coordinate $r$ is much larger than the scale $\ell$\,. 

The same discussion also applies to the `multicritical' case where double BPS decoupling limits are performed at the asymptotic infinity. In general, in flat spacetime, applying simultaneously an M$p$T and M$q$T limiting prescription, respectively parametrised by $\omega$ and $\tilde{\omega}$\,, we are led to the following background field configurations:
\begin{subequations} \label{eq:mdtdtg}
\begin{align} 
	\dd s^2 & = \omega \, \tilde{\omega} \, \dd x^a \, \dd x^{}_a + \frac{\omega}{\tilde{\omega}} \, \dd x^u \, \dd x^u + \frac{\tilde{\omega}}{\omega} \, \dd x^I \, \dd x^I + \bigl( \omega \, \tilde{\omega} \bigr)^{-1} \, \dd x^i \, \dd x^i\,, \\[4pt] 
    C^{(p+1)} & = \omega^2 \, g^{-1}_\text{s} \, \dd t \wedge \dd x^1 \wedge \cdots \wedge \dd x^{q-m} \wedge \dd x^{q+1} \wedge \cdots \wedge \dd x^{q+n}\,, \\[4pt]
    C^{(q+1)} & = \tilde{\omega}^2 \, g^{-1}_\text{s} \, \dd t \wedge \dd x^1 \wedge \cdots \wedge \dd x^{q}\,, 
        \qquad%
    e^\Phi = \omega^{\frac{p-3}{2}} \, \tilde{\omega}^{\frac{q-3}{2}} \, g^{}_\text{s}\,.
\end{align}
\end{subequations}
where the indices are as in Eq.~\eqref{eq:indices}.
Note as before that $p + m = q + n$, while here we have assumed that $p \neq q$\,. When $p = q$\,, the RR potentials are replaced with
\be
    C^{(p+1)} = \dd t \wedge \dd x^1 \wedge \cdots \wedge \dd x^{p-m} \wedge \biggl( \frac{\omega^2}{g^{}_\text{s}} \, \dd x^{p-m+1} \wedge \cdots \wedge \dd x^{p} + \frac{\tilde{\omega}^2}{g^{}_\text{s}} \, \dd x^{p+1} \wedge \cdots \wedge \dd x^{p+m} \biggr).
\ee
When $m = 0$\,, the $p = q$ case reduces to the single M$p$T prescription that we already discussed earlier. In the double scaling limit with $\omega \rightarrow \infty$ and $\tilde{\omega} \rightarrow \infty$\,, one zooms in on a background D$p$-D$q$ bound state in the type II theory, which leads to a multicritical theory in the DLCQ${}^2$ orbit. The resulting flat geometry is encoded by the coordinates
$x^a, \, x^u, \, x^I$, and $x^i$\,.
The boost-like symmetries that relate different sectors in the foliated spacetime are given by
\be
    \delta^{}_\text{B} x^a = 0\,,
        \qquad%
    \delta^{}_\text{B} x^u = \Lambda^{}_a{}^u \, x^a\,,
        \qquad%
    \delta^{}_\text{B} x^I = \Lambda^{}_a{}^I \, x^a\,,
        \qquad%
    \delta^{}_\text{B} x^i = \Lambda^{}_a{}^i \, x^a\,.
\ee
There are now two analogous $T\bar{T}$ deformations that one can choose to perform, one associated with the parameter $\omega$ and the other with $\tilde{\omega}$\,. In accordance with the convention in Section~\ref{sec:hnhbpsl}, we promote $\tilde{\omega}$ to $\tilde{\mathbb{\Omega}} (\mathbb{r}) = (\mathbb{r} / \lbb)^{(7-q)/2}$\,, where $\mathbb{r}$ is the radial coordinate within the $x^u$ or $x^i$ sector. The deformed geometry is of the M$p$T type and is in form the same as in Eq.~\eqref{eq:mptsl02}, with
\begin{subequations} \label{eq:mptsl022}
\begin{align}
    \tau^a & = \tilde{\mathbb{\Omega}}^\frac{1}{2} \, \dd x^a\,, 
        &%
    E^{I} & = \tilde{\mathbb{\Omega}}^{\frac{1}{2}} \, \dd x^{I}\,, 
        &%
    e^{\varphi} & = \tilde{\mathbb{\Omega}}^{\frac{q-3}{2}} \, g_\text{s} \,,\\[4pt]
    \tau^u & = \tilde{\mathbb{\Omega}}^{-\frac{1}{2}} \, \dd x^u\,, 
        &%
    E^{i} & = \tilde{\mathbb{\Omega}}^{-\frac{1}{2}} \, \dd x^{i} \,,
        &%
    c^{(q+1)} & = \frac{\tilde{\mathbb{\Omega}}^2}{g_\text{s}} \, \dd t \wedge \dd x^1 \dots \wedge \dd x^{q} \,.
\end{align}
\end{subequations}
This general form captures the near-horizon M$p$T geometries found in Section~\ref{sec:hnhbpsl}. This $p$-brane $T\bar{T}$-deformed geometry provides a more generic form for constructing possible holographic duals: take~\eqref{eq:mptsl022} as an ansatz and then plug it into the M$p$T supergravity equations of motion, we may solve for $\tilde{\mathbb{\Omega}}$ and determine whether there exists such a bulk geometry. This provides us with a concrete starting point for further constructing a holographic dual. This proposal may pave the way towards a more exhaustive classification of intersecting brane solutions that are relevant for mapping out new holographic dualities, which still remains as an open question.

A similar $T\bar{T}$ deformation also applies to the case for a bulk intersecting brane geometry that we discussed in Section~\ref{sec:dlcqmn}. We continue to use the D1-D5 system as a demonstrating example. At the asymptotic infinity, the geometry arises from taking \eqref{eq:mdtdtg} with $p=1$, $q=5$ and $n=0$ (so no $x^u$ coordinates) and sending $\omega$ and $\tilde{\omega}$ to infinity. Unlike the other examples that we discussed earlier in this subsection, where we only promote a single parameter $\omega$ (or $\tilde{\omega}$) to be dependent on the bulk radial direction $r$\,, in the AdS${}_3$/CFT${}_2$ correspondence we are required to promote both the $\omega$ and $\tilde{\omega}$ parameters to be $r$ dependent, by replacing $\omega$ with $\mathbb{\Omega} (r) = r / \ell_1$ and replacing $\tilde{\omega}$ with $\tilde{\mathbb{\Omega}} (r) = r / \ell_5$\,, with $r = \|x^i\|$\,. As expected, this version of multiple $T\bar{T}$ deformations map the non-Lorentzian geometry at the asymptotic infinity to the desired bulk AdS${}_3 \times S^3 \times$M${}^4$ geometry in Eq.~\eqref{eq:adstg}.

\section{\texorpdfstring{$p$}{p}-Brane \texorpdfstring{$T\bar{T}$}{TTbar} Flow Equations in Various Dimensions} \label{sec:pbttfe}

In Section \ref{sec:TTbar}, we reviewed how the $T \bar T$ deformation takes non-relativistic string theory back to string theory in the conventional Lorentzian setting.
We explained how non-relativistic string theory is S-dual to M1T, which is then related by T-duality to the general M$p$T limits we have been studying.
We asserted that this means that the deformation from turning on the non-zero parameter $1/\omega^{-2}$ takes M$p$T back to the usual type II string theory, and this deformation should be viewed as a $p$-brane generalisation of the $T \bar T$ deformation.
We showed that this deformation appears geometrically in the context of near-horizon geometries. 
However, we have not yet constructed the explicit flow equations governing these new deformations at the worldvolume level. In particular, we are interested in finding the $p$-brane analogue of the $T\bar{T}$ flow equation~\eqref{eq:ttbarfe} for the Lagrangian. 
In this final section, we will turn our attention to deriving these equations. 

Such a $p$-brane generalisation of the $T\bar{T}$ deformation gives rise to irrelevant operators, which undo the BPS decoupling limit from the type II theory to M$p$T. In this way, SYM is deformed to the non-abelian DBI theory.  In this section, we explore this connection in the U(1) case, focusing on the bosonic sector. Below, we will present the complete derivation for the flow equations using branes in curved backgrounds. 
At the end of this section, we will also give a summary of the main results in flat spacetime, adapted towards potential applications in a field-theoretical setup.

We present a uniform approach to deriving $p$-brane generalisations of the $T\bar T$ flow equations, by considering systematically the worldvolume actions for branes and strings under their respective decoupling limits.
We start with the action for a single D$p$-brane,
\be
    S = - \int \dd^{p+1} \sigma \, e^{-\Phi} \sqrt{-\det \Bigl(g_{\alpha \beta} + \mathcal{F}_{\alpha \beta} \Bigr)} 
    + \int C \wedge e^{\CF} \Big|_{p+1}\,,
\ee
where $\mathcal{F} = F +B$ and $C \equiv \prod_q C^{(q)}$ denotes the RR polyform.
For simplicity we drop the explicit factors of brane tension in this section.
Reparametrised using the M$p$T prescription~\eqref{eq:rpgcfbmpt}, we find
\be
\begin{split} 
S(\lambdap) & = \int \dd^{p+1} \sigma \, e^{-\varphi} \, \mathcal{L} (\lambdap) 
+ \int c \wedge e^{\mathcal{F}} \Big|_{p+1}\,,
\end{split} 
\label{DpForTTbar}
\ee
where $c \equiv \Pi_q c^{(q)}$ denotes polyform of M$p$T RR fields, and we define the following Lagrangian (density) function
\be
    \mathcal{L} (\lambdap) = \frac{1}{\lambdap} \, \biggl[
    \sqrt{-\det \tau} - 
    \sqrt{-\det \Bigl( \tau + \lambdap^{\frac{1}{2}} \, \mathcal{F} + \lambdap \,  E \Bigr)}\, \biggr],
\label{defLF}
\ee
where we let $\lambdap = \omega^{-2}$, and 
\begin{align}
    \tau^{}_{\alpha\beta} = \p^{}_\alpha X^\mu \, \p^{}_\beta X^\nu \, \tau^{}_{\mu\nu}\,, %\\[4pt]
    \qquad E^{}_{\alpha\beta} = \p^{}_\alpha X^\mu \, \p^{}_\beta X^\nu \, E^{}_{\mu\nu}\,.
\end{align}
By definition $\tau_{\mu\nu}$ has rank $(p+1)$ and $E_{\mu\nu}$ has rank $9-p$. 
Setting $\mathcal{F} = 0$ in the expression \eqref{defLF}, one obtains a Lagrangian applicable to limits of $(p+1)$-dimensional extended objects which do not carry gauge fields.
In particular, for $p=1$ and $\mathcal{F}=0$, the same Lagrangian describes the worldsheet action of a fundamental string in the non-relativistic string limit, as follows from the discussion in Section \ref{sec:TTbarNG}.
For $p=2$\,, $\mathcal{F}=0$\,, and with $E_{\mu\nu}$ of rank 8, we can use \eqref{defLF} to describe the worldvolume action of the 11-dimensional membrane in the limit leading to non-relativistic M-theory, which was discussed in Section \ref{sec:adsdlcq}, \emph{i.e.} in that case we have
\be
S_{\text{M2}} = \int \dd^{3} \sigma \, \mathcal{L}(\lambdap)\Big|_{\mathcal{F}=0} + \int a^{(3)} \,,\quad
\ee
where $\lambdap \rightarrow 0$ gives rise to the decoupling limit.

In the following discussions, we will assume that the longitudinal metric $\tau^{}_{\alpha\beta}$ and the dilaton $\varphi$ are independent of the embedding coordinates. 
The Chern-Simons terms in all these examples are topological and so will be unaffected by a deformation based on the energy-momentum tensor: we will therefore focus our attention entirely on the Lagrangian \eqref{defLF} in which the deformation parameter $\lambdap$ appears.
Note that the overall factor of $1/\lambdap$ in \eqref{defLF} shows that when the factors of brane tension $T$ are restored, the dimensionful deformation parameter should be $\lambdap/T$, and so have dimensions of $(\text{length})^{p+1}$.

For $\lambdap=0$\,, we have
\be
\begin{split} 
\mathcal{L} (0) & = \sqrt{-\det \tau}  \Bigl(
- \tfrac{1}{2} \, \tau^{\alpha \beta} E_{\alpha \beta}  - \tfrac{1}{4}\, \tau^{\alpha \beta} \tau^{\gamma \delta} \mathcal{F}_{\alpha \gamma} \, \mathcal{F}_{\beta \delta} \Bigr)\,. 
\end{split} 
\ee
where $\tau^{\alpha \beta}$ denotes the inverse of $\tau_{\alpha \beta}$.
We treat $E_{\alpha \beta} = \partial_\alpha X^\mu \, \partial_\beta X^\nu \, E_{\mu\nu}$ as describing the scalar fields on the worldvolume, while regarding $\tau_{\alpha \beta}$ as the metric on the brane worldvolume.
To make contact with the usual $T\bar T$ interpretation, $\tau_{\alpha\beta}$ should ultimately be fixed in static gauge to the flat worldvolume metric.
For instance, considering the simplest possible case, one can take the undeformed theory to be the U(1) version of the D$p$-brane action~\eqref{eq:mptdpnona} in M$p$T, in the flat background defined by $\tau^{}_{\alpha \beta} = \eta^{}_{\alpha \beta}$, $E_{A'B'} = \delta_{A'B'}$, $E_{A A'} =E_{AB} = 0$ and without $p$-form gauge fields. 
Then the initial Lagrangian is:
\be \label{eq:undel}
    \CL(0) = - \tfrac{1}{2} \, \p^{\alpha} X^{A'} \, \p^{}_\alpha X^{A'} - \tfrac{1}{4} \, F_{\alpha \beta}\,F^{\alpha\beta}\,. 
\ee
The $p$-brane $T\bar{T}$-deformed theory will be the DBI action reparametrised using the M$p$T prescription~\eqref{eq:rpgcfb}. The associated deformed Lagrangian is
\be
    \CL(\lambdap) = - \biggl[ \omega^{\frac{3-p}{2}} \sqrt{-\det \Bigl(\omega \, \eta^{}_{\alpha\beta} + \omega^{-1} \, \p^{}_{\alpha} X^{A'} \, \p^{}_\beta X^{A'} + F_{\alpha \beta} \Bigr)} - \omega^2 \biggr]\,, 
        \qquad%
    \omega^2 = \lambdap^{-1}\,.
\label{eq:defundel}
\ee
We can then seek to construct the analogue of the $T\bar{T}$ flow equation~\eqref{eq:ttbarfe} for the D$p$-brane. 
The advantage of using the M$p$T framework is that both the undeformed and deformed Lagrangians are already known, with the undeformed action arising from a BPS decoupling limit of the deformed one. It is then conceptually straightforward to derive the associated $p$-brane $T\bar{T}$ flow equation: denote the stress-energy tensor associated with the $\lambdap$-dependent Lagrangian~\eqref{defLF} as $T_{\alpha\beta} (\lambdap)$\,, then one simply needs to express $\p \CL(\lambdap) / \p t$ in terms of $T_{\alpha\beta} (\lambdap)$\,. However, the presence of the gauge field makes the derivation technically complicated: the flow equation may not only depend on $T_{\alpha\beta}(\lambdap)$ but also the field strength $F_{\alpha\beta}$\,. In the following, we will derive the complete flow equations for all $p$'s in $(p+1)$-dimensions when $F_{\alpha\beta} = 0$\,. Moreover, we will also present the nonperturbative flow equations with nonzero $F_{\alpha\beta}$ for $p \leq 2$ and the lowest-order terms in $\lambdap$ for $p > 2$\,. 

We now return to the general expression  \eqref{defLF} in order to derive these $p$-brane flow equations, following the approach of appendix A of \cite{Blair:2020ops}.
We define the energy-momentum tensor with respect to the constant worldvolume metric $\tau^{}_{\alpha \beta}$\,, so that
\be
T_{\alpha \beta} (\lambdap) = - \frac{2}{\sqrt{-\det\tau}} \frac{\partial \mathcal{L}(\lambdap)}{\partial \tau^{\alpha \beta}}\,.
\label{defT}
\ee
Note that, in terms of the full D$p$-brane action \eqref{DpForTTbar}, this pulls out a factor of the dilaton. 
With this definition, we obtain
\be
    T_{\alpha \beta} (\lambdap) = \frac{1}{\lambdap} \left[ \tau_{\alpha \beta} - \frac{\sqrt{- \det M}}{\sqrt{-\det \tau}} \, \bigl(M^{-1} \bigr){}^{\gamma \delta} \, \tau_{\gamma(\alpha} \tau_{\beta) \delta}    
    \right],
        \qquad%
    M_{\alpha\beta} \equiv \tau_{\alpha\beta} +\lambdap^{\frac{1}{2}} \, \mathcal{F}_{\alpha\beta} + \lambdap \, E_{\alpha\beta}\,.
\label{getT}
\ee
We can then compute the derivative of the Lagrangian \eqref{defLF} with respect to the flow parameter. 
This leads to a preliminary form of the flow equation given by
\be
\begin{split} 
    \frac{\partial \mathcal{L} (\lambdap)}{\partial \lambdap}  
        &=
    \frac{1}{2 \, \lambdap} \, \Bigl[ \bigl( p-1 \bigr) \, \mathcal{L} (\lambdap) - 
    \sqrt{-\det \tau} \, \tau^{\alpha \beta} \, T_{\alpha \beta}  \Bigr]
    - \frac{1}{4 \, \lambdap^{3/2}} \, \sqrt{-\det M} \, \bigl( M^{-1} \bigr)^{\alpha \beta} \, \mathcal{F}_{\alpha \beta} \,.
\end{split}
\label{flow_equation_general}
\ee
Our goal now is to massage this into a more meaningful form.
In particular, we wish to rewrite it solely in terms of the energy-momentum tensor $T_{\alpha \beta}$ and, as we will see, the field strength $\mathcal{F}_{\alpha \beta}$. In the following, we first consider the case where $\CF_{\alpha\beta} = 0$ and then move on to the more general case with $\CF_{\alpha\beta} \neq 0$\,. 

\vspace{3mm}

\noindent $\bullet$~\emph{Deforming scalar field theories.} We start with 
the simpler situation in which $\mathcal{F}_{\alpha\beta} = 0$\,, \emph{i.e.}~both the U(1) gauge potential and the $B$-field are set to zero. We are then dealing with scalar field theories with the scalars being the transverse embedding coordinates within $E_{\alpha\beta}$\,.
In these cases, we can take the expression for the energy-momentum tensor~\eqref{getT} and use it to solve for the combination $M_{\alpha\beta} = \tau_{\alpha\beta} + \lambdap \, E_{\alpha\beta}$ as follows:
\be
    M^{-1} = \frac{\sqrt{-\det \tau}}{\sqrt{-\det M}} \, \bigl( \mathbb{1} - \lambdap \, \CT \, \bigr) \, \tau^{-1} \,,
        \qquad%
    \CT \equiv \tau^{-1} \, T\,.
\ee
Taking the determinant on both sides, we obtain
\be
    \bigl(-\det M \bigr)^{\!\frac{p-1}{2}} = \bigl( - \det \tau \bigr)^{\!\frac{p-1}{2}}\,  \Bigl[ \det \bigl(\mathbb{1} - \lambdap \, \CT \, \bigr) \Bigr]\,.
\label{detM} 
\ee
Using this in the flow equation \eqref{flow_equation_general} with $\mathcal{F}_{\alpha \beta}=0$, we obtain the following definition of a generalised $T\bar{T}$ deformation
\be \label{genTTflow_noF}
    \frac{\partial \mathcal{L}}{\partial \lambdap} 
= \frac{\sqrt{- \det \tau}}{2 \, \lambdap^2} \left\{
\tr \bigl( \mathbb{1} - \lambdap \, \CT \, \bigr) 
- (p-1) \Bigl[ \det \bigl( \mathbb{1} - \lambdap \, \CT \, \bigr) \Bigr]^{\!\frac{1}{p-1}} - 2
\right\}.
\ee
Let us inspect the cases for $p = 0\,,\, 1\,, \, 2$\,, where the flow equations take simpler forms. 
\begin{enumerate}[(1)]

\item \emph{One dimension: particle $T\bar{T}$ flow equation.} For $p=0$, the flow equation \eqref{genTTflow_noF} becomes 
\be
    \frac{\partial \mathcal{L}}{\partial \lambdap} 
    = \frac{\sqrt{- \tau^{}_{00}}}{2} \frac{\bigl[ T^0{}_0 (\lambdap) \bigr]^2}{1 - \lambdap \, T^0{}_0(\lambdap)}\,.
\ee
Defining $\lambdap=4 \, t$ and going to static gauge to set $\tau^{}_{00} = -1$, this flow equation can be seen to correspond to the
one-dimensional version of the $T\bar T$ flow equation obtained in \cite{Gross:2019ach}. By construction, given the undeformed Lagrangian $\mathcal{L}(0) = \frac{1}{2} \dot{X}^i \dot{X}^i$ describing a non-relativistic particle, the solution to this flow equation will be the relativistic particle action \eqref{eq:pa}, coupled to a gauge field $A_\mu$ with $A_0 = \lambdap^{-1}$\,.

In this case, it is also straightforward to obtain a description of the deformation analogous to the Polyakov formulation described for the non-relativistic string theory in Section \ref{sec:TTbarPoly}.
This can be obtained by starting with the Hamiltonian formulation of the relativistic D0-brane action, parametrised using the M0T prescription~\eqref{eq:rpgcfb} as 
\be
    S = \int \dd \tau \, \biggl\{ \dot{X}^\mu \, \CP_\mu - \frac{e}{2} \Bigl[ \CP_\mu \, \CP_\nu \, \bigl( E^{\mu\nu} + \lambdap \, \tau^{\mu\nu} \bigr) + 2 \, e^{-\varphi} \,  \tau^\mu \, \CP_\mu \Bigr] + c^{}_\mu \, \dot{X}^\mu \biggr\}\,,
\ee
where $\dot{X}^\mu = \partial_\tau X^\mu$, $\CP^{}_\mu \equiv P^{}_\mu - c^{}_\mu$ with $P_\mu$ the canonical momenta and $c^{}_\mu$ the RR one-form in M0T and, again, $\lambdap=\omega^{-2}$\,. 
Note that $e$ is a Lagrange multiplier enforcing the Hamiltonian constraint.
Focusing now on this action in the M0T limit with $\lambdap=0$, 
and letting $\mathcal{P}_\mu = \mathcal{P}_{A'} E^{A'}_\mu + \lambda \, \tau_\mu$\,, we can integrate out the momenta $\mathcal{P}_{A'}$ to obtain 
\be \label{SD0_poly}
    S = \int \dd\tau \left[ \frac{1}{2 \, e} \, \dot{X}^\mu \, \dot{X}^\nu \, E_{\mu\nu} + \lambda \, \bigl( \dot{X}^\mu \, \tau_\mu - e  \,e^{-\varphi} \bigr)+ c_\mu \dot{X}^\mu \right] \,.
\ee
Integrating out $\lambda$ gives the expected non-relativistic particle action~\eqref{eq:m0td0ba} in M0T. The action \eqref{SD0_poly} can be viewed as a Polyakov-like action for the D0-brane in M0T, with $e$ being the worldline einbein.
The deformation that we add to this Polyakov-like action~\eqref{SD0_poly} in order to realise the effect of the $T\bar T$ deformation is then 
\be
    - \frac{1}{2} \int \dd \tau \, e \, \lambdap \, \lambda^2 \,,
\ee
sharing the same structure as the string marginal deformation \eqref{eq:llbd}.

\item \emph{Two dimensions: conventional $T\bar{T}$ flow equation.} 
When $p=1$, \eqref{detM} implies that $\det \bigl(\mathbb{1} - \lambdap \, \CT \bigr) = 1$.
The second term in the flow equation \eqref{genTTflow_noF} is then eliminated, and the first can be rewritten using $\tr\, \CT = \lambdap \, \det \CT$.
Hence one obtains:
\be \label{eq:tdttb}
    \frac{\partial \mathcal{L}}{\partial \lambdap} 
    =
    -\frac{\sqrt{-\det \tau}}{2 \, \lambdap} \, \tr \, \CT 
    =
    - \frac{\sqrt{-\det \tau}}{2} \, \det \CT\,,
\ee
which reproduces the usual $T \bar T$ flow equation for two-dimensional theories.

\item \emph{Three dimensions: membrane flow equation.} 
In the $p=2$ case, we obtain the flow equation 
\be \label{M2flow}
    \frac{\partial \mathcal{L}}{\partial \lambdap} 
    =
    \frac{\sqrt{- \det \tau}}{4} \Bigl[
    \tr \bigl( \CT^2 \bigr) - \bigl( \tr \, \CT \bigr)^2 
    + 2 \, \lambdap \, \det \CT 
    \Bigr] 
    \,.
\ee
Observe that the first terms coincide with $\det T$ in two dimensions.\footnote{The author of \cite{Blair:2020ops} obtained these leading terms and regrets his failure (due, perhaps, to an unaccountable bias towards deformations quadratic in $T$) to find the complete expression despite having the right idea and all the information at hand to do so...}
This membrane flow equation is a strikingly simple generalisation of the usual $T\bar T$ deformation.

\end{enumerate}

Let us comment at this point on the relationship of our flow equation~\eqref{genTTflow_noF} to various expressions in the literature.

Firstly, in the special case where there is only one scalar, due to the special form of the stress-energy tensor, the flow equation can be written equivalently (in flat spacetime) as
\be \label{eq:rootttb}
    \frac{\p\CL}{\p \lambdap} = \frac{1}{2 \, d} \, \tr (\CT^{\,2}) - \frac{1}{d^2} \, \bigl[\tr(\CT) \bigr]^2 - \frac{d-2}{2 \, d^{\,3/2} \, \sqrt{d-1}} \, \tr (\CT) \, \sqrt{\tr(\CT^{\,2}) - \frac{1}{d} \bigl[\tr(\CT) \bigr]^2}\,,
\ee
where $d=p+1$, with the restriction $p>0$.\footnote{When $p = 0$\,, the flow equation~\eqref{eq:rootttb} vanishes identically. However, Eq.~\eqref{genTTflow_noF} still holds.}
This flow equation was written down in \cite{Ferko:2023sps}. However, when one considers the more general case where there are multiple scalar fields, the flow equation takes the rather distinct form that we have found, and the deformation~\eqref{eq:rootttb} does not deform the free action towards the desired Nambu-Goto action  as one might have na\"{i}vely expected. Instead, by construction, the correct flow equation has to be Eq.~\eqref{genTTflow_noF}, which has an explicit dependence on $\lambdap$\,. In fact, the discrepancy between the two flow equations already arises at the zeroth order in $\lambdap$\,.

In the study of massive gravity generalisations of $T\bar{T}$ deformations, equivalent flow equations arise~\cite{Tsolakidis:2024wut}, which can involve multiple scalars. For example, the terms within the curly brackets of Eq.~(4.9) in~\cite{Tsolakidis:2024wut}, after plugging in the data from Eq.~(4.12) therein, can be shown to be equal to Eq.~\eqref{genTTflow_noF} in flat spacetime when $p=2$ (without using the special form of the stress-energy tensor in Eq.~\eqref{getT}).\footnote{We thank Evangelos Tsolakidis for private communications on this point.} It would be interesting to further study the relations for $p > 2$\,.

Finally, a related flow equation is shown in \cite{Morone:2024ffm} to deform the Einstein-Hilbert action to the modified Eddington-inspired Born-Infeld gravity, see Eq.~(B.29) of that reference.

\vspace{3mm}

\noindent $\bullet$~\emph{Deforming gauge theories.}
When $\CF_{\alpha\beta} \neq 0$\,, the theories that we are considering are U(1) gauge theories. The flow equations now become more complicated. However, some progress can still be made at least for the D1-string and D2-brane case, where we are able to construct the exact flow equations. For the ease of presentation, below we will use $\CF$ to denote $\CF^\alpha{}^{}_\beta$ with a raised index, instead of $\CF_{\alpha\beta}$\,, similar to how we are using $\CT^\alpha{}_\beta \equiv \tau^{\alpha\gamma} \, T_{\gamma\beta}$\,.   

\begin{enumerate}[(1)]

\item

\emph{Two dimensions:} When $p = 1$\,, it follows from Eq.~\eqref{getT} that
\be \label{eq:detmdetf}
    \det \Bigl( \mathbb{1} + \lambdap^{\frac{1}{2}} \, \CF + \lambdap \, \tau^{-1} \, E \Bigr) =  \frac{\det \CF}{\tr \,\CT - \lambdap \, \det \CT}\,. 
\ee
Plugging Eq.~\eqref{eq:detmdetf} into the preliminary form~\eqref{flow_equation_general} of the flow equation, we find
\be
    \frac{\partial \mathcal{L}}{\partial \lambdap}
    =- \frac{\sqrt{-\det \tau}}{2 \, \lambdap}  \left( 
        \tr \, \CT
        +
        \sqrt{ - \det \mathcal{F}} \, \sqrt{- \tr \, \CT + \lambdap \, \det \CT}
        \,\right).
\ee
By construction, this is the flow equation that takes a theory with a Maxwell vector field in $(1+1)$-dimensions to the DBI action. 
This explains why the usual $T\bar T$ deformation applied to such a theory does not lead to DBI but to a different endpoint~\cite{Conti:2018jho,Brennan:2019azg}.

\item \emph{Three dimensions:} For $p=2$ it turns out that the D2-brane flow equation has the same structure as for the membrane with no gauge field, thus Eq.~\eqref{M2flow} holds in terms of the Lagrangian $\mathcal{L} (\lambdap)$ of the D2-brane case, except that now the accompanying energy-momentum tensor is dependent on $\CF$.
At first sight this may be surprising, but is ultimately due to the fact that a one-form in three-dimensions is dual to a scalar. 

\item \emph{Perturbative results beyond three dimensions:} 
For $p>2$, the expressions involved increase in complexity and we have been unable to find closed form analytic results.
As they may be of future use, we collect here some perturbative results, obtained by comparing the expansion of the Lagrangian in $\lambdap$ with candidate terms built from the energy-momentum tensor and the field strength.
Firstly, for
$p = 3$\,,
\begin{align} \label{eq:fept}
    \frac{\p \CL}{\p \lambdap} = & \left( \frac{\p \CL}{\p \lambdap} \right)_\text{\!\!pre} 
    - \frac{\sqrt{-\det \tau}}{8} \, \Bigl[ \tr \bigl( \CF^{\,4} \bigr) - \tfrac{1}{4} \, \tr\bigl( \CF^2 \bigr)^2 \Bigr] \Bigl[ 1 - \tfrac{1}{2} \, \lambdap \, \tr\bigl( \CF^2 \bigr) \Bigr] \\[4pt]
    & - \frac{\lambdap \, \sqrt{-\det \tau}}{64} \, \Bigl[ \tr(\CT) \,  \tr\bigl(\CF^2\bigr)^2 + 4 \, \tr(\CT) \, \tr\bigl(\CF^{\,4}\bigr) - 8 \, \tr\bigl(\CT \, \CF^2\bigr) \, \tr\bigl(\CF^2\bigr) \Bigr] + O\bigl(\lambdap^2\bigr)\,. \notag
\end{align}
Here, $(\p \CL / \p \lambdap )_\text{pre}$ is in form the same as $\p \CL / \p \lambdap$ in Eq.~\eqref{genTTflow_noF}, except that now the energy-momentum includes the $U(1)$ field strength as in \eqref{getT}.
When $p \geq 4$\,, we record that the $\CF$-dependent flow equation at the lowest order in $\lambdap$ is
\begin{align} \label{eq:fehp}
\begin{split}
    \frac{\p \CL}{\p \lambdap} &= \frac{\sqrt{-\det \tau}}{4} \Bigl[ \tr\bigl(\CT^{\,2}\bigr) - \tfrac{1}{p-1} \tr(\CT)^2 - \tfrac{1}{2} \, \tr\bigl(\CF^4\bigr) + \tfrac{1}{4 \, (p-1)} \, \tr\bigl(\CF^2\bigr)^2 \Bigr] + O(\lambdap)\,.
\end{split}
\end{align}
This zeroth-order result in fact holds for all $p \neq 1$\,. 

\end{enumerate}

\subsection*{Summary of Flow Equations} 

In order to provide a more succinct reference for field-theoretical applications, we now summarise the above flow equations in flat spacetime, setting $\tau_{\alpha \beta} = \eta_{\alpha \beta}$.

\vspace{3mm}

\noindent $\bullet$~\emph{Without field strength terms.} 
We obtained the following flow equations defined for $(p+1)$-dimensional field theories purely in terms of the energy-momentum tensor $T_{\alpha \beta}(\lambdap)$:
\be \label{eq:ffe}
    \frac{\p \CL}{\p \lambdap} = \frac{1}{2 \, \lambdap^2} \, \biggl\{ \tr \, (\mathbb{1} - \lambdap\, \CT)  - \bigl(p-1\bigr) \, \Bigl[ \det (\mathbb{1} - \lambdap\, \CT)\Bigr]^{\frac{1}{p-1}} - 2 \biggr\}\,,
\ee
where $\CT$ denotes $\CT^\alpha{}_\beta \equiv \eta^{\alpha \gamma} \, T_{\gamma \beta}$\,. Note that $\det \CT = -\det T$\,.

Our construction demonstrates the following example of an explicit solution to this equation in arbitrary dimensions.
Given the initial Lagrangian corresponding to $D$ free scalar fields in ($p+1$) dimensions:
\be \label{eq:lzfa}
    \CL(0) = - \frac{1}{2} \, \p_\alpha X^{A'} \p^\alpha X^{A'},
\ee
the solution of the above flow equation is
\be \label{eq:lt}
    \mathcal{L} (\lambdap) = \frac{1}{\lambdap} \, \left[
    1  - 
    \sqrt{-\det \Bigl( \eta_{\alpha \beta}  + \lambdap \, \partial_\alpha X^{A'} \partial_\beta X^{A'} \Bigr)}\, \right],
\ee
corresponding to a gauge fixed $p$-brane in $D+p+1$ dimensions.

\vspace{3mm}

\noindent $\bullet$~\emph{With field strength terms.} 
We also obtained flow equations for $p=1$ and $p=2$ that can be applied to theories containing scalars and a gauge field with field strength $F^\alpha{}^{}_{\beta}$\,. 
These equations were
\begin{subequations}
\label{eq:polfd}
\begin{align}
    p & = 1: 
        \quad%
    \frac{\p \CL}{\p \lambdap}  =- \frac{1}{2 \, \lambdap}  \left( 
       \,\tr \, \CT
        +
        \sqrt{- \det {F}} \, \sqrt{- \tr\,\CT + \lambdap \det \CT}
        \,\right) \,, \\[4pt]
    p & = 2: 
        \quad%
    \frac{\p \CL}{\p \lambdap} = \frac{1}{4} \Bigl[ \tr \bigl( \CT^2 \bigr) - \bigl( \tr \, \CT \bigr)^2 \Bigr] +\frac{\lambdap}{2} \det \CT\,. \label{eq:ptttbfe}
\end{align}
\end{subequations}
Note that the $p=1$ equation reduces to the standard form of $T\bar{T}$ deformation~\eqref{eq:ttbarfe} in two dimensions when $F$ is set to zero. When $p = 3$\,, we gave the flow equation up to the linear order in $\lambdap$ in Eq.~\eqref{eq:fept}. For $p > 3$\,, the zeroth-order terms in the flow equations are given in Eq.~\eqref{eq:fehp}. 
In general, given the initial Lagrangian~\eqref{eq:undel} corresponding to $D$ free scalar fields and a free Maxwell gauge field, the solution to such flow equations will generate the DBI Lagrangian \eqref{eq:defundel} corresponding to a gauge fixed D$p$-brane in $D+p+1$ dimensions.

\section{Outlook} \label{sec:ol}

In this paper we developed a framework that brings together a wide range of topics including matrix theory, holography, non-Lorentzian geometry, and $T \bar T$ deformations. We have discussed how they are all linked to the ongoing program of studying the space of BPS decoupling limits in string theory and M-theory.
This is the first of a series of papers where we revisit matrix theory, with the focus on the BPS nature of the associated decoupling limit and its implication for holography. In an upcoming paper~\cite{MxReloadedII}, we will discuss further geometric aspects of the decoupling limits that we have considered here, as well as the finer details of the (T-, S- and U-) duality relationships between different limits.
Further follow-up papers will focus on the dynamics of string theory in multicritical decoupling limits (involving both brane-brane and string-brane limits), and on decoupling limits that lead to tensionless and Carroll strings, as set out in~\cite{Blair:2023noj} and further probed using the string worldsheet in~\cite{Gomis:2023eav}. 

In the rest of this Outlook, we comment on a few topics directly related to what we have discussed in the current paper.

\vspace{3mm}
\noindent $\bullet$ \emph{Correspondence between supergravity and matrix quantum mechanics.}
The BFSS conjecture states that matrix theory \eqref{eq:BFSS} in the limit $N\rightarrow \infty$, $R \rightarrow \infty$ describes M-theory in eleven-dimensional spacetime \cite{Banks:1996vh}.
This follows as a limit of the stronger claim that matrix theory at finite $N$ describes M-theory in the DLCQ, with $N$ units of momentum on the null circle \cite{Susskind:1997cw, Seiberg:1997ad, Sen:1997we}.

The classic checks of the BFSS conjecture consisted of computing scattering amplitudes in matrix theory and comparing them to supergraviton scattering in eleven-dimensional supergravity, see \cite{Taylor:2001vb} for a detailed survey.  
Even though there is impressive agreement for particular amplitudes, generically the matrix theory computations do not agree with supergravity. For instance, there are known discrepancies that arise from comparing higher-point and higher-loop amplitudes in matrix theory with supergravity, especially when higher-curvature terms are included (see~\emph{e.g.}~\cite{Helling:1999js}).
This can however be understood by noting that comparing matrix theory with the \emph{supergravity} regime is not guaranteed to be in correspondence.
Supergravity is a good effective description of DLCQ M-theory when the radius of the null compactification is large, while perturbative matrix theory is only valid when the radius of the null compactification is small. Therefore, one would only expect that the results match when there exists a nonrenormalisation theorem, such that the amplitude is protected against the RG flow between two different energy scales~\cite{Helling:1999js}. The recent supersymmetric index computation in~\cite{Herderschee:2023bnc} of three-point amplitudes in matrix theory is one of such examples where a protected object can be identified. The conclusion is then that general amplitudes will only match at large $N$, which is difficult as it requires nonperturbative control on the matrix theory side. 

In contrast, at finite $N$ it appears to be more doable to understand the Berenstein-Maldacena-Nastase (BMN) conjecture relating M-theory on a pp-wave background and matrix quantum mechanics with a mass deformation~\cite{Berenstein:2002jq}. Recently, it was shown in~\cite{Komatsu:2024vnb} that the spectrum of eleven-dimensional supergravity on the pp-wave background can be exactly matched to the spectrum of the BMN model at finite $N$. 

Returning to the original proposal of the correspondence between BFSS matrix quantum mechanics and eleven-dimensional supergravity on flat spacetime, our framework seems to provide a complementary perspective at finite $N$:\,\footnote{At finite $N$, BFSS matrix theory corresponds to M-theory in the DLCQ, where there is a null compactification with a finite effective radius. While the DLCQ of a QFT is normally not well-defined quantum mechanically~\cite{Hellerman:1997yu, Chapman:2020vtn}, it can be a sensible procedure in string and M-theory \cite{Seiberg:1997ad,Bilal:1998vq}. 
Furthermore, compactifying DLCQ M-theory over an extra spatial circle leads to the DLCQ of type IIA superstring theory. This is related to type IIB non-relativistic string theory via a T-duality transformation, giving a first principles definition of the DLCQ. See~\cite{Gomis:2000bd, Danielsson:2000gi} for early works and~\cite{Bergshoeff:2018yvt, Gomis:2020izd, Yan:2021hte} for recent progress. As non-relativistic string theory is unitary and UV-complete~\cite{Gomis:2000bd}, \emph{no} pathology should be present at least perturbatively. See also~\cite{Harmark:2017rpg, Kluson:2018egd,  Harmark:2018cdl, Harmark:2019upf} from the perspective of null reduction and~\cite{Ko:2015rha, Morand:2017fnv} for connections with double field theory.} it is true that the eleven-dimensional Lorentzian supergravity is no longer valid when the null compactification in DLCQ M-theory is small; however, the singular behaviour that this supergravity develops turn out to be describable in terms of the non-Lorentzian geometric data in M0T. Furthermore, as we have noted in Section~\ref{sec:matrixstringtheory}, M0T is dual to non-relativistic string theory, where the latter is a perturbative string theory. The first quantisation of such a non-relativistic string has been studied in~\cite{Gomis:2000bd} from first principles using standard CFT techniques, while the dynamics of the target space (super)gravity was later obtained from beta-function calculations~\cite{Gomis:2019zyu, Gallegos:2019icg, Bergshoeff:2019pij, Yan:2019xsf, Yan:2021lbe}.\,\footnote{The dilatation symmetries that we discussed around Eq.~\eqref{eq:dilatsmpt} are manifest in the worldsheet formalism~\cite{Bergshoeff:2018yvt, Bergshoeff:2021bmc}, which underlies their role in non-Lorentzian supergravity from first principles. Also see~\cite{Gomis:2020fui, Gomis:2020izd, Ebert:2021mfu} for first principles studies of D-branes in non-relativistic string theory.} The second-quantised version of non-relativistic string theory is matrix string theory~\cite{Motl:1997th, Dijkgraaf:1997vv}, which is for example used in~\cite{Herderschee:2023bnc} to relate the three-point amplitude in BFSS matrix theory to a supersymmetric index. One would therefore like to map the physical observables of non-relativistic string theory to BFSS matrix theory, and therefore better understand the latter in a general setting. In particular, at finite $N$\,, BFSS matrix theory should describe scattering of particles on a non-Lorentzian background, which can be mapped to the scattering between winding strings on a dual non-Lorentzian background in non-relativistic string theory. 
It would be interesting to revisit the correspondence between supergravity and matrix quantum mechanics taking into account this non-Lorentzian perspective. 

\vspace{3mm}
\noindent $\bullet$ \emph{Elements of the AdS/CFT correspondence.}
We proposed that a DLCQ${}^n$/DLCQ${}^m$ correspondence captures the notion of AdS/CFT in string/M-theory as set out in Section~\ref{sec:chdlcq}.
We particularly focused on how versions of holography arising from stacks of a single type of brane can be viewed as the case DLCQ${}^n$/DLCQ${}^{n+1}$.
The additional DLCQ at infinity corresponds in the bulk to the near-horizon limit of the brane.
These examples include those based on D-branes discussed throughout Section \ref{sec:hnhbpsl} and on the M2- and M5-brane discussed briefly in Section \ref{sec:adsdlcq}. 
For $n=0$\,, these examples include the classic AdS/CFT correspondences arising from the D3, M2 and M5. 

We further illustrated how DLCQ${}^n$/DLCQ${}^m$ with $m>n+1$ can arise by focusing on the well-known example of the D1-D5 near-horizon limit, which leads to an AdS${}_3$/CFT${}_2$ correspondence, in Section \ref{sec:dlcqmn}.
It would be interesting to explore more systematically possible decoupling limits that are linked to more general intersecting brane configurations (including the ones with three or more brane charges). 
It is clear that there are deep links between the classification of supersymmetric brane intersections and the allowed M$p$T limits of D$q$-branes, as can be inferred from a scan of the possibilities shown (for solutions with a single D$q$-brane stack) in Table \ref{tab:IIbranes}.
This points the way to a deeper web of relationships between brane configurations and possible decoupling limits.

We have shown that, in all examples, the near-horizon geometry becomes non-Lorentzian at the asymptotic infinity. However, there are a few subtleties to this claim. One subtlety is that, when one approaches the asymptotic infinity, non-geometric effects (such as quantum corrections) may dominate~\cite{Itzhaki:1998dd}, which could obscure the geometric interpretation that we gave. This subtlety can be argued around by going to the matrix theory side with a stack of D-branes in flat spacetime, where backreactions are suppressed when the 't Hooft coupling is sufficiently small. In this string picture, the M$p$T geometry is valid. 
Also note that in the cases with $p>3$, we see from \eqref{eq:dsco} that we go to strong coupling at asymptotic infinity.
For $p=4$, this means that we should pass from the description in terms of an M4T limit in ten dimensions to one in eleven dimensions involving the M5-brane decoupling limit. 
For $p=5$, we should similarly exchange the description in terms of an M5T limit to an S-dual one involving an NS5-brane decoupling limit.

It would be also interesting to understand how indispensable this non-Lorentzian description is at the asymptotic infinity, and to which degree the singular behaviour of the metric here is coordinate-dependent.\,\footnote{For example, it was shown in~\cite{Gomis:2023eav} that the fundamental string worldsheet in M$p$T naturally acquires a non-Lorentzian geometric description, while it is topologically equivalent to a nodal Riemann sphere. This means that the two-dimensional non-Lorentzian geometry can be recast in terms of a metric on a Riemannian manifold except on a discrete set of points.} In particular, our discussion is made only on the (analogue of the) Poincar\'{e} patch, and it would be important to understand how to phrase such non-Lorentzian geometries at the asymptotic infinity when global coordinates are taken for the bulk near-horizon geometry.

\vspace{3mm}
\noindent $\bullet$ 
\emph{Non-Lorentzian holography.} 
In Section~\ref{sec:hnhbpsl}, we presented a systematic approach for classifying possible holographic duals using the BPS decoupling limit. In particular, we identified the near-horizon geometries (including examples analogous to AdS geometries) for curved M$p$T backgrounds, where the bulk geometry is non-Lorentzian. A natural and important calculation here would be to compute the isometry group of this type of non-Lorentzian geometries. 
Moreover, it would also be interesting to study non-Lorentzian holography at a finite temperature in this context.  

In Section~\ref{sec:sln}, we generated non-Lorentzian bulk geometries using the trick (following \cite{Avila:2023aey,Lambert:2024uue,Lambert:2024yjk,Fontanella:2024rvn}) of introducing by hand an extra rescaling of the characteristic length $\ell$ appearing in the harmonic function $H=1+(\ell/r)^{7-q}$ in terms of the parameter $\omega$.
This is essentially a rescaling of the number $N$ of the D-branes and thus implies a large $N$ limit when $\omega$ is sent to infinity. It would be important to understand how one should interpret this rescaling of $N$\,, and its interpretation in terms of physical sources for the electric and magnetic charges of the brane solution. 
Furthermore, we showed that while this rescaling leads to D$q$-brane geometries localised in the longitudinal directions of M$p$T, alternative D$q$-brane geometries could be found by smearing on these directions, leading to solutions localised in the transverse directions of M$p$T.
A more thorough analysis of the properties of these solutions -- in particular their (super)symmetries and charges -- is needed to understand which are physically relevant for holography. 
We also did not address the crucial point of verifying that these geometries solve the equations of motion
and torsional constraints of the M$p$T limit of supergravity---although extrapolating from the M2-brane geometry in non-relativistic M-theory analysed in \cite{Lambert:2024uue, Bergshoeff:2024nin}, one would expect this should be the case. It is important to further clarify the status of these geometries and look for new examples, by explicitly formulating M$p$T supergravity and its equations of motion.

\vspace{3mm}
\noindent $\bullet$ 
\emph{New $p$-brane generalisations of the $T \bar T$ deformation.}
In Section~\ref{sec:TTbarNG}, we discussed how the bulk geometry in AdS/CFT is generated from a deformation of the flat non-Lorentzian geometry at the asymptotic infinity, and showed that this deformation is dual to the $T\bar{T}$ deformation in matrix/non-relativistic string theory. This observation reveals a general philosophy in the construction of holographic duals. We start with (multiple) BPS decoupling limits of type II superstring theory in flat spacetime, which lead to a self-contained corner of string theory with a foliated target space. The open string excitations on a flat D-brane in this foliated flat spacetime correspond to the closed string states associated with the $p$-brane $T\bar{T}$ deformation of the same foliated flat spacetime. Moreover, we have seen that this is an open/closed string duality only if we pass on to the dual frame described by matrix/non-relativistic string theory. More precisely, the usual lore of open/closed string duality is typically recast in the form of open/closed brane duality, as the fundamental string is only perturbative in the matrix/non-relativistic string frame.  

The way in which the (generalised) $T \bar T$ deformation appears in this paper is distinct from its previous roles in holography.
The usual $T \bar T$ deformation is a `double-trace' operator and adding it to a dual CFT modifies the boundary conditions for the AdS metric \cite{McGough:2016lol,Guica:2019nzm}. 
There also exists a `single-trace' version of the $T \bar T$ deformation in the context of the holographic dual of the near-horizon limit of the F1-NS5 supergravity solution \cite{Giveon:2017nie}. 
In the bulk, this deformation serves to undo the F1 near-horizon limit, taking the geometry from AdS to an asymptotic linear dilaton spacetime (\emph{i.e.}~we stay in the near-horizon limit of the NS5).
Meanwhile our observation in this paper is that, with a fixed dual field theory (associated with an asymptotic non-Lorentzian geometry), the bulk geometry itself can be generated by a geometric realisation of the generalised $p$-brane $T \bar T$ deformations.
It would be interesting to seek to combine these different perspectives.
In this paper, while we mostly considered holographic geometries arising from configurations with a single stack of D$p$-branes, we saw in Section \ref{sec:dlcqmn} how the near-horizon geometry of the D1-D5 solution was generated by a combined M1T-M5T decoupling limit. Partially undoing this near-horizon limit via the D1-brane generalised $T \bar T$ deformation would be the S-dual of the $T \bar T$ deformation appearing in F1-NS5 case of \cite{Giveon:2017nie}. It would therefore be interesting to explore the role of our new deformations in understanding the holographic landscape in this and other examples.

A potentially diverting question is whether one can find other geometric realisations of the generalised $p$-brane flow, leading to geometries where \emph{e.g.}~an M$p$T limit is realised as one approaches an apparently singular locus. One unusual class of examples, following from the string case noted in~\cite{Blair:2020ops}, is provided by geometries sourced by negative tension branes~\cite{Dijkgraaf:2016lym}.

In Section~\ref{sec:pbttfe}, we derived a collection of $p$-brane generalisations of the $T\bar T$ deformation. 
The essential trick to obtain the relevant flow equations was to consider the appropriate BPS decoupling limits of various branes (including both D$p$-branes and the M-theory membrane) and read off the dependence of the action on the parameter $\lambdap = \omega^{-2}$. 
Our examples were all bosonic and abelian, but it would be interesting to consider supersymmetric (see \emph{e.g.}~\cite{Baggio:2018rpv,Chang:2018dge,Ferko:2023sps}) and non-abelian cases. 
We could approach the non-abelian case either by seeking to apply our new flow equations directly to SYM as the initial Lagrangian, or by starting with the desired answer, \emph{i.e.}~the action~\eqref{Myers} parametrised in terms of the M$p$T prescription.

For $p=2$, we saw that the same flow equation~\eqref{M2flow} appeared for the abelian D2-brane (with a worldvolume gauge field) and the M2-brane (with only worldvolume scalars). This three-dimensional flow equation presents a strikingly simple generalisation of the usual $T\bar T$ deformation in two dimensions. 
Focusing on the M2-brane application, it would be intriguing to see 
whether the three-dimensional flow equation~\eqref{M2flow} can be applied to \emph{e.g.}~the BLG and ABJM field theories, and whether the endpoint could be some M-theoretic version of the non-abelian DBI action.

If generalised to the non-abelian case, our realisation of the $p$-brane $T\bar{T}$ deformations would deform matrix theories to the DBI actions. Uplifted to M-theory, this deformation essentially undoes the infinite momentum limit that leads to the DLCQ. On the other hand, in the string case, an expansion with respect to the parameter $\lambdap$ that controls the $T\bar{T}$ deformation of non-relativistic string theory is reminiscent of recent work on the non-relativistic expansions of string theory~\cite{Hartong:2021ekg, Hartong:2022dsx, Hartong:2024ydv}, generalising the large speed of light expansion of general relativity \cite{VandenBleeken:2017rij, Hansen:2019pkl, Hansen:2020pqs}. It would be interesting to also explore the expansions of type II superstring theory with respect to the M$p$T limit using the $p$-brane $T\bar{T}$ deformation, which may eventually help us better probe M-theory in flat spacetime. 

Beyond the brane context, our flow equations are crying out to be used to study deformations of more general $p$-dimensional field theories. Furthermore, similar ideas may be applied to gravity, where it is natural to seek flow equations that deform non-Lorentzian gravities towards General Relativity. This is reminiscent of `non-Lorentzian bootstrap' as proposed in~\cite{Bergshoeff:2023ogz}, where, focusing on the bosonic sector, IIB supergravity is recovered by `bootstrapping' the non-Lorentzian supergravity in type IIB non-relativistic superstring theory. Here, the bosonic part of non-Lorentzian IIB supergravity is constructed as an effective field theory using non-Lorentzian symmetries, which have intriguing connections to invariant theory in algebra. This program of non-Lorentzian bootstrap may have synergies with the studies of expanding General Relativity~\cite{VandenBleeken:2017rij, Hansen:2019pkl, Hansen:2020pqs, Hansen:2021fxi}, \emph{e.g.}~in the context of post-Newtonian approximation~\cite{Hartong:2023ckn}.

\vspace{3mm}
\noindent $\bullet$ 
\emph{Algebraic aspects of matrix $p$-brane theory.} %
In Section \ref{sec:alg}, we exhibited an algebraic perspective on the BPS decoupling limit, focussing for definiteness on  M0T.  In particular, this decoupling sector is governed by a  maximally extended version of the super-Bargmann algebra which we call the M0T superalgebra. This novel superalgebra can also be obtained from an appropriate $\dot{\text{I}}$nönü-Wigner contraction of the type IIA superalgebra, adapted to the D0-brane BPS limit. Similar algebras exist for each M$p$T (see~\cite{Brugues:2006yd} for a precursor), as well as for other corners of the larger web of decoupling limits~\cite{Blair:2023noj, Gomis:2023eav}, and are obtainable in principle via U-duality transformations and further $\dot{\text{I}}$nönü-Wigner contractions (one for each DLCQ). 
In particular, this should allow one to generate the supersymmetry algebra for type IIA/IIB non-relativistic string theory. 
We note that \cite{Bidussi:2021ujm}
obtained the fundamental string Galilean algebra, relevant to non-relativistic strings and their coupling to torsional string Newton-Cartan geometry. The fundamental string Galilean algebra is an $\dot{\text{I}}$nönü-Wigner contraction of the string-Poincar\'e algebra and the M0T algebra we presented should thus be U-dual to a (putative) supersymmetric extension of this bosonic algebra.
Furthermore, algebras related to further $\dot{\text{I}}$nönü-Wigner contractions, connected to longitudinal Galilean and Carrollian limits of non-relativistic strings, have appeared in \cite{Bidussi:2023rfs}. It would be interesting to study further extensions of these algebras in the larger duality web of decoupling limits along the lines of~\cite{Blair:2023noj, Gomis:2023eav}, as well as other Carroll-like algebras for M$p$T with $p < -1$ (see below).

\vspace{3mm}

\noindent $\bullet$ 
\emph{Extensions: tensionless, Carrollian, heterotic, etc.} We have been focusing on the BPS decoupling limits of type II superstring theory in this paper. A series of generalisations to other string theories are anticipated. 

In~\cite{Blair:2023noj, Gomis:2023eav}, it is shown how the decoupling limits of type II superstring theory are related to the ones in type II${}^*$ superstring theory~\cite{Hull:1998vg} via a timelike T-duality transformation. In particular, performing a timelike T-duality transformation of M0T maps the theory to matrix (-1)-brane theory (M(-1)T), where the fundamental degrees of freedom are instantons described by Ishibashi-Kawai-Kitazawa-Tsuchiya (IKKT) matrix theory~\cite{Ishibashi:1996xs}. Moreover, the worldsheet theory describing the fundamental string in M(-1)T~\cite{Gomis:2023eav} coincides with the one associated with tensionless string theory~\cite{Lindstrom:1990qb, Isberg:1993av}. Further spacelike T-duality transformations of M(-1)T lead to M$p$T with $p < -1$~\cite{Blair:2023noj, Gomis:2023eav}, where the fundamental degrees of freedom are Euclidean (or spacelike) branes~\cite{Hull:1998vg, Gutperle:2002ai}. The dynamics of such Euclidean branes is supposedly described by a new matrix theory dual to IKKT matrix theory. In M$p$T with $p < -1$\,, the target space is Carroll-like and has a foliation with $SO(1\,,\,10+p) \times SO(-p-1)$ symmetry, in which the Carrollian boost symmetry relates the longitudinal and transverse directions (as opposed to Galilean boosts)~\cite{Blair:2023noj,Gomis:2023eav}. 
See~\cite{Cardona:2016ytk,Bagchi:2023cfp,Harksen:2024bnh,Bagchi:2024rje} for related work on Carroll strings. A comprehensive understanding of these novel corners of type II${}^*$ superstring theory may shed light on a top-down formulation of Carrollian and celestial holography~\cite{Donnay:2022aba, Donnay:2023mrd}. These lines of considerations also open up the possibility of holographic duals with a Carrollian bulk geometry. 

Finally, it would be interesting to apply the techniques that we have developed to heterotic string theory and its DLCQ~\cite{Danielsson:1996es, Kachru:1996nd, Banks:1997it, Motl:1997tb}, where new structures may arise. Initial efforts along these lines are made in~\cite{Bergshoeff:2023fcf,Lescano:2024url, hete}.

\section*{Acknowledgements}

We would like to thank David Berenstein, Eric Bergshoeff, Jan de Boer, Joaquim Gomis, Troels Harmark, Jelle Hartong, Emil Have, Elias Kiritsis, Shota Komatsu, Neil Lambert, Juan Maldacena, Jeong-Hyuck Park, Ronnie Rodgers, Jan Rosseel and Erik Verlinde for useful discussions. N.O. would like to thank Robbert Dijkgraaf, Jan de Boer and Troels Harmark for collaboration on~\cite{bpslimits}. 
Z.Y. would like to thank Kevin T. Grosvenor for useful discussions and collaborations on an early stage of the development for the holographic aspects in this paper. Z.Y. would also like to thank the organizers and participants of the workshop on \emph{Matrix quantum mechanics for M-theory revisited} at CERN (January, 2024) for stimulating discussions that were useful for this work. 
The work of C.B. is supported through the grants CEX2020-001007-S and PID2021-123017NB-I00, funded by MCIN/AEI/10.13039/501100011033 and by ERDF A way of making Europe. 
The work of N.O. is supported in part by VR project Grant 2021-04013 and Villum Foundation Experiment Project No.~00050317. 
The work of Z.Y. is supported in part by the European Union’s Horizon 2020 research and innovation programme under the Marie Sklodowska-Curie Grant Agreement No.~31003710, and in part by VR project Grant 2021-04013. Nordita is supported in part by NordForsk.  

\bibliographystyle{JHEP}
\bibliography{mtr}

\end{document}